\documentclass[12pt]{article}
\usepackage{amsfonts,amssymb,epsfig,amsmath}
\usepackage{color}

\renewcommand{\baselinestretch}{1.2}

\def\det{{\rm det}}

\newcommand{\be}{\begin{eqnarray}}
\newcommand{\ee}{\end{eqnarray}}

\newcommand{\bn}{\begin{enumerate}}
\newcommand{\en}{\end{enumerate}}

\newcommand{\chisixteen}[1]{\chi_{\mathbf{#1}}^{\text{SO}(16)}}

\begin{document}

\makeatletter \@addtoreset{equation}{section} \makeatother
\renewcommand{\theequation}{\thesection.\arabic{equation}}
\renewcommand{\thefootnote}{\alph{footnote}}

\begin{titlepage}

\begin{center}
\hfill {\tt SNUTP14-010}\\
\hfill {\tt KIAS-P14063}\\

\vspace{2cm}

{\Large\bf Elliptic Genus of E-strings}

\vspace{2cm}

\renewcommand{\thefootnote}{\alph{footnote}}

{\large Joonho Kim$^1$, Seok Kim$^1$, Kimyeong Lee$^2$, Jaemo Park$^3$,
Cumrun Vafa$^4$}

\vspace{0.7cm}

\textit{$^1$Department of Physics and Astronomy \& Center for
Theoretical Physics,\\
Seoul National University, Seoul 151-747, Korea.}\\

\vspace{0.2cm}

\textit{$^2$School of Physics, Korea Institute for Advanced Study,
Seoul 130-722, Korea.}\\

\vspace{0.2cm}

\textit{$^3$Department of Physics, Postech, Pohang 790-784, Korea.}\\

\vspace{0.2cm}

\textit{$^4$ Jefferson Physical Laboratory, Harvard University, Cambridge,
MA 02138, USA.}\\

\vspace{0.7cm}

E-mails: {\tt joonho0@snu.ac.kr, skim@phya.snu.ac.kr,
klee@kias.re.kr, jaemo@postech.ac.kr, vafa@physics.harvard.edu}

\end{center}

\vspace{1cm}

\begin{abstract}

We study a family of 2d $\mathcal{N}=(0,4)$ gauge theories which describes
at low energy the dynamics of E-strings, the M2-branes suspended between a
pair of M5 and M9 branes. The gauge theory is engineered using a duality with
type IIA theory, leading to the D2-branes suspended between
an NS5-brane and 8 D8-branes on an O8-plane. We compute the elliptic
genus of this family of theories, and find agreement with the known results
for single and two E-strings. The partition function can in principle be
computed for arbitrary number of E-strings, and we compute them explicitly for
low numbers.  We test our predictions against the partially known results from
topological strings, as well as from the instanton calculus of 5d $Sp(1)$
gauge theory.  Given the relation to topological strings, our computation provides
the all genus partition function of the refined topological strings on
the canonical bundle over ${1\over 2}K3$.

\end{abstract}

\end{titlepage}

\renewcommand{\thefootnote}{\arabic{footnote}}

\setcounter{footnote}{0}

\renewcommand{\baselinestretch}{1}

\tableofcontents

\renewcommand{\baselinestretch}{1.2}

\section{Introduction}

Six dimensional superconformal theories with (2,0) and (1,0) supersymmetry
enjoy a special status among all superconformal theories:  they are at the highest
possible dimension.  They play a key role in various aspects of string dualities
as well as in obtaining lower dimensional supersymmetric systems upon compactification.
They are rather enigmatic as they include tensionless self-dual strings as their
building blocks.

The study of these theories has recently intensified, leading to computations of their
superconformal indices \cite{Kim:2012ava,Lockhart:2012vp,Kim:2012qf,Kim:2013nva}, the
elliptic genera of the self-dual strings in the Coulomb branch
\cite{Haghighat:2013gba,Haghighat:2013tka,Haghighat:2014pva} (see \cite{Kim:2011mv}
for an earlier work), as well as a partial classification of 6d superconformal theories
\cite{Heckman:2013pva,Gaiotto:2014lca,DelZotto:2014hpa}.
The aim of this paper is to take a step forward in this direction, in particular focusing
on one of the most basic $(1,0)$ superconformal theories.  The theory is known
to arise in heterotic strings for small $E_8$ instantons
\cite{Witten:1995gx,Ganor:1996mu,Seiberg:1996vs}, and also when an M5 brane approaches
the M9 brane boundary \cite{Ganor:1996mu,Seiberg:1996vs}. It also has an F-theory dual
description given by blowing up a point on $\mathbb{C}^2$ base of F-theory \cite{Witten:1996qb,Morrison:1996na,Morrison:1996pp}.   This superconformal theory has an
$E_8$ global symmetry.  It also has a one dimensional Coulomb branch, parameterized by a
real scalar in the (1,0) tensor multiplet. In the M-theory setup, the scalar parameterizes
the distance between M5 and M9 branes \cite{Horava:1995qa}.
In F-theory setup, it parameterizes the size of the $\mathbb{P}^1$ obtained by blowing up a point.
On the Coulomb branch this theory has light strings, known as E-strings \cite{Klemm:1996hh}.
In the M-theory setup they arise by M2 branes stretched between M5 brane and M9 brane.
In F-theory setup they arise by wrapping D3 branes on the blown up
$\mathbb{P}^1$. It is natural to ask whether one can find a nice description of E-strings.
The main aim of this paper is to find such a description and use it to compute the twisted
partition function of such strings on $T^2$.  More precisely we would be computing the
elliptic genus of E-strings on $T^2$. Knowing the elliptic genus of E-strings is useful in
its own right, as well as for uncovering aspects of the superconformal theory. For example,
a basic quantity one may wish to compute for a superconformal theory is its superconformal
index, which involves the computation of its partition function on $S^1\times S^5$ with
suitable fugacities turned on along $S^1$. As was argued in \cite{Lockhart:2012vp,Kim:2012qf} (see also \cite{Qiu:2013aga,Qiu:2014oqa}), the computation of the superconformal index reduces to an integral over the Coulomb branch where the integrand consists of three copies of
elliptic genus of the corresponding strings.

If one is computing supersymmetry protected quantities, such as elliptic genus, we can
change parameters to make the computation easy.  In particular one can change parameters
and use string dualities to find a suitable description of the resulting strings.  This
strategy was employed in particular for M-strings and their orbifolds \cite{Haghighat:2013gba,Haghighat:2013tka}.  Two basic ways were used to compute
the elliptic genus of the M-strings: one was to use string dualities to map the 2d theory
to a super-Yang-Mills type gauge theory and use the technique developed recently
\cite{Gadde:2013dda,Benini:2013nda,Benini:2013xpa} to compute their elliptic genera.
The other way was to use the relation of the elliptic genus to BPS quantities upon circle
compactification of these theories, that can in principle be computed using topological strings.

In the context of E-strings we employ the former method, and identify the gauge theory
which captures their low energy physics.  This is done by considering the duality of
M-theory with type IIA, by introducing a circle transverse to M5 brane, leading to a
system involving NS5-brane and where the M9 brane is replaced by O8 plane with 8
D8 branes on it. The M2 branes suspended between M5 and M9 branes map to D2 branes
suspended between NS5-brane and O8-D8 pair. We find a simple $(0,4)$ supersymmetric
quiver describing this system with $O(n)$ gauge symmetry, where $n$ denotes the number
of suspended M2 branes.  We use it to compute the elliptic genus of $n$ E-strings by
employing the techniques developed in \cite{Benini:2013nda,Benini:2013xpa}.

The other method of computing the elliptic genus of E-string involves the F-theory picture.  Namely, we compactify the theory on a circle leading to an M-theory description, and consider the BPS states of wrapped M2 branes, which correspond to E-strings wound around $S^1$
\cite{Vafa:1996xn}.  M-theory geometry involves the canonical bundle over ${1\over 2} K3$.
As is well known, the BPS states of M2 branes wrapped on it, are captured by topological string amplitudes \cite{Gopakumar:1998ii,Gopakumar:1998jq}.  In this context
the (refined) topological string for ${1\over 2}K3$ has been computed to a high genus \cite{Hosono:1999qc,Huang:2013yta}, though an all genus answer is not available.
So our method leads to a complete answer for refined topological string on ${1\over 2} K3$.
Our answer can also be related to ${\cal N}=4$ Yang-Mills in $d=4$ in two different ways.
In the F-theory setup, E-strings arise by wrapping D3 branes on a $\mathbb{P}^1$.  From this perspective the elliptic genus of $n$ E-strings gets mapped to the study of $n$ D3 branes on
$T^2\times \mathbb{P}^1$, i.e. the partition function of ${\cal N}=4$ $U(n)$ Yang-Mills on this geometry.  Except that the coupling constant of Yang-Mills $\tau$ is not a constant and
varies over $\mathbb{P}^1$ according to the complex
structure of the elliptic curve given by
$$y^2=x^3+f_4(z)x +g_6(z)$$
where $z$ parameterizes the $\mathbb{P}^1$ and $f_4$ and $g_6$ are polynomials of degree
$4$ and $6$ respectively.
Note that this takes into account the S-duality of $U(n)$ Yang-Mills.  Moreover lifting this to M-theory
leads to $n$ M5 branes on $T^2\times {1\over 2}K3$, which gets mapped to $U(n)$ ${\cal N}=4$ Yang-Mills on ${1\over 2} K3$ \cite{Minahan:1998vr}  (for the $SU(2)$ case see \cite{Yoshioka:1998ti} and for computations in related cases see \cite{Manschot:2014cca}).

Explicit computations for the elliptic genus are now straightforward, but somewhat cumbersome.  Nevertheless we carry it out explicitly for the case of $n$ E-strings for $n=1,2,3,4$, and also explain the concrete procedures needed to compute the elliptic genus in the case with
general $n$.  The case with $n=1$ was already known in \cite{Klemm:1996hh}, and the case
with $n=2$ was recently found in \cite{Haghighat:2014pva}.  For the other two cases we
check our results against partial results from topological strings on ${1\over 2}K3$ (where
low genus answer is known). We also check them at $n=4$ against a recent proposal of
\cite{Sakai:2014hsa}, where the elliptic genus was proposed at a special value of
$E_8$ fugacities with reduced symmetry $SO(8)\times SO(8)\subset E_8$.
In all these cases we find agreements with our computations.

Finally, we explain an alternative method to compute the E-string elliptic genus, from
the instanton calculus of 5d SYM theories with $Sp(1)$ gauge group and $8$ fundamental
hypermultiplets. The index for $k$ instantons captures the $k$'th order coefficient of
the elliptic genus expanded in the modular parameter, but keeps the information on all
higher E-strings' spectrum at this order. It was recently shown in \cite{Hwang:2014uwa}
how to compute this index. Making double expansions of the indices of
our 2d gauge theory and the instanton quantum mechanics, we confirm that the indices
computed from the two approaches agree with each other.

The organization of this paper is as follows:  In section 2 we describe the basic
type IIA brane setup.  In section 3 we use this to compute the elliptic genera
of E-strings.  We give the explicit details for $1,2,3,4$ E-strings and indicate how the
higher case works.  We also compare with (partial) known results.  In section 4 we also formulate how the E-string partition function can be computed using 5 dimensional
Yang-Mills instantons, and compare the results with those obtained in section 3.  In section 5 we
present some concluding remarks.
Some technical details are relegated to the appendices.

\section{The brane setup and the 2d $(0,4)$ gauge theories}

\begin{figure}[t!]
  \begin{center}
    \includegraphics[width=10cm]{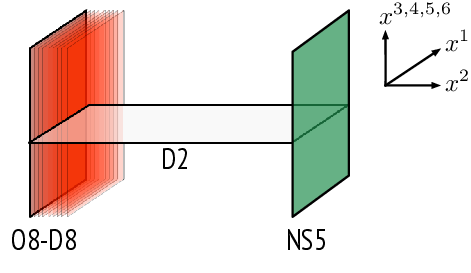}
\caption{The type IIA brane configuration for the E-strings.}\label{brane-figure}
  \end{center}
\end{figure}

We construct a brane system in the type IIA string theory, which at low energy
engineers the 6d $E_8$ SCFT and the 2d CFT for E-strings. We first take
an NS5-brane to wrap the $013456$ directions, located at $x^2=L$ $(>0)$,
$x^7=x^8=x^9=0$. An O8-plane and $8$ D8-branes (or $16$ D8-branes in the covering space
of orientifold) wrap $013456789$ directions, located at $x^2=0$. To describe E-strings,
$n$ D2-branes are stretched between the NS5 and 8-brane system ($0<x^2<L$),
occupying $012$ directions. This brane
system has $SO(4)\sim SU(2)_L\times SU(2)_R$ and
$SO(3)\sim SU(2)_I$ symmetries which
rotate $3456$ and $789$ directions. We denote by $\alpha,\beta,\cdots=1,2$,
$\dot\alpha,\dot\beta,\cdots=1,2$ and $A,B,\cdots=1,2$ the doublet indices
of these three $SU(2)$ symmetries. See Table \ref{branes} and Fig. \ref{brane-figure}.
\begin{table}[h!]
$$
\begin{array}{c|cccccccccc}
  \hline &0&1&2&3&4&5&6&7&8&9\\
  \hline {\rm NS5}&\bullet&\bullet&&\bullet&\bullet&\bullet&\bullet&&&\\
  \textrm{D8-O8}&\bullet&\bullet&&\bullet&\bullet&\bullet&\bullet&\bullet&\bullet&\bullet\\
  {\rm D2}&\bullet&\bullet&\bullet&&&&&&&\\
  \hline
\end{array}
$$
\caption{Brane configuration for the E-strings}\label{branes}
\end{table}

The M-theory uplift of this brane configuration, with extra circle direction
labeled by $x^{10}$, is given as follows. The NS5-brane lifts to the M5-brane
transverse to the $x^{10}$ direction. The D8-O8 system uplifts to an M9-plane, or
the Horava-Witten wall \cite{Horava:1995qa}, longitudinal in $x^{10}$ direction.
In order to get a weakly-coupled type IIA string theory at low energy, one has to
turn on suitable $E_8$ Wilson line along $x^{10}$ to break $E_8\rightarrow SO(16)$
\cite{Ganor:1996mu}. See our section 4 for more details. D2-branes uplift to M2-branes
transverse in $x^{10}$. In the strong coupling limit of the type IIA theory, the radius
of the M-theory circle becomes large. The geometry $\mathbb{R}^3\times S^1$ transverse to
the 5-brane is replaced by $\mathbb{R}^4$. So the brane configuration contains the M5-M9 system,
in the Coulomb branch of the 6d $E_8$ CFT. M2-branes suspended between them are the
E-strings.

At an energy scale much lower than $L^{-1}$, one obtains a 2d QFT living at the
intersection of these branes. At $g_{\rm YM}\ll E\ll L^{-1}$ with
$g_{\rm YM}^2\sim\frac{g_s}{L\ell_s}$, where $\ell_s,g_s$ are the string scale and
the coupling constant, one obtains a weakly coupled 2d Yang-Mills description
with coupling constant $g_{YM}$. (One can take $g_s$ to be sufficiently small, and
$L$ to be sufficiently larger than $\ell_s$.) When $E\ll g_{\rm YM}$, the
2d Yang-Mills theory is strongly coupled and is expected to flow to an interacting SCFT.
In terms of the Planck scale $\ell_P\sim g_s^{1/3}\ell_s$ of M-theory and the radius
$R\sim g_s\ell_s$ of the $x^{10}$ circle, the strong coupling regime of the 2d Yang-Mills
theory is $E\ll \frac{R}{L^{1/2}\ell_P^{3/2}}$. $L$ is related to the VEV $v$ of the scalar
in the 6d tensor multiplet by $L\sim v\ell_P^3$. So the low energy limit is
$E\ll \frac{R}{v^{1/2}\ell_P^{3}}$. In the Coulomb branch with fixed $v$,
this low energy limit of the 2d theory is obtained by taking the M-theory
limit $R\rightarrow\infty$, in which case the system describes E-strings
as explained in the previous paragraph. Thus our 2d gauge theory describes
E-strings at its strong coupling fixed point.

Let us comment on the enhanced IR symmetries. We first consider the $SO(3)\times U(1)$
acting on $\mathbb{R}^3\times S^1$. In the M-theory limit, this enhances to
$SO(4)\sim SU(2)_l\times SU(2)_r$ of $\mathbb{R}^4$. $SO(3)\sim SU(2)_I$
is identified as
the diagonal combination of $SU(2)_r$ and $SU(2)_l$. On the other hand,
from the viewpoint of 6d superconformal symmetry, $SU(2)_r$ is the
R-symmetry of the 6d $(1,0)$ SCFT and $SU(2)_l$ is a flavor symmetry.
So it might appear that our 2d gauge theory is probing only a combination of
the R-symmetry and a flavor symmetry. However, in the rank $1$ system
with only one M5-brane, the extra flavor $SU(2)_l$ completely
decouples with the 6d CFT. For instance, these can be seen by studying the instanton
partition functions of circle reduced 5d SYM \cite{Hwang:2014uwa}, which will also be
the subject of our section 4. Thus we can identify $SO(3)$ visible in our 2d UV
theory as the superconformal R-symmetry of the 6d CFT. E-strings of the higher rank
6d SCFTs which see $SU(2)_l$ are discussed in \cite{Kim:2015fxa,Gadde:2015tra}.

We also discuss the $E_8$ global symmetry. The 2d UV theory exhibits $SO(16)$ symmetry only. This should enhance to $E_8$ in the IR,
which is naturally expected from the brane perspective. Namely, the type IIA brane system
is obtained by compactifying M-theory brane system with an $E_8$ Wilson line which breaks $E_8$ to $SO(16)$. The IR limit on the 2d gauge theory is the strong coupling limit, which
is the decompactification limit of the M-theory circle. So in this limit, the information
on the Wilson line will be invisible, making us to expect an IR $E_8$ enhancement. In
section 3, we shall compute the elliptic genera of these gauge theories at various
values of $n$, which will be invariant under the $E_8$ Weyl symmetry and support the
$E_8$ enhancement.

Let us study the SUSY of this system. The D2, D8 SUSY are associated with the
projectors $\Gamma^{012}$ and $\Gamma^{013456789}\Gamma^{11}\sim\Gamma^2$ respectively, while the NS5-brane projector is $\Gamma^{01}\Gamma^{3456}$. Various
combinations of branes share different SUSY. We list the following projectors which
should assume definite eigenvalues for the type IIA SUSY parameter $\epsilon$,
for various combinations of branes:
\begin{eqnarray}
  \textrm{D2-D8-NS5}&:&\Gamma^{01}\ ,\ \ \Gamma^2\ ,\ \ \Gamma^{3456}\label{SUSY}\\
  \textrm{D2-NS5}&:&\Gamma^{01}\Gamma^2\ ,\ \ \Gamma^{01}\Gamma^{3456}\label{boundary1}\\
  \textrm{D2-D8-O8}&:&\Gamma^{01}\ ,\ \ \Gamma^2\ .\label{boundary2}
\end{eqnarray}
The projectors (\ref{SUSY}) will yield the SUSY preserved by our system.
The SUSY given by (\ref{boundary1}) and (\ref{boundary2}) will constrain the boundary
conditions of the 3d D2-brane fields at the two ends of the segment along $x^2$.
Let us investigate them in more detail. The type IIA supercharges with
$32$ components can be arranged to be eigenstates of $\Gamma^{01},\Gamma^{3456},\Gamma^2$.
The eigenspinors of $\Gamma^{01}$ are 2d chiral spinors, while those of
$\Gamma^{3456}$ belong to either $({\bf 2}, {\bf 1})$ or $({\bf 1},{\bf 2})$ representations
of $SU(2)_L\times SU(2)_R$. The $32$ supercharges decompose into the sum of the
$({\bf 2},{\bf 1},{\bf 2})_{\pm\pm}\oplus({\bf 1},{\bf 2},{\bf 2})_{\pm\pm}$ representations
of $SU(2)_L\times SU(2)_R\times SU(2)_I$ with all four possible choices of $\pm\pm$, where
the first/second $\pm$ subscripts denote 2d chirality and $\Gamma^2$ eigenvalues, respectively. The SUSY preserved by various combinations of branes are given by
\begin{eqnarray}
  \textrm{D2-D8-NS5}&:&({\bf 1},{\bf 2},{\bf 2})_{-+}\label{(0,4)}\\
  \textrm{D2-NS5}&:&({\bf 2},{\bf 1},{\bf 2})_{+-}\oplus({\bf 1},{\bf 2},{\bf 2})_{-+}\label{(4,4)}\\
  \textrm{D2-D8-O8}&:&({\bf 2},{\bf 1},{\bf 2})_{-+}\oplus({\bf 1},{\bf 2},{\bf 2})_{-+}\ .\label{(0,8)}
\end{eqnarray}
(\ref{(0,4)}) yields the 2d $(0,4)$ SUSY,
which we write as $Q^{\dot\alpha A}_-$. (\ref{(4,4)}) yields 2d $(4,4)$ SUSY
$Q^{\alpha A}_+$, $Q^{\dot\alpha A}_-$. (\ref{(0,8)}) yields 2d $(0,8)$ SUSY
$Q^{\alpha A}_-$, $Q^{\dot\alpha A}_-$. $\pm$ subscripts of $Q$ denote 2d
left/right chiral spinors.

We study the field contents of the 2d $\mathcal{N}=(0,4)$ gauge theory.
This is obtained by starting from the 3d field theory living on D2-branes,
together with the boundary degrees of freedom at $x^2=0,L$, and then taking
a 2d limit when $E\ll L^{-1}$. The 3d fields living in the region $0<x^2<L$ are
\begin{eqnarray}
  \textrm{D2-D2}&:&A_{\mu}\ \ (\mu=0,1,2)\ ;\ \
  X^I\sim\varphi^{\alpha\dot\beta}\ \ (I=3,4,5,6)\ ;\ \ X^{I^\prime}\ \ (I^\prime=7,8,9)\nonumber\\
  &&\lambda\ \ (\textrm{has 16 components, satisfying}\ \
  \Gamma^{11}\lambda=-\lambda)\ .
\end{eqnarray}
The D2-D2 fields are in adjoint representation of $U(n)$.
One also finds boundary degrees at the
brane intersections. At the intersection of D2-D8, open strings provide
2d Fermi multiplet fields which we write as $\Psi_l$ ($l=1,\cdots,16$).
They will be in the bi-fundamental representation of
$O(n)\times SO(16)$ (after introducing the $O8^-$ orientifold).
$\Psi_l$ are left-moving Majorana-Weyl spinors. The maximal supersymmetry
on D2-brane worldvolume is parameterized by $\frac{1+\Gamma^{11}}{2}\epsilon$,
where $\epsilon$ is an eigenvector of $\Gamma^{012}$ (and further projection
conditions listed above at the boundaries).

Let us consider the boundary conditions of the 3d fields. At the two ends $x^2=0,L$,
we shall find separate boundary conditions. As our goal is to obtain the 2d theory,
we shall only keep the zero modes of the 3d fields along the $x^2$ direction. This
means that we shall keep the bosonic fields satisfying the Neumann boundary conditions
on both ends, and the fermionic fields which survive suitable projection conditions at
both ends. The SUSY conditions for the D2-D2 fields at $x^2=0,L$ take the form of
\begin{equation}\label{supercurrent}
 (x^2\textrm{ component of supercurrent})\sim
 {\rm tr}\left(\bar\epsilon(1+\Gamma^{11})
 \Gamma^{MN}F_{MN}\Gamma_2\lambda\right)=0
\end{equation}
in the 10d notation with $M,N=0,\cdots,9$. $\epsilon$
is chosen to be $(4,4)$ on D2-NS5 ($x^2=L$), and $(0,8)$ on D2-D8 ($x^2=0$).
One can follow the strategy of \cite{Gaiotto:2008sa} to obtain the SUSY boundary
conditions. With given SUSY $\epsilon$, one first imposes suitable bosonic boundary
condition, depending on which branes D2's are ending on. Then the condition (\ref{supercurrent}) would determine the boundary condition for the fermions $\lambda$.

We study the D2-NS5 boundary condition first, for which $\bar{\epsilon}$
is taken to be $({\bf 2},{\bf 1},{\bf 2})_{+-}\oplus({\bf 1},{\bf 2},{\bf 2})_{-+}$.
The D2-D2 fermion $\lambda$ satisfies $\lambda=-\Gamma^{11}\lambda$, where
$\Gamma^{11}\sim\Gamma^{01}\Gamma^{3456}\Gamma^{78}\Gamma^{29}$. So
depending on the eigenvalues of $\Gamma^{01}$, $\Gamma^{3456}$, $\Gamma^{78}$
(the spin of $SU(2)_I$), $\lambda$ can be decomposed into
\begin{equation}
  (SU(2)_L,SU(2)_R,SU(2)_I)_{\Gamma^{01}}=
  ({\bf 2},{\bf 1},{\bf 2})_+\oplus({\bf 2},{\bf 1},{\bf 2})_-\oplus
  ({\bf 1},{\bf 2},{\bf 2})_+\oplus({\bf 1},{\bf 2},{\bf 2})_-\ ,
\end{equation}
and $\Gamma^{29}$ eigenvalues are determined from $\Gamma^{11}\lambda=-\lambda$.
Unlike $\epsilon$, the $\Gamma^2$ eigenvalue cannot be specified for
$\lambda$, since it does not commute with $\Gamma^{29}$.
We start from the boundary conditions for the bosonic fields that we know
for D2-NS5:
\begin{equation}\label{NS5-bdy-boson}
  F_{\mu 2}=0\ ,\ \ D_2X^I=0\ ,\ \ X^{I^\prime}=0
\end{equation}
with $\mu=0,1$, $I=3,4,5,6$, $I^\prime=7,8,9$. This provides the following
constraints on $\lambda$:
\begin{equation}
  0=\bar\epsilon\lambda=\bar\epsilon\Gamma^{\mu 2 I}\lambda
  =\bar\epsilon\Gamma^{IJ}\Gamma^2\lambda=\bar\epsilon\Gamma^{I^\prime}\lambda\ .
\end{equation}
This requires $\lambda$ to be in
\begin{equation}\label{NS5-bdy-fermion}
  (SU(2)_L,SU(2)_R,SU(2)_I)_{\Gamma^{01}}=
  ({\bf 2},{\bf 1},{\bf 2})_{-}\oplus({\bf 1},{\bf 2},{\bf 2})_{+}\ ,
\end{equation}
namely, with a right mover $\lambda^{\alpha A}_-$ and a left mover
$\lambda^{\dot\alpha A}_+$.
(The former will belong to a 2d $(0,4)$ hypermultiplet
and the latter will belong to a 2d $(0,4)$ vector multiplet.)

Now we consider the D2-D8-O8 boundary conditions. The effect of having $8$ D8-branes
is simply adding Fermi multiplet fields as explained above. So we focus on
the effect of the O8-plane. Following \cite{Gaiotto:2008sa}, we consider the covering
space of $x^2>0$ and consider the 3d SYM on $\mathbb{R}^{2,1}$. The reflection
$x^2\rightarrow-x^2$ of space is accompanied by an outer automorphism $\tau$
acting on $G=U(n)$ gauge group. The algebra $g$ of $G$ decomposes into
$g^{(+)}\oplus g^{(-)}$, where $\tau$ acts on $g^{(\pm)}$ as $\pm 1$. In our case,
$g^{(+)}$ is the algebra of $O(n)\subset U(n)$, and $g^{(-)}$ forms a rank $2$
symmetric representation of $O(n)$.
So any adjoint-valued field $\Phi$ can be written as $\Phi=\Phi^{(+)}+\Phi^{(-)}$. The reflection is further accompanied by $X^I\rightarrow-X^I$
for $I=3,\cdots,9$. This is because odd number of scalars should flip sign
for the net reflection to preserve the orientation of $\mathbb{R}^{9,1}$, e.g.
to preserve $\Gamma^{11}$ projection conditions in the 3d maximal SYM. Since
the D2-D8-O8 boundary condition preserves $SO(7)$ which rotates
$I=3,\cdots,9$, all $X^I$'s should be flipped.
So the fields are required to be invariant under the
net reflection:
\begin{equation}
  A_\mu(x^2)=A^\tau_\mu(-x^2),\ A_2(x^2)=-A_2^\tau(-x^2),\
  X_I(x^2)=-X_I^\tau(-x^2)
\end{equation}
where $\Phi^\tau=\tau\Phi\tau^{-1}$, $\mu=0,1$ and $I=3,\cdots,9$. So at the
fixed plane $x^2=0$, the boundary condition is given by
\begin{equation}\label{O8-bdy-boson}
  F_{\mu 2}^{(+)}=0\ ,\ \ F_{\mu\nu}^{(-)}=0\ ,\ \
  D_2X_I^{(-)}=0\ ,\ \ X_I^{(+)}=0\ \ \ \ (I=3,\cdots,9)\ .
\end{equation}
$A_2(x^2)$ can be gauged away using $x^2$ dependent gauge transformation along
the interval. We can again find the fermionic boundary conditions from
(\ref{supercurrent}). This requires
\begin{equation}
  0=\bar\epsilon\lambda^{(+)}=\bar\epsilon\Gamma^I\lambda^{(+)}=
  \bar\epsilon\Gamma^{IJ2}\lambda^{(+)}
  \ ,\ \ 0=\bar\epsilon\Gamma^\mu\lambda^{(-)}=\bar\epsilon
  \Gamma^{\mu I2}\lambda^{(-)}
\end{equation}
with $\mu=0,1$ and $I,J=3,\cdots,9$. $\bar\epsilon$ is chosen
to be (\ref{(0,8)}). Solving these constraints, the $O(n)$ adjoint fermion
$\lambda^{(+)}$ and the $O(n)$ symmetric fermion $\lambda^{(-)}$ are required to be in
\begin{eqnarray}\label{O8-bdy-fermion}
  \lambda^{(+)}&:&(SU(2)_L,SU(2)_R,SU(2)_I)_{\Gamma^{01}}=
  ({\bf 2},{\bf 1},{\bf 2})_{+}\oplus({\bf 1},{\bf 2},
  {\bf 2})_{+}\nonumber\\
  \lambda^{(-)}&:&(SU(2)_L,SU(2)_R,SU(2)_I)_{\Gamma^{01}}=
  ({\bf 2},{\bf 1},{\bf 2})_{-}\oplus({\bf 1},{\bf 2},{\bf 2})_{-}\ .
\end{eqnarray}

\begin{figure}[t!]
  \begin{center}
    \includegraphics[width=10cm]{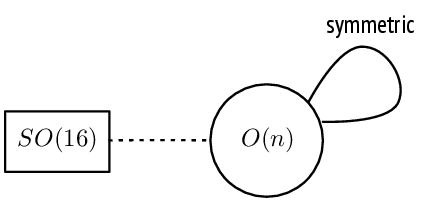}
\caption{The quiver diagram of the 2d $\mathcal{N}=(0,4)$ gauge theory for E-strings: solid/dotted lines denote hyper/Fermi multiplets, respectively.}\label{quiver}
  \end{center}
\end{figure}

We combine the D2-NS5 and D2-O8 boundary conditions to read off the 2d field contents.
For bosons, requiring (\ref{NS5-bdy-boson}) and (\ref{O8-bdy-boson}) yields the following 2d fields:
\begin{equation}
  A_{\mu}^{(+)}\ ,\ \ X_I^{(-)}\sim\varphi_{\alpha\dot\beta}\ \ \ (I=3,4,5,6)\ .
\end{equation}
For fermions,
requiring (\ref{NS5-bdy-fermion}) and (\ref{O8-bdy-fermion}) together, one finds that
$\lambda^{\alpha A}_-\sim({\bf 2},{\bf 1},{\bf 2})_{-}$ is in the symmetric representation of $O(n)$, while
$\lambda^{\dot\alpha A}_+\sim({\bf 1},{\bf 2},{\bf 2})_{+}$ is in the
adjoint (i.e. antisymmetric) representation. So from the D2-D2 modes, we obtain
the $(0,4)$ vector multiplet $A_\mu$, $\lambda^{\dot\alpha A}_+$ of $O(n)$, and also
a $(0,4)$ hypermultiplet $\varphi_{\alpha\dot\beta}$, $\lambda^{\alpha A}_-$ in the
symmetric representation of $O(n)$.
So to summarize, one obtains the following 2d $\mathcal{N}=(0,4)$ field contents:
\begin{eqnarray}
  \textrm{vector}&:&O(n)\ {\rm antisymmetric}\ \ \ (A_\mu,\ \lambda^{\dot\alpha A}_+)\nonumber\\
  \textrm{hyper}&:&O(n)\ {\rm symmetric}\ \ \ (\varphi_{\alpha\dot\beta}, \
  \lambda^{\alpha A}_-)\nonumber\\
  \textrm{Fermi}&:&O(n)\times SO(16)\ {\rm bifundamental}\ \ \ \Psi_l\ .
\end{eqnarray}
Fig. \ref{quiver} shows the quiver diagram of this gauge theory.
One can check the $SO(n)$ gauge anomaly cancelation of this chiral matter content.
Note that we have no twisted hypermultiplets, whose scalars form doublets
of $SU(2)_I$ and fermions form doublets of $SU(2)_R$.

We also explain how to get the full Lagrangian of this system.
Viewing this as a special case of $\mathcal{N}=(0,2)$
supersymmetric system, it suffices to determine the two holomorphic functions
$E_{\Psi}(\Phi_i)$, $J^{\Psi}(\Phi_i)$ for each Fermi multiplet $\Psi$, depending
on the $(0,2)$ chiral multiplet fields $\Phi_i$. We choose $Q\equiv Q^{\dot{1}}_1$
and $Q^\dag$ as the $(0,2)$ subset.
To have $(0,4)$ SUSY, the $E$, $J$ functions for the adjoint $(0,2)$ Fermi multiplet
$\Theta\equiv(\lambda_+^{\dot{1}2},\lambda^{\dot{2}1}_+)$
in the $(0,4)$ vector multiplet are required to be \cite{Tong:2014yna}
\begin{equation}\label{J-E}
  J_\Theta=\varphi\tilde\varphi-\tilde\varphi\varphi\ ,\ \ E_{\Theta}=0\ ,
\end{equation}
where $\varphi\equiv\varphi_{1\dot{1}}$, $\tilde\varphi\equiv\varphi_{2\dot{1}}$
are $(0,2)$ chiral multiplet scalars which transform under $Q\equiv Q^{\dot{1}}_1$.
Note that, if the $(0,4)$ theory has both hypermultiplets and twisted hypermultiplets,
the full interaction has to be more complicated \cite{Tong:2014yna}. Without
twisted hypermultiplets in our system, (\ref{J-E}) provides the full
interactions associated with $\Theta$. This induces a bosonic
potential of the form $|J_{\Theta}|^2$, as well as the Yukawa interaction.
Extra Fermi multiplets in the $(0,2)$ viewpoint are $\Psi_l$ from D2-D8-O8 modes,
so we should also determine their $E,J$. $E_{\Psi_l}$, $J^{\Psi_l}$ are simply zero,
from $SO(16)$ symmetry. With all the $E$, $J$ functions determined, the
supersymmetric action can be written down if $E^aJ_a=0$, where the index $a$ runs over
all $(0,2)$ Fermi multiplets. This condition is clearly met.
With these data, the full action can be written down in a standard manner: see, for
instance, \cite{Gadde:2014ppa,Tong:2014yna}. In our case, the bosonic potential consists
of $|J_\Theta|^2$ and the usual D-term potential, making the D-term potential from
the `$SU(2)_R$ triplet' of D-terms. The classical Higgs branch moduli space, given by
nonzero $\varphi,\tilde\varphi$, is real $4n$ dimensional. Semi-classically,
these are the positions of $n$ E-strings.

One can also compute the central charges of the IR CFT from our UV gauge theory.
Once we know the correct superconformal R-symmetry of the IR SCFT, the (right-moving)
central charge of the IR CFT can be computed in UV by the anomaly of the superconformal
R-symmetry. We closely follow \cite{Benini:2012cz,Gadde:2014ppa,Tong:2014yna},
which use the $(0,2)$ superconformal R-symmetry to determine the central charges.

In our $(0,4)$ system, a semi-classical description is allowed when
$\varphi^{\alpha\dot\beta}$ scalars are large. This is the CFT associated
with the classical Higgs branch \cite{Witten:1997yu}. In this CFT, the superconformal
R-symmetry can only come from $SU(2)_I$ in the UV theory. This is because
the right sector contains the $O(n)$ symmetric scalar $\varphi_{\alpha\dot\beta}$,
and the superconformal R-symmetry should not act on it \cite{Witten:1997yu}. Following
\cite{Tong:2014yna}, let
us choose the supercharge $Q\equiv Q^{\dot{1}2}$ and use the $(0,2)$ superconformal symmetry
to determine the central charge. The right-moving central charge $c_R$ is given by
\begin{equation}\label{right-central}
  c_R=3{\rm Tr}(\gamma^3R^2)\ ,
\end{equation}
with $\gamma^3=\pm 1$ for the right/left moving fermions, respectively, and the trace
acquires an extra $\frac{1}{2}$ factor for real fermions. The $(0,2)$ R-charge $R$ is
normalized so that $R[Q]=-1$. In the Higgs branch CFT, this should be proportional to
the Cartan of $SU(2)_I$, so we set $R=2J_I$. Collecting the contribution from $O(n)$
symmetric $\lambda^{\alpha A}$ in the right sector and adjoint $\lambda^{\dot\alpha A}$ in the left sector, one obtains
\begin{equation}
  c_R=3\times\frac{1}{2}\times \frac{n^2+n}{2}\times (4\times 1^2)-
  3\times\frac{1}{2}\times\frac{n(n-1)}{2}\times(4\times 1^2)=6n\ .
\end{equation}
The left moving central charge
$c_L$ is determined from $c_R$ by the gravitational anomaly \cite{Gadde:2014ppa}:
\begin{equation}
  c_R-c_L={\rm Tr}(\gamma^3)=\frac{1}{2}\times 4\frac{n^2+n}{2}
  -\frac{1}{2}\times 4\frac{n^2-n}{2}-\frac{1}{2}\times 16n=-6n\
  \rightarrow\ \ c_L=12n\ .
\end{equation}
$c_L=12n$ is consistent with the result obtained in \cite{Minahan:1998vr}
(where $c_L=12n-4$ was found after eliminating $4$ from the decoupled center-of-mass
degrees of freedom.) One can semiclassically understand some of these results, by
studying the region with large value of the Higgs scalar $\varphi^{\alpha\dot\beta}$.
$c_R=6n$ comes from the $n$ pairs of $4$ scalars and $4$ fermions
for $n$ E-strings. As for $c_L=12n$, the $4n$ scalars in the left moving sector
accounts for $4n$, and the $16n$ real fermions $\Psi_l$ accounts for $8n$.
For $n=1$, we know that the last $8$ is given by the $G=E_8$ current
algebra at level $k=1$ (with dual Coxeter number $c_2=30$) \cite{Ganor:1996mu,Klemm:1996hh},
whose central charge is indeed $\frac{k|G|}{k+c_2}=\frac{248}{1+30}=8$.

\section{E-string elliptic genera from 2d gauge theories}

We consider the elliptic genus of the 2d $(0,4)$ $O(n)$ gauge theory, constructed in
the previous section. We pick the same $(0,2)$ SUSY as before, and define the elliptic genus as follows:
\begin{equation}\label{elliptic}
  Z_n(q,\epsilon_{1,2},m_l)={\rm Tr}_{\rm RR}\left[(-1)^Fq^{H_L}\bar{q}^{H_R}
  e^{2\pi i\epsilon_1(J_1+J_I)}
  e^{2\pi i\epsilon_2(J_2+J_I)}\prod_{l=1}^8e^{2\pi im_lF_l}\right]\ .
\end{equation}
$J_1,J_2$ are the Cartans of $SO(4)\sim SU(2)_L\times SU(2)_R$ which rotate
the $34$ and $56$ orthogonal 2-planes, and $J_I$ is the Cartan of $SU(2)_I$.
$F_l$ are the Cartans of $SO(16)$, which we expect to be the Cartans of
enhanced $E_8$ in IR. Note that $H_R\sim\{Q,Q^\dag\}$ with $Q=Q^{\dot{1}}_1$
and $Q^\dag=-Q^{\dot{2}}_2$, and the remaining factors inside the trace commute
with $Q,Q^\dag$. Note also that, the 2d gauge theory itself has a noncompact Higgs
branch spanned by $\varphi^{\alpha\dot\beta}$. They are given nonzero masses by
turning on $\epsilon_1,\epsilon_2$, so that the path integral for this index does
not have any noncompact zero modes. The interpretation of the zero modes from
$\varphi^{\alpha\dot\beta}$ at $\epsilon_1,\epsilon_2=0$ is clearly the multi-particle
positions, so by keeping nonzero $\epsilon_{1,2}$ we are computing the multi-particle
index, as usual. The single particle spectrum can be extracted from the
multi-particle index.

The index (\ref{elliptic}) for $\mathcal{N}=(0,2)$ gauge theories
was studied in \cite{Benini:2013nda,Benini:2013xpa}, by computing the path integral of
the gauge theory on $T^2$. There appear compact zero modes from the path
integral, coming from the flat connections on $T^2$. \cite{Benini:2013nda,Benini:2013xpa}
first fix the flat connections, integrate over the nonzero modes, and
then integrate (or sum) over the flat connections to obtain their final
expression for the index.

Let us first explain the possible flat connections of our $O(n)$ gauge theories
on $T^2$. These are given by two commuting $O(n)$ group elements $U_1$, $U_2$, the Wilson
lines along the temporal and spatial circles of $T^2$. Note that $O(n)$ is a disconnected
group so that $U_1$ and $U_2$ can each have two disconnected sectors, depending on whether their determinants are $1$ or $-1$. The general $O(n)$ holonomies on $T^2$,
up to conjugation, can be derived using a D-brane picture
\cite{Witten:1997bs}.\footnote{If the gauge group is not $O(n)$ but, say Spin($n$)
as in \cite{Witten:1997bs}, one has to make a variation of the simple D-brane
argument that we shall present here.}
The $O(n)$ flat connections are the zero energy configurations of the $n$ D2-branes
and an O2-plane wrapping $T^2$. By T-dualizing twice along the torus, one obtains
$n$ D0-branes moving along the
$T^2/\mathbb{Z}_2$ orientifold. The flat connections T-dualize
to the positions of D0-branes on $T^2/\mathbb{Z}_2$. There are four O0-plane fixed points
on the covering space $T^2$. It suffices for us to classify all possible positions of
D0-branes.
When two D0-branes on the covering space are paired as $\mathbb{Z}_2$ images of each other,
they have one complex parameter $u$ as their position. Some D0-branes can also be stuck
at the $\mathbb{Z}_2$ fixed points without a pair: they are fractional branes on
$T^2/\mathbb{Z}_2$, whose positions are freezed at the fixed points. So the classification
of $O(n)$ flat connections reduces to classifying the possible fractional brane
configurations.

When $n=2p$ is even, one can first have all $2p$ D0-branes to make $p$ pairs.
In this branch, one finds $p$ complex moduli $u_i$ ($i=1,\cdots,p$).
Another possibility
is to form $p-1$ pairs to freely move, while having $2$ fractional D-branes stuck at two
of the $4$ fixed points. Note that the two fractional branes have to be stuck at different
fixed points: otherwise they can pair and leave the fixed point,
being a special case of the first branch. There are $6$ ways of choosing $2$ fixed points
among $4$, so we obtain $6$ more sectors. Finally, one finds a sector in which $p-2$ pairs
freely move, while $4$ fractional D-branes are stuck at $4$ different fixed points (when
$p\geq 2$). After T-dualizing, $U_1,U_2$ are exponentials of the D0-brane positions.
The above $8$ sectors are summarized by the following
pairs of Wilson lines $U_1,U_2$, for $O(2p)$ with $p\geq 2$:
\begin{eqnarray}\label{flat-even}
  (\rm{ee})&:&U_1={\rm diag}(e^{iu_{1i}\sigma_2})_p\ ,\ \
  U_2={\rm diag}(e^{iu_{2i}\sigma_2})_p\ ;\nonumber\\
  &&U_1={\rm diag}(e^{iu_{1i}\sigma_2},1,-1,-1,1)_{p-2}\ ,\ \
  U_2={\rm diag}(e^{iu_{2i}\sigma_2},1,1,-1,-1)_{p-2};\nonumber\\
  ({\rm eo})&:&U_1={\rm diag}(e^{iu_{1i}\sigma_2},1,1)_{p-1}\ ,\ \
  U_2={\rm diag}(e^{iu_{2i}\sigma_2},1,-1)_{p-1}\ ;\nonumber\\
  &&U_1={\rm diag}(e^{iu_{1i}\sigma_2},-1,-1)_{p-1}\ ,\ \
  U_2={\rm diag}(e^{iu_{2i}\sigma_2},1,-1)_{p-1};\nonumber\\
  ({\rm oe})&:&U_1={\rm diag}(e^{iu_{1i}\sigma_2},1,-1)_{p\!-\!1}\ ,\ \
  U_2={\rm diag}(e^{iu_{2i}\sigma_2},1,1)_{p-1}\ ;\nonumber\\
  &&U_1={\rm diag}(e^{iu_{1i}\sigma_2},1,-1)_{p-1}\ ,\ \
  U_2={\rm diag}(e^{iu_{2i}\sigma_2},-1,-1)_{p\!-\!1};\nonumber\\
  ({\rm oo})&:&U_1={\rm diag}(e^{iu_{1i}\sigma_2},1,-1)_{p\!-\!1}\ ,\ \
  U_2={\rm diag}(e^{iu_{2i}\sigma_2},1,-1)_{p\!-\!1}\ ;\nonumber\\
  &&U_1={\rm diag}(e^{iu_{1i}\sigma_2},1,-1)_{p\!-\!1}\ ,\ \
  U_2={\rm diag}(e^{iu_{2i}\sigma_2},-1,1)_{p\!-\!1}\ .
\end{eqnarray}
(ee), (eo), (oe), (oo) are for $U_1,U_2$ in the even or odd elements of $O(n)$.
The symbol `diag' denotes a block-diagonalized matrix.
The subscripts are the number of independent complex parameters. The parameters
live on $u_i=u_{1i}+\tau u_{2i}\in\mathbb{C}/(\mathbb{Z}+\tau\mathbb{Z})$, where
$\tau$ is related to our fugacity $q$ by $q=e^{2\pi i\tau}$. For odd $n=2p+1$ with
$n\geq 3$, one can make a similar analysis. There are $4$ cases in which one has $1$
fractional brane stuck at one of the $4$ fixed points, and $4$ more cases
(when $p\geq 1$) in which $3$ fractional branes are stuck at three of the $4$ fixed
points. So one obtains the following $8$ sectors, for $p\geq 1$:
\begin{eqnarray}\label{flat-odd}
  (\rm{ee})&:&U_1={\rm diag}(e^{iu_{1i}\sigma_2},1)_p\ ,\ \
  U_2={\rm diag}(e^{iu_{2i}\sigma_2},1)_p\ ;\nonumber\\
  &&U_1={\rm diag}(e^{iu_{1i}\sigma_2},-1,-1,1)_{p-1}\ ,\ \
  U_2={\rm diag}(e^{iu_{2i}\sigma_2},1,-1,-1)_{p-1};\nonumber\\
  ({\rm eo})&:&U_1={\rm diag}(e^{iu_{1i}\sigma_2},1)_{p}\ ,\ \
  U_2={\rm diag}(e^{iu_{2i}\sigma_2},-1)_{p}\ ;\nonumber\\
  &&U_1={\rm diag}(e^{iu_{1i}\sigma_2},-1,-1,1)_{p-1}\ ,\ \
  U_2={\rm diag}(e^{iu_{2i}\sigma_2},1,-1,1)_{p-1};\nonumber\\
  ({\rm oe})&:&U_1={\rm diag}(e^{iu_{1i}\sigma_2},-1)_{p}\ ,\ \
  U_2={\rm diag}(e^{iu_{2i}\sigma_2},1)_{p}\ ;\nonumber\\
  &&U_1={\rm diag}(e^{iu_{1i}\sigma_2},1,-1,1)_{p-1}\ ,\ \
  U_2={\rm diag}(e^{iu_{2i}\sigma_2},-1,-1,1)_{p-1};\nonumber\\
  ({\rm oo})&:&U_1={\rm diag}(e^{iu_{1i}\sigma_2},-1)_{p}\ ,\ \
  U_2={\rm diag}(e^{iu_{2i}\sigma_2},-1)_{p}\ ;\nonumber\\
  &&U_1={\rm diag}(e^{iu_{1i}\sigma_2},1,1,-1)_{p-1}\ ,\ \
  U_2={\rm diag}(e^{iu_{2i}\sigma_2},1,-1,1)_{p-1}\ .
\end{eqnarray}
There are two exceptional cases. For $O(1)$, the four sectors in (\ref{flat-odd})
with rank $p-1$ are absent. So we only have four rank $0$ sectors
\begin{equation}
  (U_1,U_2)=(1,1),\ (1,-1),\ (-1,1),\ (-1,-1)\ .
\end{equation}
For $O(2)$, the second sector in (\ref{flat-even}) with rank $p-2$ is absent.
So we have seven sectors
\begin{equation}\label{flat-O(2)}
  (U_1,U_2)=(e^{iu_1\sigma_2},e^{iu_2\sigma_2}),\ (1,\sigma_3),\ (-1,\sigma_3),\
  (\sigma_3,1),\ (\sigma_3,-1),\ (\sigma_3,\sigma_3),\ (\sigma_3,-\sigma_3)\ .
\end{equation}
The Wilson lines can be more conveniently labeled by their exponents,
which we call $u=(u_1,\cdots,u_{n})$ for $O(n)$. In the $2\times 2$ blocks $e^{iu_{1i}\sigma_2},e^{iu_{2i}\sigma_2}$ with
continuous elements, the associated two $u$ parameters are given by
the two eigenvalues $\pm(u_{1i}+\tau u_{2i})$. In the blocks with discrete
numbers, we assign $u_i=0$ for an eigenvalue pair $(1,1)$ of $U_1,U_2$,
$u_i=\frac{1}{2}$ for an eigenvalue pair $(-1,1)$, $u_i=\frac{\tau}{2}$ for $(1,-1)$,
and $u_i=\frac{1+\tau}{2}$ for $(-1,-1)$. For the above $8$ sectors, one thus obtains
\begin{eqnarray}\label{holonomy-even}
  \textrm{(ee)}&:&u=(\pm u_1,\cdots,\pm u_{p})\ ;\ \
  u=(\pm u_1,\cdots,\pm u_{p-2},0,\frac{1}{2},\frac{1+\tau}{2},\frac{\tau}{2})\nonumber\\
  \textrm{(eo)}&:&u=(\pm u_1,\cdots,\pm u_{p-1},0,\frac{\tau}{2})\ ;\ \
  u=(\pm u_1,\cdots,\pm u_{p-1},\frac{1}{2},\frac{1+\tau}{2})\nonumber\\
  \textrm{(oe)}&:&u=(\pm u_1,\cdots,\pm u_{p-1},0,\frac{1}{2})\ ;\ \
  u=(\pm u_1,\cdots,\pm u_{p-1},0,\frac{\tau}{2},\frac{1+\tau}{2},\frac{\tau}{2})\nonumber\\
  \textrm{(oo)}&:&u=(\pm u_1,\cdots,\pm u_{p-1},0,\frac{1+\tau}{2})\ ;\ \
  u=(\pm u_1,\cdots,\pm u_{p-1},\frac{\tau}{2},\frac{1}{2})
\end{eqnarray}
for $O(2p)$, and
\begin{eqnarray}\label{holonomy-odd}
  \textrm{(ee)}&:&u=(\pm u_1,\cdots,\pm u_{p},0)\ ;\ \
  u=(\pm u_1,\cdots,\pm u_{p-1},\frac{1}{2},\frac{1+\tau}{2},\frac{\tau}{2})\nonumber\\
  \textrm{(eo)}&:&u=(\pm u_1,\cdots,\pm u_{p},\frac{\tau}{2})\ ;\ \
  u=(\pm u_1,\cdots,\pm u_{p-1},\frac{1}{2},\frac{1+\tau}{2},0)\nonumber\\
  \textrm{(oe)}&:&u=(\pm u_1,\cdots,\pm u_{p},\frac{1}{2})\ ;\ \
  u=(\pm u_1,\cdots,\pm u_{p-1},\frac{\tau}{2},\frac{1+\tau}{2},0)\nonumber\\
  \textrm{(oo)}&:&u=(\pm u_1,\cdots,\pm u_{p},\frac{1+\tau}{2})\ ;\ \
  u=(\pm u_1,\cdots,\pm u_{p-1},0,\frac{\tau}{2},\frac{1}{2})
\end{eqnarray}
for $O(2p+1)$. These $u$ couple minimally to the matters in the fundamental
representation. The parameters coupling to a field in a different representation
of $SO(n)$ are given by $\rho(u)$, where $\rho$ runs over the weights of
the representation of the field.

With the Wilson line backgrounds identified, we study the subgroup of $O(n)$
gauge symmetry which acts within the $U_1,U_2$ specified above. This is the
`Weyl group,' defined in each disconnected sector of $(U_1,U_2)$. When $U_1,U_2$
are given by $r$ $2\times 2$ blocks and an $s\times s$ diagonal matrix with $\pm 1$
eigenvalues (with $2r+s=n$ and $s\leq 4$), the Weyl group is given by
\begin{equation}
  \left[\textrm{Weyl group of }O(2r)\right]\times\left[O(s)\textrm{\ elements commuting with the }
  s\times s\textrm{\ block}\right]\ .
\end{equation}
The former part has order $2^{r}r!$, and the latter has order $2^{s}$
coming from the $O(s)$ transformations
${\rm diag}_{s\times s}(\pm 1,\pm 1,\cdots,\pm 1)$. So the order
of the Weyl group $W(O(n))_s$, acting within a given connected sector of
$U_1,U_2$, is given by
\begin{eqnarray}\label{weyl}
  &&|W(O(2p))_0|=2^pp!\ ,\ \ |W(O(2p))_2|=2^{p+1}(p-1)!\ ,\ \
  |W(O(2p))_4|=2^{p+2}(p-2)!\nonumber\\
  &&|W(O(2p+1))_1|=2^{p+1}p!\ ,\ \
  |W(O(2p+1)_3)|=2^{p+2}(p-1)!\ ,
\end{eqnarray}
where the subscript denotes the value of $s$ for $U_1,U_2$.

In the above background, the Gaussian path integral of non-zero modes
yields $Z_{\textrm{1-loop}}$, which is the product of the following 1-loop determinants
for various supermultiplets
\cite{Benini:2013xpa}:\footnote{One difference from \cite{Benini:2013xpa} is that we put
a factor $i$ in the denominator of the contribution $\frac{\theta_1(q,z)}{i\eta(q)}$ from
each Fermi multiplet. Of course this only affects the overall sign of the index, which is
ambiguous in 2d without knowing the spin-statistics relation inherited from
higher dimensional physics. We shall see that our choice is compatible with the
physics of circle compactified 6d CFT, by comparing with some known results.
Collecting all the factors of $i$ in $Z_{\textrm{1-loop}}$, one obtains $(-1)^n$.}
\begin{eqnarray}\label{determinant}
  Z_{\rm sym.\ hyper}&=&\prod_{\rho\in{\rm sym}}
  \frac{i\eta(\tau)}{\theta_1(\tau,\epsilon_1+\rho(u))}\cdot
  \frac{i\eta(\tau)}{\theta_1(\tau,\epsilon_2+\rho(u))}\nonumber\\
  Z_{SO(16)\ {\rm Fermi}}&=&\prod_{\rho\in{\rm fund}}\prod_{l=1}^8
  \frac{\theta_1(\tau,m_l+\rho(u))}{i\eta(\tau)}\\
  Z_{\rm vector}&=&\prod_{i=1}^r
  \left(\frac{2\pi\eta^2du_i}{i}
  \cdot\frac{\theta_1(\epsilon_1+\epsilon_2)}{i\eta}\right)\cdot
  \prod_{\alpha\in{\rm root}}
  \frac{\theta_1(\alpha(u))\theta_1(\epsilon_1+\epsilon_2+\alpha(u))}{i^2\eta^2}\ .\nonumber
\end{eqnarray}
Whenever we omit the modular parameters, like
$\theta_i(\tau,z)\rightarrow\theta_i(z)$ or $\eta(\tau)\rightarrow\eta$,
it is understood as $\tau$. See appendix A for explanations on these functions.
The `rank' $r$ is the number of continuous complex parameters in $U_1,U_2$.
$\alpha$ runs over the roots of $SO(n)$. Multiplying all these factors, one finally
has to integrate over the continuous parameters in $u$ and then sum over disconnected
sectors of flat connections. The result is
\begin{equation}\label{index-full}
  \sum_{a}\frac{1}{|W_a|}\cdot\frac{1}{(2\pi i)^r}
  \oint Z_{\textrm{1-loop}}^{(a)}\ \ ,\ \ \
  Z_{\textrm{1-loop}}^{(a)}\equiv Z_{\rm vector}^{(a)}
  Z_{\textrm{sym. hyper}}^{(a)}Z_{SO(16)\textrm{\ Fermi}}^{(a)}\ ,
\end{equation}
$a$ labels the disconnected sectors of the flat connection $U_1,U_2$.
The integral is a suitable `contour integral' over the continuous parameters
$u$, to be explained shortly. $W_a$ is the Weyl group with given $U_1,U_2$ explained
above.

Before proceeding, let us comment on the periodicity of (\ref{determinant})
in $u$. Each $u_i$ (for $i=1,\cdots,r$) lives
on $T^2/\mathbb{Z}_2$, due to large gauge transformations on $T^2$, so is a
periodic variable $u_i\sim u_i+1\sim u_i+\tau$.
However, since $\theta_1(u,\tau)$ is only a quasi-periodic function,
\begin{equation}\label{period-shift}
  \theta_1(z+1)=-\theta_1(z),\ \theta_1(z+\tau)=-q^{-1/2}y^{-1}\theta_1(z),\
  \theta_1(z+1+\tau)=q^{-1/2}y^{-1}\theta_1(z)\ ,
\end{equation}
with $y\equiv e^{2\pi iz}$, each $\frac{\theta_1}{\eta}$ factor in
(\ref{determinant}) is not invariant under these shifts. The
failure of periodicity is related to the gauge anomaly of the chiral theory.
The factors spoiling the periodicity cancel in the combination
(\ref{index-full}), due to the anomaly cancelation of our gauge theory.

Another subtlety is the determinant of the real scalars and Majorana fermions.
Each real scalar or fermion contributes to a `square-root' of $\theta_1$ factor.
Equivalently, each charge conjugate pair of fermion modes contributes a factor of
$\frac{\theta_1(z)}{i\eta}$, while such a pair of bosons contributes
$\frac{i\eta}{\theta_1(z)}$ in (\ref{determinant}). In particular, on these modes, the discrete shifts on
the holonomy (\ref{holonomy-even}), (\ref{holonomy-odd}) given by $u_i=\frac{1}{2}$,
$\frac{1+\tau}{2}$, $\frac{\tau}{2}$ has to be understood with some care.
When such a shift is made in the argument of $\theta_1$ coming from a pair of
real fields, one should understand it as $``\theta_1(z+u_i)\textrm{''}\sim\sqrt{\theta_1(z+u_i)\theta_1(z-u_i)}$.
Having this in mind, and applying
\begin{equation}\label{fractional-shift}
  \theta_1(z+\tfrac{1}{2})=\theta_2(z)\ ,\ \
  \theta_1(z+\tfrac{\tau}{2})=iq^{-1/8}y^{-1/2}\theta_4(z)\ ,\ \
  \theta_1(z+\tfrac{1+\tau}{2})=q^{-1/8}y^{-1/2}\theta_3(z)\ ,
\end{equation}
one can replace $\theta_1(z+\frac{1}{2})$, $\theta_1(z+\frac{\tau+1}{2})$,
$\theta_1(z+\frac{\tau}{2})$ by $\theta_2(z)$, $\theta_3(z)$, $\theta_4(z)$,
respectively, apart from the extra factors appearing in (\ref{fractional-shift}).
These extra factors in (\ref{index-full}) again cancel to $1$. So the theta function
$\theta_1$ with a half-period shift can be
replaced by one of $\theta_2,\theta_3,\theta_4$ without the shift.

Now we finally explain the meaning of the `contour integral' in (\ref{index-full}),
following \cite{Benini:2013nda,Benini:2013xpa}. The `contour integral' is defined
by providing a prescription for the residue sum which replaces the integral,
whenever one encounters a pole on the parameter space of $(U_1,U_2)$. The prescription
is derived in \cite{Benini:2013xpa}, using the so-called Jeffrey-Kirwan residues.
At each pole $u=u_\ast$ on the $r$ complex dimensional $u$ space,
there are $r$ or more hyperplanes of the form $\rho_i(u)+z_i=0$ (mod $\mathbb{Z}+\tau\mathbb{Z}$)
which passes through it, where $i=1,\cdots, d$ ($\geq r$). $z_i$ is a linear
combination of the chemical potentials that appears in
$\theta_1(\rho_i(u)+z_i)$ in the denominator of $Z_{\textrm{1-loop}}$.
In our problem, $z_i$ is either $\epsilon_1$ or $\epsilon_2$. When exactly $r$ hyperplanes intersect
at a point $u=u_\ast$ (mod $\mathbb{Z}+\tau\mathbb{Z}$), this pole is called non-degenerate.
When $d>r$, the pole is called degenerate.

Before explaining the Jeffrey-Kirwan residues (or JK-Res) of our integrand at $u=u_\ast$,
let us first note that the results of \cite{Benini:2013xpa} apply when the pole at
$u_\ast$ is `projective.' The pole is called projective when all the weight vectors
$\rho_i$ associated with the hyperplanes meeting at $u=u_\ast$ are contained in a half
space. Namely, the projective condition requires that there is a vector $v$ in the Cartan
$\mathfrak{h}$ so that $\rho_i(v)>0$. Note that all non-degenerate
poles are projective. In our problem, even for degenerate poles, one can generally show
that all poles should be projective, thus allowing us to use the results of
\cite{Benini:2013xpa}. To see this, first note that
\begin{equation}
  \rho_i(u_\ast)=-z_i+m_i+n_i\tau\ ,
\end{equation}
for suitable integers $m_i,n_i$. In our problem, since $\rho_i$ is chosen among
the weight system of the $O(n)$ symmetric representation, it is either $\pm 2e_I$ or
$\pm e_I\pm e_J$ with $I,J=1,\cdots,\left[\frac{n}{2}\right]$. Thus, we can take all
$m_i,n_i$ to be either $0$ or $1$ to find all possible solutions for $u_\ast$,
mod $\mathbb{Z}+\tau\mathbb{Z}$. Also, $z_i$ is either $\epsilon_1$ or $\epsilon_2$
for all $i$'s. Then, taking a solution $u_\ast(\epsilon_1,\epsilon_2)$ which depends
on $\epsilon_{1,2}$, one deforms the solution to the regime in which $\epsilon_1$,
$\epsilon_2$ are real and negative, taken to be $-\epsilon_{1,2}\gg 1$ and
$-\epsilon_{1,2}\gg|{\rm Re}(\tau)|$.
Then one finds that $\rho_i\cdot{\rm Re}(u_\ast)>0$, fulfilling the projective
condition. In fact, one can always provide this kind of argument on the
projectivity of poles when the system has independent flavor symmetry for each
matter supermultiplet. The $\mathcal{N}=(2,2)$
or $(0,2)$ models may exhibit non-projective poles if there are nonzero superpotentials so that
flavor symmetries are restricted. In $\mathcal{N}=(0,4)$ models, independent flavor symmetry can
be found for each hypermultiplet.
This is why it is easier to apply the results of \cite{Benini:2013xpa} to $(0,4)$ theories. For
instance, the quantum mechanical version of this index formula is well applicable to the ADHM
instanton quantum mechanics \cite{Hwang:2014uwa}, as these systems always have $(0,4)$ SUSY. (The results of \cite{Hwang:2014uwa} will be used in
our section 4.)

\cite{Benini:2013xpa} finds that the integral in (\ref{index-full}) is
given by
\begin{equation}\label{final-JK}
  \frac{1}{(2\pi i)^r}\oint Z^{(a)}_{\textrm{1-loop}}=
  \sum_{u_\ast}\textrm{JK-Res}_{u_\ast}({\bf Q}_\ast,\eta)
  Z_{\textrm{1-loop}}^{(a)}\ ,
\end{equation}
where $u_\ast$ runs over all the poles in the integrand. The JK-Res appearing in
this expression is given as follows. JK-Res is a linear functional which refers
to an auxiliary vector $\eta$ in the charge space, and also to the set of charge vectors
${\bf Q}_\ast=(Q_1,\cdots,Q_d)$ for the hyperplanes crossing $u_\ast$. The defining property
of JK-Res$_{u_\ast}({\bf Q}_\ast,\eta)$ is
\begin{equation}
  \textrm{JK-Res}_{u_\ast}(\!{\bf Q}_\ast,\eta)
  \frac{dQ_{j_1}(u)\wedge \cdots\wedge dQ_{j_r}(u)}{Q_{j_1}(u\!-\!u_\ast)\cdots Q_{j_r}(u\!-\!u_\ast)}
  =\left\{\begin{array}{ll}{\rm sign}\ \det(Q_{j_1},\cdots,Q_{j_r})&
  \textrm{if }\eta\in{\rm Cone}(Q_{j_1},\cdots, Q_{j_r})\\
  0&{\rm otherwise}\end{array}\right.,
\end{equation}
or equivalently
\begin{equation}\label{JK-Res}
  \textrm{JK-Res}_{u_\ast}(\!{\bf Q}_\ast,\eta)
  \frac{du_1\wedge \cdots\wedge du_r}{Q_{j_1}(u\!-\!u_\ast)\cdots Q_{j_r}(u\!-\!u_\ast)}
  =\left\{\begin{array}{ll} | \det(Q_{j_1},\cdots,Q_{j_r}) |^{-1}&
  \textrm{if }\eta\in{\rm Cone}(Q_{j_1},\cdots, Q_{j_r})\\
  0&{\rm otherwise}\end{array}\right..
\end{equation}
To make the condition $\eta\in {\rm Cone}(Q_{j_1},\cdots, Q_{j_r})$ unambiguous,
one has to put $\eta$ at a sufficiently generic point, as explained in \cite{Benini:2013xpa}.
These rules are giving a definite residue when the integrand takes the form of
a `simple pole.' Although this definition apparently overdetermines JK-Res
due to many relations among the forms $\bigwedge_{i=1}^r\frac{dQ_{j_i}(u)}{Q_{j_i}(u)}$,
it turns out to be consistent (see \cite{Benini:2013xpa} and references
therein). As one expands the integrand $Z_{\textrm{1-loop}}^{(a)}$ around $u=u_\ast$, one
will encounter not just simple poles, but also multiple poles and less singular homogeneous
expressions in $u-u_\ast$, multiplied by $du_1\wedge\cdots\wedge du_r$. The JK-Res of the
last two classes of monomials are all (naturally) zero: this is also consistent with the
alternative `constructive definition,' which expresses JK-Res as an iterated integral over
a cycle. Using this definition to compute
the integral is especially simple for non-degenerate poles, in which case one can directly read off a unique integral of the form (\ref{JK-Res}) at a given $u=u_\ast$. The case with degenerate poles require some more work, but of course coming with a clear rule. The final
result (\ref{final-JK}) is independent of the choice of $\eta$ \cite{Benini:2013xpa}.

In the remaining part of this section, we first analyze the elliptic genera
for $n=1,2,3,4$ E-strings in great detail. In section 3.5, we then illustrate the
structure of the higher E-string indices. In particular, degenerate poles start to
appear from $n\geq 6$. The residue evaluations are almost as simple as the non-degenerate
poles for $n=6,7$, all coming from simple poles. Their residues are simply given by
combinations of theta functions. For $n\geq 8$, we explain that there start to appear
degenerate poles which are also multiple poles. Their residues are
given by theta functions and their derivatives in the elliptic parameters.

\subsection{One E-string}

We consider the elliptic genus for the $O(1)$ theory.
Since $O(1)=\mathbb{Z}_2$, there are four different flat connections $(1,1)$,
$(1,-1)$, $(-1,1)$, $(-1,-1)$. The indices in the four sectors are given by
\begin{equation}
  Z_{1(i)}=-\left[1\right]_{\rm vec}\cdot
  \left[\frac{\eta^2}{\theta_1(\epsilon_1)\theta_1(\epsilon_2)}\right]_{\rm sym\ hyper}
  \cdot\left[\prod_{l=1}^{8}\frac{\theta_i(m_l)}{\eta}\right]_{\rm Fermi}\ ,
\end{equation}
where $i=1,2,3,4$ for the Wilson line $(1,1), (-1,1), (-1,-1), (1,-1)$,
respectively. Combining all four contributions, and dividing by
the Weyl group order $|W|=2$ in each sector, the full index is given by
\begin{equation}
  Z_1=\sum_{i=1}^4\frac{Z_{1(i)}}{2}=
  -\frac{\Theta(q,m_l)}{\eta^6\theta_1(\epsilon_1)\theta_1(\epsilon_2)}\ ,
\end{equation}
where the $E_8$ theta function $\Theta$ is given by
\begin{equation}
  \Theta(\tau,m_l)=\frac{1}{2}\sum_{n=1}^4\prod_{l=1}^8\theta_n(\tau,m_l)\ .
\end{equation}
Physically, $\frac{Z_{1(1)}+Z_{1(2)}}{2}$ simply imposes the $O(1)=\mathbb{Z}_2$
singlet condition, while the remainder $\frac{Z_{1(3)}+Z_{1(4)}}{2}$ is the
contribution from the twisted sector.

In \cite{Klemm:1996hh}, the above result was derived using topological strings and was explained using an effective
free string theory calculus, in which the left moving sector consists of the
$E_8$ current algebra at level $1$ and the right moving sector consists of
a $(0,4)$ supersymmetric string with target space $\mathbb{R}^4$. The four
terms of $\Theta(\tau,m_i)$ can be understood as coming from the Ramond and
Neveu-Schwarz sectors of the left-moving fermions, and then truncating
the Hilbert space by a GSO projection. In our UV gauge theory calculus,
the twisting and GSO projection come from the $O(1)$ gauge symmetry.
These summation and projection will generalize curiously to higher $O(n)$ gauge
theories below. It will be interesting to see if one can provide a CFT
interpretation, extending the notions of twisted sectors and GSO projection.

Since $\Theta(q,m_l)$ is given by the summation over the $E_8$
root lattice, $Z_1$ has a manifest $E_8$ symmetry, and is expanded
as the sum of $E_8$ characters. This supports the IR enhancement
$SO(16)\rightarrow E_8$ of global symmetry in our gauge theory.

\subsection{Two E-strings}

Now we consider the $O(2)$ theory. There are $7$ sectors of
$O(2)$ Wilson lines given by (\ref{flat-O(2)}). One in the (ee) sector
has a complex modulus, while the other six are all discrete. We name the sectors
as follows, where $(a_+,a_-)$ are the two eigenvalues of $u$ in the discrete sectors
which act on the fundamental representation \cite{Benini:2013nda}:
\begin{eqnarray}
  (0)\equiv(\textrm{ee})&:&(U_1,U_2)=(e^{iu_1\sigma_2},e^{iu_2\sigma_2})\nonumber\\
  (1),(2)\equiv(\textrm{oe})_{\pm}&:&(\sigma_3,\pm 1)\ \rightarrow\ \
  (a_v,a_+,a_-)=(\tfrac{1}{2},0,\tfrac{1}{2})\ ,\ \ (\tfrac{1}{2},\tfrac{\tau}{2},\tfrac{1+\tau}{2})\nonumber\\
  (3),(4)\equiv(\textrm{eo})_{\pm}&:&(\pm 1,\sigma_3)\ \rightarrow\ \ (a_v,a_+,a_-)=(\tfrac{\tau}{2},0,\tfrac{\tau}{2})\ ,\ \
  (\tfrac{\tau}{2},\tfrac{1}{2},\tfrac{1+\tau}{2})\nonumber\\
  (5),(6)\equiv(\textrm{oo})_{\pm}&:&(\pm \sigma_3,\sigma_3)\ \rightarrow\ \
  (a_v,a_+,a_-)=(\tfrac{1+\tau}{2},0,\tfrac{1+\tau}{2})\ ,\
  (\tfrac{1+\tau}{2},\tfrac{1}{2},\tfrac{\tau}{2})\ .\nonumber
\end{eqnarray}
All eigenvalues $a_+,a_-$ are defined mod $\mathbb{Z}+\tau\mathbb{Z}$.
$a_v=a_++a_-$ is the eigenvalue acting on the $O(2)$ adjoint (antisymmetric) representation.
The discrete holonomy eigenvalues acting on the $O(2)$
symmetric representation are $a_v=a_++a_-$, $2a_+$, $2a_-$.
The contributions $Z_{2(a)}$ (with $a=0,\cdots,6$) are given by
\begin{eqnarray}
  Z_{2(0)}&=&\oint\left[\eta^2du\cdot\frac{\theta_1(2\epsilon_+)}{i\eta}
  \right]_{\rm vec}\cdot\left[\frac{\eta^6}{\theta_1(\epsilon_1)\theta_1(\epsilon_2)
  \theta_1(\epsilon_1\pm 2u)\theta_1(\epsilon_2\pm 2u)}\right]_{\rm sym}
  \cdot\left[\prod_{l=1}^8\frac{\theta_1(m_l\pm u)}{\eta^2}\right]_{\rm Fermi}\nonumber\\
  Z_{2(a)}&=& \left[\frac{\theta_1(a_v)\theta_1(2\epsilon_++a_v)}{\eta^2}
  \right]_{\rm vec}\cdot
  \left[\frac{\eta^6}{\theta_1(\epsilon_1+a_v)\theta_1(\epsilon_2+a_v)
  \theta_1(\epsilon_1+2a_\pm)\theta_1(\epsilon_2+2a_\pm)}\right]_{\rm sym}\nonumber\\
  &&\cdot\left[\prod_{l=1}^8\frac{\theta_1(m_l+a_+)\theta_1(m_l+a_-)}{\eta^2}\right]_{\rm Fermi}
  \ \ \ \ \ (a=1,\cdots,6)\ ,
\end{eqnarray}
where we defined $\epsilon_+ = \frac{\epsilon_1 + \epsilon_2}{2}$. As explained after (\ref{fractional-shift}), $\theta_1(z+a_v)$ factors should be
understood as $\theta_i$, with $i=1,2,3,4$ for $a_v=0,\frac{1}{2},\frac{1+\tau}{2},\frac{\tau}{2}$, respectively.

The contour integral in $Z_{2(0)}$ can be done by taking residues from poles
with positive $SO(2)$ electric charge only: this is the simple rule for the rank $1$
theory obtained by taking $\eta=1$ \cite{Benini:2013nda}. The relevant poles are at $\theta_1(\epsilon_1+2u)=0$ and
$\theta_1(\epsilon_2+2u)=0$. Using
\begin{equation}
  \frac{1}{2\pi i}\oint_{u=a+b\tau}\frac{du}{\theta_1(\tau|u)}=
  \frac{(-1)^{a+b}e^{i\pi b^2\tau}}{\theta_1^\prime(\tau|0)}=
  \frac{(-1)^{a+b}e^{i\pi b^2\tau}}{2\pi\eta^3}\ ,
\end{equation}
one should pick the residues at
$u=-\frac{\epsilon_{1,2}}{2}$, $-\frac{\epsilon_{1,2}}{2}+\frac{1}{2}$,
$-\frac{\epsilon_{1,2}}{2}+\frac{1+\tau}{2}$,
$-\frac{\epsilon_{1,2}}{2}+\frac{\tau}{2}$. The residue sum is
\begin{equation}
    Z_{2(0)} = \frac{1}{2\eta^{12}\theta_1 (\epsilon_1)\theta_1 (\epsilon_2)}
    \sum_{i=1}^{4} \left[ \frac{\prod_{l=1}^{8}\theta_i (m_l \pm \frac{\epsilon_1}{2})}
    {\theta_1(2\epsilon_1)\theta_1 (\epsilon_2- \epsilon_1)}+
    \frac{\prod_{l=1}^{8} \theta_i (m_l \pm \frac{\epsilon_2}{2})}
    {\theta_1(2\epsilon_2)\theta_1 (\epsilon_1- \epsilon_2)}\right].
\end{equation}
Expressions with $\pm$ signs mean $\theta_i(x\pm y)\equiv\theta_i(x+y)\theta_i(x-y)$.
The contributions from the other six sectors are
\begin{eqnarray}
    \hspace*{-1.2cm}&&Z_{2(1)}= \frac{\theta_2 (0) \theta_2 (2 \epsilon_+)
    \prod_{l=1}^{8}\theta_1 (m_l) \theta_2 (m_l)}{\eta^{12}\theta_1(\epsilon_1)^2 \theta_1(\epsilon_2)^2 \theta_2(\epsilon_1) \theta_2(\epsilon_2)}\ \ ,\ \
    Z_{2(2)}= \frac{\theta_2 (0) \theta_2 (2 \epsilon_+)\prod_{l=1}^{8}
    \theta_3 (m_l) \theta_4 (m_l)}
    {\eta^{12}\theta_1(\epsilon_1)^2 \theta_1(\epsilon_2)^2 \theta_2(\epsilon_1) \theta_2(\epsilon_2)}\ \ , \nonumber\\
    \hspace*{-1.2cm}&&Z_{2(3)}= \frac{\theta_4 (0) \theta_4 (2 \epsilon_+)\prod_{l=1}^{8}
    \theta_1 (m_l) \theta_4 (m_l)}
    {\eta^{12}\theta_1(\epsilon_1)^2 \theta_1(\epsilon_2)^2 \theta_4(\epsilon_1) \theta_4(\epsilon_2)}\ \ ,\ \
    Z_{2(4)}= \frac{\theta_4 (0) \theta_4 (2 \epsilon_+) \prod_{l=1}^{8}
    \theta_2 (m_l) \theta_3 (m_l)}
    {\eta^{12}\theta_1(\epsilon_1)^2 \theta_1(\epsilon_2)^2 \theta_4(\epsilon_1) \theta_4(\epsilon_2)}\ \ , \nonumber \\
    \hspace*{-1.2cm}&&Z_{2(5)}= \frac{\theta_3 (0) \theta_3 (2 \epsilon_+)\prod_{l=1}^{8}
    \theta_1 (m_l) \theta_3 (m_l)}
    {\eta^{12}\theta_1(\epsilon_1)^2 \theta_1(\epsilon_2)^2 \theta_3(\epsilon_1) \theta_3(\epsilon_2)}\ \ ,\ \
    Z_{2(6)}= \frac{\theta_3 (0) \theta_3 (2 \epsilon_+)\prod_{l=1}^{8}
    \theta_2 (m_l) \theta_4 (m_l)}
    {\eta^{12}\theta_1(\epsilon_1)^2 \theta_1(\epsilon_2)^2 \theta_3(\epsilon_1) \theta_3(\epsilon_2)}\ \ .
\end{eqnarray}
The two E-string elliptic genus is given by
\begin{equation}\label{2E-gauge}
  Z_2(\tau,\epsilon_{1,2},m_l)=\frac{1}{2}Z_{2(0)}+\frac{1}{4}\sum_{a=1}^6Z_{2(a)}\ ,
\end{equation}
dividing each $Z_{2(a)}$ by the order of the `Weyl group,' given by (\ref{weyl}).

Recently, \cite{Haghighat:2014pva} obtained the 2 E-string elliptic genus. This was
done by constraining its form with its modularity, the `domain wall' ansatz of
\cite{Haghighat:2013gba}, and a few low order coefficients in the genus expansion
known from the topological string calculus. The result of \cite{Haghighat:2014pva}
is given by
\begin{eqnarray}\label{2E-E8}
  Z_2&=&\frac{1}{576\eta^{12}\theta_1(\epsilon_1)\theta_1(\epsilon_2)
  \theta_1(\epsilon_2-\epsilon_1)\theta_1(2\epsilon_1)}\left[4A_1^2(\phi_{0,1}(\epsilon_1)^2
  \!-\!E_4\theta_{-2,1}(\epsilon_1)^2)\frac{}{}\right.\\
  &&\hspace{-1cm}\left.\frac{}{}+3A_2(E_4^2\phi_{-2,1}(\epsilon_1)^2
  \!-\!E_6\phi_{-2,1}(\epsilon_1)\phi_{0,1}(\epsilon_1))
  +5B_2(E_6\phi_{-2,1}(\epsilon_1)^2-E_4\phi_{-2,1}(\epsilon_1)
  \phi_{0,1}(\epsilon_1))\right] + (\epsilon_1\leftrightarrow\epsilon_2)\nonumber
\end{eqnarray}
where $E_4(\tau)$, $E_6(\tau)$ are the Eisenstein series, summarized in appendix A,
\begin{equation}
  \phi_{-2,1}(\epsilon,\tau)=-\frac{\theta_1(\epsilon,\tau)^2}{\eta(\tau)^6}\ ,\ \
  \phi_{0,1}(\epsilon,\tau)=4\left[
  \frac{\theta_2(\epsilon,\tau)^2}{\theta_2(0,\tau)^2}+
  \frac{\theta_3(\epsilon,\tau)^2}{\theta_3(0,\tau)^2}+
  \frac{\theta_4(\epsilon,\tau)^2}{\theta_4(0,\tau)^2}\right]\ ,
\end{equation}
and $A_1(m_l)$, $A_2(m_l)$, $B_2(m_l)$ are three of the nine Jacobi forms
which are invariant under the Weyl group of $E_8$. See, for instance, the appendix
of \cite{Huang:2013yta} for the full list. $A_1$
is simply the $E_8$ theta function $A_1=\Theta(m_l,\tau)$, and
\begin{eqnarray}\label{eq:e8weyl-inv}
  A_2&=&\frac{8}{9}\left[\Theta(2m_l,2\tau)
  +\frac{\Theta(m_l,\frac{\tau}{2})+\Theta(m_l,\frac{\tau+1}{2})}{16}\right]\\
  B_2&=&\frac{8}{15}\left[(\theta_3^4+\theta_4^4)\Theta(2m_l,2\tau)
  -\frac{1}{16}(\theta_2^4+\theta_3^4)\Theta(m_l,\tfrac{\tau}{2})
  +\frac{1}{16}(\theta_2^4-\theta_4^4)\Theta(m_l,\tfrac{\tau+1}{2})\right]\ ,\nonumber
\end{eqnarray}
where $\theta_i\equiv \theta_i(0)$.
We made a full analytic proof, at $\epsilon_1=-\epsilon_2$ for simplicity
(but keeping all $E_8$ masses and $\epsilon_-=\frac{\epsilon_1-\epsilon_2}{2}$),
that (\ref{2E-gauge}) and (\ref{2E-E8}) agree with each other. See appendix C
for our proof. On one side, this agreement shows that the `domain wall ansatz'
of \cite{Haghighat:2014pva} is at work. On the other hand, it also shows that our
gauge theory index exhibits the Weyl symmetry of $E_8$, which is manifest in (\ref{2E-E8}).
So this supports the IR $E_8$ symmetry enhancement of our gauge theory.

\subsection{Three E-strings}

There are eight sectors of $O(3)$ holonomies on $T^2$, which we label as follows:
\begin{eqnarray}
  ({\rm ee})&:&{\rm diag}(e^{iu_1\sigma_2},1),\ {\rm diag}(e^{iu_2\sigma_2},1)
  \rightarrow(1)\ \ ;\ \
  {\rm diag}(-1,-1,1),\ {\rm diag}(1,-1,-1)\rightarrow(1)^\prime\ ;\nonumber\\
  ({\rm eo})&:&{\rm diag}(e^{iu_1\sigma_2},1),\ {\rm diag}(e^{iu_2\sigma_2},-1)
  \rightarrow(4)\ \ ;\ \
  {\rm diag}(-1,-1,1),\ {\rm diag}(1,-1,1)\rightarrow(4)^\prime\ ;\nonumber\\
  ({\rm oe})&:&{\rm diag}(e^{iu_1\sigma_2},-1),\ {\rm diag}(e^{iu_2\sigma_2},1)
  \rightarrow(2)\ \ ;\ \
  {\rm diag}(1,-1,1),\ {\rm diag}(-1,-1,1)\rightarrow(2)^\prime\ ;\nonumber\\
  ({\rm oo})&:&{\rm diag}(e^{iu_1\sigma_2},-1),\ {\rm diag}(e^{iu_2\sigma_2},-1)
  \rightarrow(3)\ \ ;\ \
  {\rm diag}(1,1,-1),\ {\rm diag}(1,-1,1)\rightarrow(3)^\prime\ .\nonumber
\end{eqnarray}
The indices in various sectors are given as follows. Firstly,
\begin{eqnarray}
  Z_{3(1)}&=&-\oint\left[\eta^2 du\cdot
  \frac{\theta_1(2\epsilon_+)\theta_1(2\epsilon_+\pm u)\theta_1(\pm u)}{i\eta^5}\right]_{\rm vec}
  \cdot\left[\frac{\eta^{12}}{\theta_1(\epsilon_{1,2})^2\theta_1(\epsilon_{1,2}\pm u)
  \theta_1(\epsilon_{1,2}\pm 2u)}\right]_{\rm sym}\nonumber\\
  &&\cdot\left[\prod_{l=1}^8\frac{\theta_1(m_l)\theta_1(m_l+u)\theta_1(m_l-u)}{\eta^3}\right]_{\rm Fermi}\\
  Z_{3(1)^\prime}&=&-\left[\frac{\theta_2(0)\theta_3(0)\theta_4(0)
  \theta_2(2\epsilon_+)\theta_3(2\epsilon_+)\theta_4(2\epsilon_+)}{\eta^6}\right]_{\rm vec}\cdot
  \left[\frac{\eta^{12}}{\theta_1(\epsilon_{1,2})^3\theta_2(\epsilon_{1,2})\theta_3(\epsilon_{1,2})
  \theta_4(\epsilon_{1,2})}\right]_{\rm sym}\nonumber\\
  &&\cdot\left[\prod_{l=1}^8\frac{\theta_2(m_l)\theta_3(m_l)\theta_4(m_l)}{\eta^3}\right]_{\rm Fermi}\ .
\end{eqnarray}
$Z_{3(1)^\prime}$ is obtained with discrete holonomy $(a_1,a_2,a_3)=(\frac{1}{2},\frac{1+\tau}{2},\frac{\tau}{2})$ acting on the fundamental, $(a_1+a_2,a_2+a_3,a_3+a_1)=(\frac{\tau}{2},\frac{1}{2},\frac{1+\tau}{2})$ on adjoint,
and $(2a_1,2a_2,2a_3,a_1+a_2,a_2+a_3,a_3+a_1)$ on symmetric representations. Similarly,
one obtains
\begin{eqnarray}
  Z_{3(4)}&=&-\oint\left[\eta^2 du\cdot
  \frac{\theta_1(2\epsilon_+)\theta_4(2\epsilon_+\pm u)\theta_4(\pm u)}{i\eta^5}\right]_{\rm vec}
  \cdot\left[\frac{\eta^{12}}{\theta_1(\epsilon_{1,2})^2\theta_4(\epsilon_{1,2}\pm u)
  \theta_1(\epsilon_{1,2}\pm 2u)}\right]_{\rm sym}\nonumber\\
  &&\cdot\left[\prod_{l=1}^8\frac{\theta_4(m_l)\theta_1(m_l+u)\theta_1(m_l-u)}{\eta^3}\right]_{\rm Fermi}\\
  Z_{3(4)^\prime}&=&-\left[\frac{\theta_2(0)\theta_3(0)\theta_4(0)
  \theta_2(2\epsilon_+)\theta_3(2\epsilon_+)\theta_4(2\epsilon_+)}{\eta^6}\right]_{\rm vec}\cdot
  \left[\frac{\eta^{12}}{\theta_1(\epsilon_{1,2})^3\theta_2(\epsilon_{1,2})\theta_3(\epsilon_{1,2})
  \theta_4(\epsilon_{1,2})}\right]_{\rm sym}\nonumber\\
  &&\cdot\left[\prod_{l=1}^8\frac{\theta_1(m_l)\theta_2(m_l)\theta_3(m_l)}{\eta^3}\right]_{\rm Fermi}
\end{eqnarray}
from the (eo) sectors with $(a_1,a_2,a_3)=(\frac{1}{2},\frac{1+\tau}{2},0)$
for $Z_{3(4)^\prime}$,
\begin{eqnarray}
  Z_{3(2)}&=&-\oint\left[\eta^2 du\cdot
  \frac{\theta_1(2\epsilon_+)\theta_2(2\epsilon_+\pm u)\theta_2(\pm u)}{i\eta^5}\right]_{\rm vec}
  \cdot\left[\frac{\eta^{12}}{\theta_1(\epsilon_{1,2})^2\theta_2(\epsilon_{1,2}\pm u)
  \theta_1(\epsilon_{1,2}\pm 2u)}\right]_{\rm sym}\nonumber\\
  &&\cdot\left[\prod_{l=1}^8\frac{\theta_2(m_l)\theta_1(m_l+u)\theta_1(m_l-u)}{\eta^3}\right]_{\rm Fermi}\\
  Z_{3(2)^\prime}&=&-\left[\frac{\theta_2(0)\theta_3(0)\theta_4(0)
  \theta_2(2\epsilon_+)\theta_3(2\epsilon_+)\theta_4(2\epsilon_+)}{\eta^6}\right]_{\rm vec}\cdot
  \left[\frac{\eta^{12}}{\theta_1(\epsilon_{1,2})^3\theta_2(\epsilon_{1,2})\theta_3(\epsilon_{1,2})
  \theta_4(\epsilon_{1,2})}\right]_{\rm sym}\nonumber\\
  &&\cdot\left[\prod_{l=1}^8\frac{\theta_1(m_l)\theta_3(m_l)\theta_4(m_l)}{\eta^3}\right]_{\rm Fermi}
\end{eqnarray}
from the (oe) sectors with $(a_1,a_2,a_3)=(\frac{\tau}{2},\frac{1+\tau}{2},0)$
for $Z_{3(2)^\prime}$, and
\begin{eqnarray}
  Z_{3(3)}&=&-\oint\left[\eta^2 du\cdot
  \frac{\theta_1(2\epsilon_+)\theta_3(2\epsilon_+\pm u)\theta_3(\pm u)}{i\eta^5}\right]_{\rm vec}
  \cdot\left[\frac{\eta^{12}}{\theta_1(\epsilon_{1,2})^2\theta_3(\epsilon_{1,2}\pm u)
  \theta_1(\epsilon_{1,2}\pm 2u)}\right]_{\rm sym}\nonumber\\
  &&\cdot\left[\prod_{l=1}^8\frac{\theta_3(m_l)\theta_1(m_l+u)\theta_1(m_l-u)}{\eta^3}\right]_{\rm Fermi}\\
  Z_{3(3)^\prime}&=&-\left[\frac{\theta_2(0)\theta_3(0)\theta_4(0)
  \theta_2(2\epsilon_+)\theta_3(2\epsilon_+)\theta_4(2\epsilon_+)}{\eta^6}\right]_{\rm vec}\cdot
  \left[\frac{\eta^{12}}{\theta_1(\epsilon_{1,2})^3\theta_2(\epsilon_{1,2})\theta_3(\epsilon_{1,2})
  \theta_4(\epsilon_{1,2})}\right]_{\rm sym}\nonumber\\
  &&\cdot\left[\prod_{l=1}^8\frac{\theta_1(m_l)\theta_2(m_l)\theta_4(m_l)}{\eta^3}\right]_{\rm Fermi}
\end{eqnarray}
from the (oo) sectors with $(a_1,a_2,a_3)=(0,\frac{\tau}{2},\frac{1}{2})$
for $Z_{3(3)^\prime}$.
The contour integrals in $Z_{3(i)}$ acquire residue contributions from
poles $u_\ast=-\frac{\epsilon_{1,2}}{2},-\frac{\epsilon_{1,2}}{2}+\frac{1}{2},
-\frac{\epsilon_{1,2}}{2}+\frac{\tau}{2},-\frac{\epsilon_{1,2}}{2}+\frac{1+\tau}{2}$ and
$u_\ast=-\epsilon_{1,2}+\cdots$, where $\cdots$ part is decided by
$\theta_i(u+\epsilon_{1,2})=0$. The residue sums are given by
\begin{eqnarray}
  Z_{3(i)}&=&-\frac{\eta^4}{\theta_1(\epsilon_1)^2\theta_1(\epsilon_2)^2}
  \left[\frac{\eta^2\theta_1(\epsilon_1)\theta_1(\epsilon_2)}
  {\theta_1(2\epsilon_1)\theta_1(\epsilon_2-\epsilon_1)\theta_1(3\epsilon_1)
  \theta_1(\epsilon_2-2\epsilon_1)}\prod_{l=1}^{8}\frac{\theta_i(m_l)\theta_i(m_l\pm\epsilon_1)}
  {\eta^3}\right.\\
  &&\left.+\frac{1}{2}\sum_{a=1}^4\frac{\eta^2\theta_{\sigma_i(a)}(\frac{3\epsilon_1}{2}+\epsilon_2)
  \theta_{\sigma_i(a)}(-\frac{\epsilon_1}{2})}
  {\theta_1(2\epsilon_1)\theta_1(\epsilon_2-\epsilon_1)
  \theta_{\sigma_i(a)}(\frac{3\epsilon_1}{2})\theta_{\sigma_i(a)}(\epsilon_2-\frac{\epsilon_1}{2})}
  \prod_{l=1}^8\frac{\theta_i(m_l)\theta_a(m_l\pm\frac{\epsilon_1}{2})}{\eta^3}
  +(\epsilon_1\leftrightarrow\epsilon_2)\right]\nonumber
\end{eqnarray}
where the permutations are defined by
\begin{eqnarray}\label{permutation}
  &&\sigma_1(1,2,3,4)=(1,2,3,4)\ ,\ \ \sigma_2(1,2,3,4)=(2,1,4,3),\nonumber\\ &&\sigma_3(1,2,3,4)=(3,4,1,2)\ ,\ \ \sigma_4(1,2,3,4)=(4,3,2,1)\ .
\end{eqnarray}
The full index is given by
\begin{equation}\label{eq:elliptic-genus-3E}
  Z_3=\sum_{i=1}^4\left(\frac{1}{4}Z_{3(i)}+\frac{1}{8}Z_{3(i)^\prime}\right)\ ,
\end{equation}
after dividing by the Weyl factors (\ref{weyl}).

For simplicity, we study the indices at $m_l=0$,
$\epsilon_1=-\epsilon_2\equiv\epsilon$ in more detail, which are
\begin{eqnarray}
  Z_{3(i)}&=&\frac{\eta^4}{\theta_1(\epsilon)^4}\left[\frac{2
  \theta_1(\epsilon)^2\theta_i(0)^8\theta_i(\epsilon)^{16}}
  {\eta^{22}\theta_1(2\epsilon)^2\theta_1(3\epsilon)^2}
  +\sum_{a=1}^4\frac{\theta_{\sigma_i(a)}(\frac{\epsilon}{2})^2\theta_i(0)^8
  \theta_a(\frac{\epsilon}{2})^{16}}
  {\eta^{22}\theta_1(2\epsilon)^2\theta_{\sigma_i(a)}(\frac{3\epsilon}{2})^2}\right]
\end{eqnarray}
and
\begin{eqnarray}
  Z_{3(1)^\prime}&=&\frac{\theta_2(0)^{10}\theta_3(0)^{10}\theta_4(0)^{10}}
  {\eta^{18}\theta_1(\epsilon)^6\theta_2(\epsilon)^2\theta_3(\epsilon)^2\theta_4(\epsilon)^2}
  =\frac{4\theta_2(0)^{8}\theta_3(0)^{8}\theta_4(0)^{8}}
  {\eta^{18}\theta_1(\epsilon)^4\theta_1(2\epsilon)^2}\ ,
\end{eqnarray}
with $Z_{3(2)^\prime}=Z_{3(3)^\prime}=Z_{3(4)^\prime}=0$.
We consider the genus expansion of $Z_3$, where genus is defined for the topological
string amplitudes on the CY$_3$ which engineers our 6d CFT in the F-theory context.
Namely, we expand
\begin{equation}\label{genus-expansion}
  F_3\equiv Z_3-Z_1Z_2+\frac{1}{3}Z_1^{\ 3}=\sum_{n\geq 0,g\geq 0}(\epsilon_1+\epsilon_2)^n
  (\epsilon_1\epsilon_2)^{g-1}F^{(n,g,3)}(\tau)\ .
\end{equation}
Taking $\epsilon_+=0$, some known results on $F^{(0,g,3)}$ are summarized in
(\ref{eq:genus-3E}), which were computed in \cite{Mohri:2001zz} up to genus $5$.
This can be compared with $F^{(0,g,3)}$ obtained from our gauge theory index.
Numerically, we checked the agreements for $g\leq 5$ up to first $10$ terms in the
$q$ expansions, starting at $q^{-3/2}$, with the last term that we checked at $q^{15/2}$. (The two coefficients at $q^{-1/2}$ and $q^{1/2}$ are zero,
due to a vanishing theorem.)

We also analytically checked the agreements for $F^{(0,0,3)}$, $F^{(0,1,3)}$, and
a refined amplitude $F^{(1,0,3)}$, against the results known from the topological
string calculus. See appendix C for the details.

\subsection{Four E-strings}

The indices from the two sectors in the (ee) part of $O(4)$ holonomy are
\begin{eqnarray}
  Z_{4(1)}&=&-\oint\left[\eta^4 du_1du_2\cdot
  \frac{\theta_1(2\epsilon_+)^2\theta_1(2\epsilon_+\pm u_1\pm u_2)
  \theta_1(\pm u_1\pm u_2)}{\eta^{10}}\right]_{\rm vec}\\
  &&\hspace{-1cm}\cdot
  \left[\frac{\eta^{20}}{\theta_1(\epsilon_{1,2})^2\theta_1(\epsilon_{1,2}\pm u_1\pm u_2)\theta_1(\epsilon_{1,2}\pm 2u_1)\theta_1(\epsilon_{1,2}\pm 2u_2)}\right]_{\rm sym}
  \cdot\left[\prod_{l=1}^8\frac{\theta_1(m_l\pm u_1)\theta_1(m_l\pm u_2)}{\eta^4}
  \right]_{\rm Fermi}\nonumber\\
  Z_{4(1)^\prime}&=&\left[\frac{\theta_2(0)^2\theta_3(0)^2\theta_4(0)^2
  \theta_2(2\epsilon_+)^2\theta_3(2\epsilon_+)^2\theta_4(2\epsilon_+)^2}{\eta^{12}}
  \right]_{\rm vec}\nonumber\\
  &&\hspace*{-1cm}\cdot
  \left[\frac{\eta^{20}}{\theta_1(\epsilon_{1,2})^4\theta_2(\epsilon_{1,2})^2
  \theta_3(\epsilon_{1,2})^2\theta_4(\epsilon_{1,2})^2}\right]_{\rm sym}\cdot\left[\prod_{l=1}^8
  \frac{\theta_1(m_l)\theta_2(m_l)\theta_3(m_l)\theta_4(m_l)}{\eta^4}\right]_{\rm Fermi}
\end{eqnarray}
where $Z_{4(1)^\prime}$ is obtained with discrete holonomy
$(a_1,a_2,a_3,a_4)=(0,\frac{1}{2},\frac{1+\tau}{2},\frac{\tau}{2})$ for
the fundamental representation. We used a shorthand notation $\theta_i(\epsilon_{1,2})\equiv
\theta_i(\epsilon_1)\theta_i(\epsilon_2)$. The indices from the two sectors in the (oe) part are
\begin{eqnarray}
  Z_{4(2)}&=&\oint\left[\eta^2 du\cdot
  \frac{\theta_1(2\epsilon_+)\theta_2(2\epsilon_+)\theta_1(2\epsilon_+\pm u)\theta_2(2\epsilon_+\pm u)
  \theta_2(0)\theta_1(\pm u)\theta_2(\pm u)}{i\eta^{11}}\right]_{\rm vec}\\
  &&\hspace*{-1cm}
  \cdot\left[\frac{\eta^{20}}{\theta_1(\epsilon_{1,2}\pm 2u)\theta_1(\epsilon_{1,2})^3
  \theta_2(\epsilon_{1,2})
  \theta_1(\epsilon_{1,2}\pm u)\theta_2(\epsilon_{1,2}\pm u)}\right]_{\rm sym}
  \cdot\left[\prod_{l=1}^8\frac{\theta_1(m_l\pm u)\theta_1(m_l)\theta_2(m_l)}{\eta^4}
  \right]_{\rm Fermi}\nonumber\\
  Z_{4(2)^\prime}&=&\oint\left[\eta^2 du\cdot
  \frac{\theta_1(2\epsilon_+)\theta_2(2\epsilon_+)\theta_3(2\epsilon_+\pm u)\theta_4(2\epsilon_+\pm u)
  \theta_2(0)\theta_3(\pm u)\theta_4(\pm u)}{i\eta^{11}}\right]_{\rm vec}\\
  &&\hspace*{-1cm}
  \cdot\left[\frac{\eta^{20}}{\theta_1(\epsilon_{1,2}\pm 2u)\theta_1(\epsilon_{1,2})^3
  \theta_2(\epsilon_{1,2})
  \theta_3(\epsilon_{1,2}\pm u)\theta_4(\epsilon_{1,2}\pm u)}\right]_{\rm sym}
  \cdot\left[\prod_{l=1}^8\frac{\theta_1(m_l\pm u)\theta_3(m_l)\theta_4(m_l)}{\eta^4}
  \right]_{\rm Fermi}\nonumber
\end{eqnarray}
where the holonomy $(a_1,a_2,a_3,a_4)=(u,-u,0,\frac{1}{2})$ and
$(u,-u,\frac{\tau}{2},\frac{1+\tau}{2})$ are used for $Z_{4(2)}$ and $Z_{4(2)^\prime}$,
respectively. The indices from the two sectors in the (oo) part are
\begin{eqnarray}
  Z_{4(3)}&=&\oint\left[\eta^2 du\cdot
  \frac{\theta_1(2\epsilon_+)\theta_3(2\epsilon_+)\theta_1(2\epsilon_+\pm u)\theta_3(2\epsilon_+\pm u)
  \theta_3(0)\theta_1(\pm u)\theta_3(\pm u)}{i\eta^{11}}\right]_{\rm vec}\\
  &&\hspace*{-1cm}
  \cdot\left[\frac{\eta^{20}}{\theta_1(\epsilon_{1,2}\pm 2u)\theta_1(\epsilon_{1,2})^3
  \theta_3(\epsilon_{1,2})
  \theta_1(\epsilon_{1,2}\pm u)\theta_3(\epsilon_{1,2}\pm u)}\right]_{\rm sym}
  \cdot\left[\prod_{l=1}^8\frac{\theta_1(m_l\pm u)\theta_1(m_l)\theta_3(m_l)}{\eta^4}
  \right]_{\rm Fermi}\nonumber\\
  Z_{4(3)^\prime}&=&\oint\left[\eta^2 du\cdot
  \frac{\theta_1(2\epsilon_+)\theta_3(2\epsilon_+)\theta_2(2\epsilon_+\pm u)\theta_4(2\epsilon_+\pm u)
  \theta_3(0)\theta_2(\pm u)\theta_4(\pm u)}{i\eta^{11}}\right]_{\rm vec}\\
  &&\hspace*{-1cm}
  \cdot\left[\frac{\eta^{20}}{\theta_1(\epsilon_{1,2}\pm 2u)\theta_1(\epsilon_{1,2})^3
  \theta_3(\epsilon_{1,2})\theta_2(\epsilon_{1,2}\pm u)\theta_4(\epsilon_{1,2}\pm u)}\right]_{\rm sym}
  \cdot\left[\prod_{l=1}^8\frac{\theta_1(m_l\pm u)\theta_2(m_l)\theta_4(m_l)}{\eta^4}
  \right]_{\rm Fermi}\nonumber
\end{eqnarray}
where the holonomy $(a_1,a_2,a_3,a_4)=(u,-u,0,\frac{1+\tau}{2})$ and
$(u,-u,\frac{\tau}{2},\frac{1}{2})$ are used for $Z_{4(3)}$ and $Z_{4(3)^\prime}$,
respectively. Finally, the indices from the two sectors in the (eo) part are
\begin{eqnarray}
  Z_{4(4)}&=&\oint\left[\eta^2 du\cdot
  \frac{\theta_1(2\epsilon_+)\theta_4(2\epsilon_+)\theta_1(2\epsilon_+\pm u)\theta_4(2\epsilon_+\pm u)
  \theta_4(0)\theta_1(\pm u)\theta_4(\pm u)}{i\eta^{11}}\right]_{\rm vec}\\
  &&\hspace*{-1cm}\cdot\left[\frac{\eta^{20}}{\theta_1(\epsilon_{1,2}\pm 2u)\theta_1(\epsilon_{1,2})^3
  \theta_4(\epsilon_{1,2})
  \theta_1(\epsilon_{1,2}\pm u)\theta_4(\epsilon_{1,2}\pm u)}\right]_{\rm sym}
  \cdot\left[\prod_{l=1}^8\frac{\theta_1(m_l\pm u)\theta_1(m_l)\theta_4(m_l)}{\eta^4}
  \right]_{\rm Fermi}\nonumber\\
  Z_{4(4)^\prime}&=&\oint\left[\eta^2 du\cdot
  \frac{\theta_1(2\epsilon_+)\theta_4(2\epsilon_+)\theta_2(2\epsilon_+\pm u)\theta_3(2\epsilon_+\pm u)
  \theta_4(0)\theta_2(\pm u)\theta_3(\pm u)}{i\eta^{11}}\right]_{\rm vec}\\
  &&\hspace*{-1cm}
  \cdot\left[\frac{\eta^{20}}{\theta_1(\epsilon_{1,2}\pm 2u)\theta_1(\epsilon_{1,2})^3
  \theta_4(\epsilon_{1,2})
  \theta_2(\epsilon_{1,2}\pm u)\theta_3(\epsilon_{1,2}\pm u)}\right]_{\rm sym}
  \cdot\left[\prod_{l=1}^8\frac{\theta_1(m_l\pm u)\theta_2(m_l)\theta_3(m_l)}{\eta^4}
  \right]_{\rm Fermi}\nonumber
\end{eqnarray}
where the holonomy $(a_1,a_2,a_3,a_4)=(u,-u,0,\frac{\tau}{2})$ and
$(u,-u,\frac{1}{2},\frac{1+\tau}{2})$ are used for $Z_{4(4)}$ and $Z_{4(4)^\prime}$,
respectively.

We also need to specify the residues which contribute to the above contour integrals.
For the rank $1$ cases, one just keeps all poles
and residues associated with positively charged chiral fields.
So for $Z_{4(i)}$ with $i=2,3,4$, the relevant poles are at
$u_\ast=-\frac{\epsilon_{1,2}}{2}+\frac{p}{2}$, where $p$ runs over $(p_1,p_2,p_3,p_4)=(0,1,1+\tau,\tau)$, and
$u_\ast=-\epsilon_{1,2}$, $-\epsilon_{1,2}+\frac{p_i}{2}$.
For $Z_{4(i)^\prime}$ with $i=2,3,4$, the
poles are at $u_\ast=-\frac{\epsilon_{1,2}}{2}+\frac{p}{2}$, again with $p$ running over $(p_1,p_2,p_3,p_4)=(0,1,1+\tau,\tau)$, and at
$u_\ast=-\epsilon_{1,2}+p_j$ with two possible values of $j\neq 1,i$.
The resulting residue sums are given by
\begin{eqnarray}
  \hspace*{-1cm}Z_{4(2)}&=&\frac{1}{2}\sum_{i=1}^4
  \frac{\theta_2(\epsilon_1\!+\!\epsilon_2)\theta_i(\frac{3\epsilon_1}{2}\!+\!\epsilon_2)
  \theta_{\sigma_2(i)}(\frac{3\epsilon_1}{2}\!+\!\epsilon_2)\theta_2(0)
  \theta_i(-\frac{\epsilon_1}{2})\theta_{\sigma_2(i)}(-\frac{\epsilon_1}{2})
  \prod_l\theta_1(m_l)\theta_2(m_l)\theta_i(m_l\!\pm\!\frac{\epsilon_1}{2})}
  {\eta^{24}\theta_1(2\epsilon_1)\theta_1(\epsilon_2-\epsilon_1)\theta_1(\epsilon_{1,2})^3
  \theta_2(\epsilon_{1,2})\theta_i(\frac{3\epsilon_1}{2})
  \theta_i(\epsilon_2-\frac{\epsilon_1}{2})\theta_{\sigma_2(i)}(\frac{3\epsilon_1}{2})
  \theta_{\sigma_2(i)}(\epsilon_2-\frac{\epsilon_1}{2})}\nonumber\\
  \hspace*{-1cm}&&+\frac{\theta_2(2\epsilon_1+\epsilon_2)
  \theta_2(\epsilon_1)\left(\prod_l\theta_1(m_l\pm\epsilon_1)+\prod_l\theta_2(m_l\pm\epsilon_1)
  \right)\prod_l\theta_1(m_l)\theta_2(m_l)}
  {\eta^{24}\theta_1(3\epsilon_1)\theta_1(\epsilon_2-2\epsilon_1)\theta_1(\epsilon_{1,2})^2
  \theta_1(2\epsilon_1)\theta_1(\epsilon_2-\epsilon_1)
  \theta_2(2\epsilon_1)\theta_2(\epsilon_2-\epsilon_1)}+(\epsilon_1\leftrightarrow\epsilon_2)\\
  \hspace*{-1cm}Z_{4(2)^\prime}&=&
  \frac{1}{2}\sum_{i=1}^4\frac{\theta_2(\epsilon_1\!+\!\epsilon_2)
  \theta_{\sigma_3(i)}(\frac{3\epsilon_1}{2}\!+\!\epsilon_2)
  \theta_{\sigma_4(i)}(\frac{3\epsilon_1}{2}\!+\!\epsilon_2)\theta_2(0)
  \theta_{\sigma_3(i)}(-\frac{\epsilon_1}{2})\theta_{\sigma_4(i)}(-\frac{\epsilon_1}{2})
  \prod_l\theta_3(m_l)\theta_4(m_l)\theta_i(m_l\!\pm\!\frac{\epsilon_1}{2})}
  {\eta^{24}\theta_1(2\epsilon_1)\theta_1(\epsilon_2-\epsilon_1)\theta_1(\epsilon_{1,2})^3
  \theta_2(\epsilon_{1,2})\theta_{\sigma_3(i)}(\frac{3\epsilon_1}{2})
  \theta_{\sigma_3(i)}(\epsilon_2-\frac{\epsilon_1}{2})
  \theta_{\sigma_4(i)}(\frac{3\epsilon_1}{2})
  \theta_{\sigma_4(i)}(\epsilon_2-\frac{\epsilon_1}{2})}\nonumber\\
  \hspace*{-1cm}&&+\frac{\theta_2(2\epsilon_1+\epsilon_2)
  \theta_2(\epsilon_1)\left(\prod_l\theta_3(m_l\pm\epsilon_1)+\prod_l\theta_4(m_l\pm\epsilon_1)\right)
  \prod_l\theta_3(m_l)\theta_4(m_l)}
  {\eta^{24}\theta_1(3\epsilon_1)\theta_1(\epsilon_2-2\epsilon_1)\theta_1(\epsilon_{1,2})^2
  \theta_1(2\epsilon_1)\theta_1(\epsilon_2-\epsilon_1)
  \theta_2(2\epsilon_1)\theta_2(\epsilon_2-\epsilon_1)}+(\epsilon_1\leftrightarrow\epsilon_2)
\end{eqnarray}
where $\sigma_i$ are defined as (\ref{permutation}). The expressions for
$Z_{4(i)}$ and $Z_{4(i)^\prime}$ with $i=3,4$ are obtained by permuting the roles
of the subscripts $2,3,4$ of the theta functions and $\sigma_i$.

The rank $2$ contour integral in $Z_{4(1)}$ can be done as follows.
The charges of the $(0,2)$ chiral multiplets, responsible for the poles in the integrand,
are $\pm 2e_I$, $\pm e_I\pm e_J$ ($I\neq J$) with $I,J=1,2$. We choose the vector $\eta$
to be in the cone between $e_1+e_2$ and $2e_2$. Then, the poles with nonzero
Jeffrey-Kirwan residues (after eliminating the fake poles due to vanishing numerators
from Fermi multiplets) are at the following $104$ positions:
\begin{eqnarray}
  (1)&:& 2u_2+\epsilon=0,\ u_1+u_2+\epsilon^\prime=0\ \rightarrow\
  u_2=-\frac{\epsilon}{2}+\frac{p_i}{2},\ u_1=-\epsilon^\prime+\frac{\epsilon}{2}+\frac{p_i}{2}\\
  (2)&:&2u_2+\epsilon=0,\ 2u_1+\epsilon=0\ \rightarrow\
  u_2=-\frac{\epsilon}{2}+\frac{p_i}{2},\
  u_1=-\frac{\epsilon}{2}+\frac{p_j}{2}\ \ \ (p_i\neq p_j)\nonumber\\
  (3)&:&2u_2+\epsilon=0,\ 2u_1+\epsilon^\prime=0\ \rightarrow\
  u_2=-\frac{\epsilon}{2}+\frac{p_i}{2},\
  u_1=-\frac{\epsilon^\prime}{2}+\frac{p_j}{2}\nonumber\\
  (4)&:&2u_2+\epsilon=0,\ u_1-u_2+\epsilon=0\ \rightarrow\
  u_2=-\frac{\epsilon}{2}+\frac{p_i}{2},\
  u_1=-\frac{3\epsilon}{2}+\frac{p_i}{2}\nonumber\\
  (5)&:&u_2-u_1+\epsilon=0,\ u_1+u_2+\epsilon=0\ \rightarrow\
  u_2=-\epsilon+\frac{p_i}{2},\
  u_1=0+\frac{p_i}{2}\nonumber\\
  (6)&:&u_2-u_1+\epsilon=0,\ u_1+u_2+\epsilon^\prime=0\ \rightarrow\
  u_2=-\frac{\epsilon+\epsilon^\prime}{2}+\frac{p_i}{2},\
  u_1=-\frac{\epsilon^\prime-\epsilon}{2}+\frac{p_i}{2}\nonumber\\
  (7)&:&u_2-u_1+\epsilon=0,\ 2u_1+\epsilon=0\ \rightarrow\
  u_2=-\frac{3\epsilon}{2}+\frac{p_i}{2},\
  u_1=-\frac{\epsilon}{2}+\frac{p_i}{2}\nonumber\\
  (8)&:&-2u_1+\epsilon=0,u_1+u_2+\epsilon=0\ \rightarrow\
  u_1=+\frac{\epsilon}{2}+\frac{p_i}{2},\
  u_2=-\frac{3\epsilon}{2}+\frac{p_i}{2}\ .\nonumber
\end{eqnarray}
We defined $(p_1,p_2,p_3,p_4)=(0,1,1+\tau,\tau)$. $\epsilon$ can be
either $\epsilon_1$ or $\epsilon_2$, and $\epsilon^\prime\neq \epsilon$
is chosen between $\epsilon_1,\epsilon_2$ at given $\epsilon$.
In the second case, the four cases with $p_i=p_j$ do
not provide poles since there are vanishing factors in the numerator. One can
check that these poles are all non-degenerate.

The residue sums from these $8$ cases are given by (the sectors labeled by
(4), (7), (8) yield same result, shown on the second line)
\begin{eqnarray}
  (1)&:&\sum_{i=1}^4\frac{\theta_1(2\epsilon_1\!+\!\epsilon_2)
  \theta_1(-\epsilon_1)\prod_l\theta_i(m_l\pm(\epsilon_1\!-\!\frac{\epsilon_2}{2}))
  \theta_i(m_l\pm\frac{\epsilon_2}{2})}{2\eta^{24}\theta_1(\epsilon_{1,2})^2
  \theta_1(2\epsilon_1)\theta_1(\epsilon_2\!-\!\epsilon_1)
  \theta_1(2\epsilon_1\!-\!\epsilon_2)\theta_1(2\epsilon_2\!-\!\epsilon_1)
  \theta_1(3\epsilon_1\!-\!\epsilon_2)\theta_1(2\epsilon_2\!-\!2\epsilon_1)}
  +(\epsilon_1\leftrightarrow\epsilon_2)\nonumber\\
  (4)&:&
  \sum_{i=1}^4\frac{\prod_l\theta_i(m_l\pm\frac{\epsilon_1}{2})
  \theta_i(m_l\pm \frac{3\epsilon_1}{2})}
  {2\eta^{24}\theta_1(\epsilon_{1,2})
  \theta_1(2\epsilon_1)\theta_1(3\epsilon_1)\theta_1(4\epsilon_1)
  \theta_1(\epsilon_2\!-\!\epsilon_1)\theta_1(\epsilon_2\!-\!2\epsilon_1)
  \theta_1(\epsilon_2\!-\!3\epsilon_1)}+(\epsilon_1\leftrightarrow\epsilon_2)
  =(7)=(8)\nonumber\\
  (5)&:&\sum_{i=1}^4\frac{
  \theta_1(2\epsilon_1+\epsilon_2)\theta_1(-\epsilon_1)
  \prod_l\theta_i(m_l)^2\theta_i(m_l\pm\epsilon_1)}
  {2\eta^{24}\theta_1(\epsilon_{1,2})^2\theta_1(2\epsilon_1)^2
  \theta_1(\epsilon_2-\epsilon_1)^2
  \theta_1(3\epsilon_1)\theta_1(\epsilon_2-2\epsilon_1)}
  +(\epsilon_1\leftrightarrow\epsilon_2)\nonumber\\
  (6)&:&\sum_{i=1}^4\frac{
  \prod_l\theta_i(m_l\pm\frac{\epsilon_1+\epsilon_2}{2})
  \theta_i(m_l\pm\frac{\epsilon_1-\epsilon_2}{2})}{\eta^{24}
  \theta_1(\epsilon_{1,2})\theta_1(2\epsilon_1)\theta_1(\epsilon_1-\epsilon_2)
  \theta_1(2\epsilon_2)\theta_1(\epsilon_2-\epsilon_1)
  \theta_1(2\epsilon_1-\epsilon_2)\theta_1(2\epsilon_2-\epsilon_1)}
\end{eqnarray}
and
\begin{eqnarray}
  (2)&:&\left[\frac{}{}\right.\!\!
  \frac{\theta_2(0)\theta_2(-\epsilon_1)\theta_2(\epsilon_1\!+\!\epsilon_2)
  \theta_2(2\epsilon_1\!+\!\epsilon_2)
  \left(\prod_l\theta_1(m_l\!\pm\!\frac{\epsilon_1}{2})\theta_2(m_l\!\pm\!\frac{\epsilon_1}{2})
  +\prod_l\theta_3(m_l\!\pm\!\frac{\epsilon_1}{2})\theta_4(m_l\!\pm\!\frac{\epsilon_1}{2})\right)}
  {2\eta^{24}\theta_1(\epsilon_{1,2})^2\theta_1(2\epsilon_1)^2\theta_1(\epsilon_2-\epsilon_1)^2
  \theta_2(\epsilon_{1,2})\theta_2(2\epsilon_1)\theta_2(\epsilon_2-\epsilon_1)}\nonumber\\
  &&\hspace{.5cm}+(2,3,4\rightarrow 3,4,2)+(2,3,4\rightarrow 4,2,3)\!\!\left.\frac{}{}\right]
  +(\epsilon_1\leftrightarrow\epsilon_2)\\
  (3)&:&\sum_{i,j=1}^4\frac{\prod_l\theta_j(m_l\pm\frac{\epsilon_1}{2})
  \theta_i(m_l\pm\frac{\epsilon_2}{2})}{2\eta^{24}\theta_1(\epsilon_{1,2})^2
  \theta_1(2\epsilon_1)\theta_1(\epsilon_2-\epsilon_1)\theta_1(2\epsilon_2)
  \theta_1(\epsilon_1-\epsilon_2)}\
  \frac{\theta_{\sigma_j(i)}(-\frac{\epsilon_1+\epsilon_2}{2})
  \theta_{\sigma_j(i)}(\frac{3(\epsilon_1+\epsilon_2)}{2})}
  {\theta_{\sigma_j(i)}(\frac{3\epsilon_1-\epsilon_2}{2})
  \theta_{\sigma_j(i)}(\frac{3\epsilon_2-\epsilon_1}{2})}\ .
\end{eqnarray}
$Z_{4(1)}$ is given by the sum of eight contributions $(1),\cdots,(8)$.
The full index is given by
\begin{equation}\label{eq:elliptic-genus-4E}
  Z_4=\frac{1}{8}\sum_{i=1}^4Z_{4(i)}+\frac{1}{8}\sum_{i=2}^4Z_{4(i)^\prime}
  +\frac{1}{16}Z_{4(1)^\prime}\ ,
\end{equation}
with the Weyl factors given by (\ref{weyl}).

We test our results against various known ones. We first consider the case
in which one sets
\begin{equation}\label{sakai}
  \epsilon_1=-\epsilon_2\equiv\epsilon,\ m_1=m_2=0, m_3=m_4=\frac{1}{2},\ m_5=m_6=-\frac{1+\tau}{2},\ m_7=m_8=\frac{\tau}{2}\ .
\end{equation}
This case was considered recently in \cite{Sakai:2014hsa}. In particular,
\cite{Sakai:2014hsa} wrote down the concrete forms of the elliptic genera in
this limit for $2$ and $4$ E-strings. The case
with $2$ E-strings is a special case of \cite{Haghighat:2014pva}, so also agrees with
our results. The index of \cite{Sakai:2014hsa} at (\ref{sakai}) is always zero for odd
number of E-strings. By plugging in (\ref{sakai}) to our
$3$ E-string indices in the previous subsection, all $Z_{3(i)},Z_{3(i)^\prime}$
are identically zero, agreeing with the results of \cite{Sakai:2014hsa}.
Now let us study our $4$ E-string index. Plugging in (\ref{sakai}),
one finds that the contributions from the seven sectors are zero,
and the only nonzero contribution is $Z_{4(1)}$. The surviving contributions are
\begin{eqnarray}
  (1)=(4)=(7)=(8)&=&\frac{4\prod_{i=1}^4\theta_i(3\epsilon/2)^{4}\theta_i(\epsilon/2)^{4}}
  {\eta^{24}\theta_1(\epsilon)^2\theta_1(2\epsilon)^2
  \theta_1(3\epsilon)^2\theta_1(4\epsilon)^2}\nonumber\\
  (2)=(3)&=&\frac{2\prod_i\theta_i(\epsilon/2)^8}
  {\eta^{24}\theta_1(\epsilon)^4\theta_1(2\epsilon)^4}
  \left[\frac{\theta_2(0)^2}{\theta_2(2\epsilon)^2}+\frac{\theta_3(0)^2}{\theta_3(2\epsilon)^2}
  +\frac{\theta_4(0)^2}{\theta_4(2\epsilon)^2}\right]
\end{eqnarray}
while $(5),(6)$ become zero. So one obtains
\begin{eqnarray}
  Z_{4(1)}&=&\frac{16\prod_{i=1}^4\theta_i(\frac{3\epsilon}{2})^{4}\theta_i(\frac{\epsilon}{2})^{4}}
  {\eta^{24}\theta_1(\epsilon)^2\theta_1(2\epsilon)^2
  \theta_1(3\epsilon)^2\theta_1(4\epsilon)^2}+
  \frac{4\prod_i\theta_i(\frac{\epsilon}{2})^8}
  {\eta^{24}\theta_1(\epsilon)^4\theta_1(2\epsilon)^4}
  \left[\frac{\theta_2(0)^2}{\theta_2(2\epsilon)^2}+\frac{\theta_3(0)^2}{\theta_3(2\epsilon)^2}
  +\frac{\theta_4(0)^2}{\theta_4(2\epsilon)^2}\right]\nonumber\\
  &=&\frac{16\theta_1(\epsilon)^2\theta_1(3\epsilon)^2}{\theta_1(2\epsilon)^2
  \theta_1(4\epsilon)^2}+\frac{4\theta_1(\epsilon)^4}{\theta_1(2\epsilon)^4}
  \left[\frac{\theta_2(0)^2}{\theta_2(2\epsilon)^2}+\frac{\theta_3(0)^2}{\theta_3(2\epsilon)^2}
  +\frac{\theta_4(0)^2}{\theta_4(2\epsilon)^2}\right]\ .
\end{eqnarray}
The four E-string index at (\ref{sakai}) is given in \cite{Sakai:2014hsa} by
\begin{equation}
  \frac{\theta_1(\epsilon)^{20}}{2\eta^{48}\theta_1(2\epsilon)^2\theta_1(4\epsilon)^2}
  \left[72(\wp^\prime)^4\wp^2-18(\wp^{\prime\prime})^2(\wp^\prime)^2\wp+2\wp^{\prime\prime}
  (\wp^\prime)^4+(\wp^{\prime\prime})^4\right]\ ,
\end{equation}
where $\wp(\tau,\epsilon)$ is the Weierstrass's elliptic function. We checked that this
agrees with our index $\frac{1}{8}Z_{4(1)}$ in a serious expansion in $q$ for the first
$11$ terms, up to and including $\mathcal{O}(q^{10})$.

We also compare our result with the genus expansion, at $m_l=0$ and
$\epsilon_1=-\epsilon_2=\epsilon$. Our indices become
\begin{eqnarray}
  Z_{4(1)}&=&\sum_{i=1}^4\left[\frac{4\theta_i(\frac{3\epsilon}{2})^{16}
  \theta_i(\frac{\epsilon}{2})^{16}}
  {\eta^{24}\theta_1(\epsilon)^2\theta_1(2\epsilon)^2\theta_1(3\epsilon)^2
  \theta_1(4\epsilon)^2}+\frac{2\theta_i(0)^{16}\theta_i(\epsilon)^{16}}
  {\eta^{24}\theta_1(\epsilon)^2\theta_1(2\epsilon)^4\theta_1(3\epsilon)^2}\right]\\
  &&+\frac{2}{\eta^{24}\theta_1(\epsilon)^4\theta_1(2\epsilon)^4}\left[\frac{\theta_2(0)^2
  (\theta_1(\frac{\epsilon}{2})^{16}\theta_2(\frac{\epsilon}{2})^{16}
  +\theta_3(\frac{\epsilon}{2})^{16}\theta_4(\frac{\epsilon}{2})^{16})}
  {\theta_2(2\epsilon)^2}+ (3,4,2)+(4,2,3)\right]\nonumber\\
  Z_{4(2)^\prime}&=&\sum_{i=1}^4\frac{\theta_2(0)^2\theta_{\sigma_3(i)}(\frac{\epsilon}{2})^2
  \theta_{\sigma_4(i)}(\frac{\epsilon}{2})^2\theta_3(0)^8\theta_4(0)^8
  \theta_i(\frac{\epsilon}{2})^{16}}
  {\eta^{24}\theta_1(2\epsilon)^2\theta_1(\epsilon)^6\theta_2(\epsilon)^2
  \theta_{\sigma_3(i)}(\frac{3\epsilon}{2})^2\theta_{\sigma_4(i)}(\frac{3\epsilon}{2})^2}
  +\frac{2\theta_2(\epsilon)^2\theta_3(0)^8\theta_4(0)^8(\theta_3(\epsilon)^{16}+
  \theta_4(\epsilon)^{16})}{\eta^{24}\theta_1(3\epsilon)^2\theta_1(2\epsilon)^2
  \theta_1(\epsilon)^4\theta_2(2\epsilon)^2}\ ,\nonumber
\end{eqnarray}
with $Z_{4(1)^\prime}=0$, $Z_{4(2)}=Z_{4(3)}=Z_{4(4)}=0$, and $Z_{4(3)^\prime}$,
$Z_{4(4)^\prime}$ are obtained from $Z_{4(2)^\prime}$ by changing the
roles of $2,3,4$ appearing in the subscripts of the theta functions and
$\sigma_2(i),\sigma_3(i),\sigma_4(i)$. We first confirmed numerically the agreement with
$F^{(0,g,4)}$ computed from topological strings for $g\leq 5$ till $q^{5}$, by checking the first $10$ terms in the serious expansion in $q$. We also exactly checked the agreements
of $F^{(0,0,4)}$, $F^{(0,1,4)}$, $F^{(0,2,4)}$. See appendix C for the details.

\subsection{Higher E-strings}

The computation of the elliptic genus using the methods of \cite{Benini:2013xpa}
quickly becomes complicated for higher rank gauge groups. In general, there could
be a fundamental complication due to some poles failing to be projective. But we showed at the beginning of this section that this does not happen in our problem.
So the computation of the elliptic genus can be done using our
methods for any number of E-strings. With higher rank,
the computational problem is that there is a large number of poles and residues to be
considered. For $U(n)$ indices, the possible poles are often completely classified by the so-called `colored Young diagrams.'
This classification first appeared in the context of instanton counting \cite{Nekrasov:2002qd,Nekrasov:2003rj},
which was reproduced recently in the context of Jeffrey-Kirwan residues \cite{Hwang:2014uwa}.
The resulting residues are often nicely arranged into a reasonably compact
form \cite{Flume:2002az,Bruzzo:2002xf}. However, for other gauge
groups, we are not aware of systematic classifications of poles.\footnote{The pole
structure of our $O(n)$ index is similar to that of the $Sp(N)$ instanton partition
function, whose ADHM quantum mechanics comes with $O(n)$ group for $n$ instantons.
The poles in our E-string index could be slightly simpler, because we only have
$O(n)$ symmetric hypermultiplets while the ADHM mechanics also has extra $N$ fundamental
hypermultiplets. In either case, we do not know the pole classification, apart from the
basic rule given by the Jeffrey-Kirwan residues.}
In this subsection, we shall illustrate the pole structures for
some higher E-strings, with $O(5)$, $O(6)$, $O(7)$, $O(8)$ gauge groups, and also make
some qualitative classifications of these poles.
Since the purpose is to illustrate the computations for higher ranks, we only consider
the branch of $O(n)$ holonomy with maximal number of continuous parameters, in the (ee)
sector.

We start by studying the $O(5)$ index, for five E-strings.
Taking $\eta=e_1+\varepsilon e_2$ with $0<\varepsilon\ll 1$, the following pair
of weights $\{\rho_1,\rho_2\}$ can potentially give nonzero JK-Res:
\begin{eqnarray}
  &&\{2e_1,2e_2\},\ \{2e_1,e_2\},\ \{2e_1,e_2\pm e_1\},\ \{e_1,2e_2\},\
  \{e_1,e_2\},\ \{e_1,e_2\pm e_1\}\\
  &&\{e_1-e_2,2e_2\},\ \{e_1-e_2,e_1+e_2\},\ \{e_1-e_2,e_2\},\
  \{e_1+e_2,-2e_2\},\ \{e_1+e_2,-e_2\}\ .\nonumber
\end{eqnarray}
These poles define the pole $u_\ast$ by hyperplanes $\rho_i(u_\ast)+z_i=0$ for suitable
$z_i$, chosen between $\epsilon_1,\epsilon_2$. Considering all possible
values of $u_\ast$, we find $142$ poles, which are all non-degenerate. The evaluation of
residue sum should be marginally more laborious than the $O(4)$ case.

Next, we consider the $O(6)$ contour integral.
The poles come from the scalar fields with charges $\pm 2e_I$, $\pm e_I\pm e_J$.
We choose $\eta$ to be $\eta=e_1+\varepsilon e_2+\varepsilon^2e_3$
with $0<\varepsilon\ll 1$.
The groups of $3$ vectors which contain $\eta$ in their cones are
\begin{eqnarray}
  \hspace*{-0.5cm}&&\{2e_1,2e_2,2e_3\},\ \{2e_1,2e_2,e_3\pm e_{1,2}\},\
  \{2e_1,2e_3,e_2\pm e_{1}\},\ \{2e_1,2e_3,e_2-e_{3}\},\ \{2e_1,-2e_3,e_2+e_3\},\nonumber\\
  \hspace*{-0.5cm}&&\{2e_1,e_2\!\pm\! e_1,e_3\!\pm\! e_1\},\
  \{2e_1,e_2\!\pm\! e_1,e_3\!\pm\! e_2\},\
  \{2e_1,e_3\!\pm\! e_1,e_2\!-\!e_3\},\ \{2e_1,-e_3\!\pm\! e_1,e_2\!+\!e_3\}\nonumber\\
  \hspace*{-0.5cm}&&\{2e_1,e_2+e_3,e_2-e_3\},\ \{2e_2,2e_3,e_1-e_{2,3}\},\
  \{2e_2,-2e_3,e_1+e_3\},\ \{2e_2,e_1-e_2,e_3\pm e_{1,2}\}\nonumber\\
  \hspace*{-0.5cm}&&\{2e_2,e_1+e_3,e_1-e_3\},\ \{2e_2,e_1+e_3,-e_2-e_3\},\
  \{2e_2,e_1-e_3,-e_2+e_3\},\ \{2e_3,-2e_2,e_1+e_2\},\ \nonumber\\
  \hspace*{-0.5cm}&&\{2e_3,e_1+e_2,e_1-e_{2,3}\},\
  \{2e_3,e_1+e_2,-e_2-e_3\},\ \{2e_3,e_1-e_2,e_2-e_3\},\ \{2e_3,e_2-e_1,e_1-e_3\},\ \nonumber\\
  \hspace*{-0.5cm}&& \{2e_3,e_1-e_3,e_2\pm e_3\},\
  \{-2e_2,e_1+e_2,e_3\pm e_{1,2}\},\ \{-2e_2,e_1+e_3,e_2-e_3\},\ \{-2e_2,e_1\!-\!e_3,e_2\!+\!e_3\},\ \nonumber\\
  \hspace*{-0.5cm}&&\{-2e_3,e_1\!+\!e_2,e_1\!+\!e_3\},\
  \{-2e_3,e_1\!+\!e_2,-e_2\!+\!e_3\},\ \{-2e_3,e_1\!-\!e_2,e_2\!+\!e_3\},\ \{-2e_3,e_2\!-\!e_1,e_1\!+\!e_3\},\ \nonumber\\
  \hspace*{-0.5cm}&&
  \{-2e_3,e_1\!+\!e_3,e_2\!\pm\!e_3\},\ \{e_1\!+\!e_2,e_1\!-\!e_2,e_3\!\pm\! e_{1,2}\},\
  \{e_1+e_2,e_1+e_3,e_1-e_3\},\ \nonumber\\
  \hspace*{-0.5cm}&&\{e_1+e_2,e_1+e_3,-e_2-e_3\},\
  \{e_1+e_2,e_1-e_3,-e_2+e_3\},\ \{e_1+e_2,e_3-e_2,-e_2-e_3\},\ \nonumber\\
  \hspace*{-0.5cm}&&\{e_1-e_2,e_1+e_3,e_2-e_3\},\
  \{e_1-e_2,e_1-e_3,e_2+e_3\},\{e_1-e_2,e_2+e_3,e_2-e_3\},\ \nonumber\\
  \hspace*{-0.5cm}&&\{e_2-e_1,e_1+e_3,e_1-e_3\},\
  \{e_1+e_3, e_1-e_3, e_2\pm e_3\},\ \{e_1+e_3,e_2-e_3,-e_2-e_3\},\ \nonumber\\
  \hspace*{-0.5cm}&& \{e_1-e_3,e_2+e_3,e_3-e_2\}\ .
\end{eqnarray}
With these chosen $\{\rho_1,\rho_2,\rho_3\}$, the hyperplanes
$\rho_i(u_\ast)+z_i=0$ with $i=1,2,3$ meet at a point $u_\ast$ with suitable choices
of $z_i$, which are either $\epsilon_1$ or $\epsilon_2$. There may exist more than the
chosen three hyperplanes which meet at the same point $u_\ast$, in which case we have
degenerate poles. Also, at some $u_\ast$ there could be some vanishing theta functions
in the numerator. Let us call the number of vanishing theta functions from the
numerator and denominator as $N_{n}(u_\ast)$ and $N_{d}(u_\ast)$, respectively. When $N_d-N_u<r=3$, then the corresponding $u_\ast$ is not a pole due to too many vanishing
terms in the numerator. The list below
covers all the poles which have nonzero JK-Res, also provided with some
illustrations on how to evaluate the residues:
\begin{enumerate}

\item When $N_d=3$, $N_n=0$, this is a non-degenerate and simple pole. We find
$1680$ poles in this class. Near $u=u_\ast$, the integrand relevant for evaluating the
residue approximately takes the form of
\begin{equation}\label{non-degenerate}
  \frac{1}{\prod_{i=1}^{r}(\rho_i(u)-\rho(u_\ast))}\cdot F(u_\ast)\ ,
\end{equation}
where $F(u)$ denotes the rest of the integrand, with $F(u_\ast)\neq 0$.
The integral of the first factor of
(\ref{non-degenerate}) can be immediately obtained from the basic definition (\ref{JK-Res}).

\item There could be degenerate poles with $N_d=N_n+r$, $N_n\neq 0$. The leading divergences
of the integrands are simple poles in this case, since $N_d-N_n=r$. Near the pole,
the integrand relevant for computing the residue approximately takes the
form of
\begin{equation}\label{degenerate-1}
  \frac{\prod_{i=1}^{N_n}(\rho_i(u)-\rho_i(u_\ast))}
  {\prod_{i=N_n+1}^{r+2N_n}(\rho_i(u)-\rho_i(u_\ast))}\cdot F(u_\ast)\ ,
\end{equation}
where $F(u)$ is the rest of the integrand. The basic rule (\ref{JK-Res})
has to be applied to the first factor of (\ref{degenerate-1}) after decomposing it into
a linear combination of the expressions appearing in (\ref{JK-Res}). In the $O(6)$
case with $r=3$, we find two subclasses. Firstly, we find $104$ poles with $N_d=4$, $N_n=1$. For all the poles in this class, we find
\begin{equation}\label{basic-JK-1}
  \textrm{JK-Res}\frac{\rho_1(u)-\rho_1(u_\ast)}
  {\prod_{i=2}^{5}(\rho_i(u)-\rho_i(u_\ast))}=\frac{1}{2}\ ,
\end{equation}
thus all with nonzero residues. We illustrate how this is
evaluated with an example among the $104$ poles, defined with $\{\rho_1,\rho_2,\rho_3,\rho_4\}=\{e_1-e_2,e_1+e_2,e_1+e_3,-e_2-e_3,-2e_2\}$:
\begin{eqnarray}
  &&\textrm{JK-Res}\frac{\bigwedge_{a=1}^3 du_a \cdot (\epsilon_1+\epsilon_2+u_1-u_2)}
  {(\epsilon_1-2u_2)(\epsilon_2+u_1+u_2)(\epsilon_2-u_2-u_3)(\epsilon_1+u_2+u_3)}\nonumber\\
  &&=\textrm{JK-Res}\frac{\bigwedge_{a=1}^3 d\tilde{u}_a}
  {(\tilde{u}_1+\tilde{u}_3)(-\tilde{u}_2-\tilde{u}_3)}
  \left(\frac{1}{\tilde{u}_1+\tilde{u}_2}+\frac{1}{-2\tilde{u}_2}\right)
  =\frac{1}{2}+0=\frac{1}{2}\  ,
\end{eqnarray}
where $\tilde{u}=u-u_\ast$. Moreover, we find $72$ poles with
$N_d=5$, $N_n=2$, in which case we find either
\begin{eqnarray}\label{basic-JK-2}
  &&\textrm{JK-Res}\frac{(\rho_1(u)-\rho_1(u_\ast))
  (\rho_2(u)-\rho_2(u_\ast))}{\prod_{i=3}^{7}(\rho_i(u)-\rho_i(u_\ast))}=\\
  &&\hspace{2cm}0\ (32\ {\rm cases}),\ -\frac{1}{4}\ (16\ {\rm cases}),\
  \frac{1}{4},\ (16\ {\rm cases})\ \frac{1}{2}\ (8\ {\rm cases})\ .\nonumber
\end{eqnarray}
Thus we find $40$ more poles.
There are no more poles in this class with larger $N_d,N_n$.

\item In general, there could be degenerate poles with $N_d>N_n+r$.
The integrand contains `multiple poles' in this case.
The integrand takes the form of
\begin{equation}
    \frac{\prod_{i=1}^{N_n}\theta_1(\rho_i(u)-\rho_i(u_\ast))}
  {\prod_{i=N_n+1}^{N_d+N_n}\theta_1(\rho_i(u)-\rho_i(u_\ast))}\cdot F(u)\ ,
\end{equation}
where $F(u)$ is a combination of $\theta_1$ functions which are nonzero at $u_\ast$.
Since
the first factor contains multiple poles, one would have to expand both first and
second factors to certain orders near $u=u_\ast$, until one obtains a linear combination
of the functions appearing in (\ref{JK-Res}). The residue will thus be expressed by
$\theta_1$ functions and their suitable derivatives at $u_\ast$. This class of poles
do not show up in the $O(6)$ case. (They will first appear in the $O(8)$ index,
explained below.)

\end{enumerate}
With the above $1680+104+40=1824$ poles and the computational rules stated in the
list, clearly the $O(6)$ elliptic genus can be computed straightforwardly,
although the resulting expression will be very long.

Let us explain the pole/residue structures of $O(7)$ index,
with rank $r=3$. The poles are again classified into the above three classes.
To be definite, we chose $\eta=e_1+\varepsilon e_2+\varepsilon^2 e_3$.
We simply list the number poles in each class.
\begin{enumerate}

\item non-degenerate poles ($N_d=3$, $N_n=0$): $2468$ cases

\item degenerate (but simple) poles: With $N_d=4$, $N_n=1$, we find
$106$ degenerate and simple poles. The relevant integrals of the form of
(\ref{basic-JK-1}) are either $\frac{1}{2}$ or $1$, depending on $u_\ast$. With
$N_d=5$, $N_n=2$, we find $72$ cases. The integral analogous to (\ref{basic-JK-2})
are either $0,-\frac{1}{4},\frac{1}{4},\frac{1}{2}$. There are $32$ cases with zero
residues. So we find $40$ poles in this class. Finally, there
are $4$ cases with $N_d=6$, $N_n=3$, and the JK-Res of the rational
functions are either
\begin{equation}\label{basic-JK-3}
  \textrm{JK-Res}\frac{\bigwedge_{a=1}^r d\tilde{u}_a\cdot\prod_{i=1}^3\rho_i(\tilde{u})}
  {\prod_{i=4}^{r+6}\rho_i(\tilde{u})}=\frac{1}{2}\ \ (2\textrm{\ cases}),\ {\rm or}
  \ \ 0\ \ (2\textrm{\ cases})\ .
\end{equation}
So we have $2$ poles in the last class. We do not find further degenerate
simple poles with larger $N_n$.

\item degenerate multiple poles ($N_d>N_n+3$): We do not find any poles
in this case.

\end{enumerate}
So we find $2468+106+40+2=2616$ poles with nonzero JK-Res.

As a final illustration, let us consider the $O(8)$ contour integral
with rank $r=4$. The number of poles quickly increases, as follows:
\begin{enumerate}

\item non-degenerate poles ($N_d=4$, $N_n=0$): $32304$ poles

\item degenerate (but simple) poles: With $N_d=5$, $N_n=1$, we find
$4424$ poles. With $N_d=6$, $N_n=2$, we find
$1696$ poles. With $N_d=7$, $N_n=3$, we find $88$ poles. Finally,
with $N_d=8$, $N_n=4$, we finds $200$ poles.

\item degenerate multiple poles ($N_d>N_n+3$): We find $72$ such poles.

\end{enumerate}
So we find $32304+4424+1696+88+200+72=38784$ poles for
the $O(8)$ contour integral.

\section{E-strings from Yang-Mills instantons}

In this section, we explain how one can alternatively compute the E-string elliptic genus
from the instanton partition function of a suitable 5 dimensional super-Yang-Mills theory
with $Sp(1)$ gauge group. The basic idea is that suitable circle reductions of 6d SCFTs
sometimes admit 5d SYM descriptions at low energy. The latter SYM, despite being non-renormalizable, remembers the 6d KK degrees in its solitonic sector as the instanton solitons
\cite{Douglas:2010iu,Lambert:2010iw}. The self-dual strings wrapping the circle
become the W-bosons, quarks or their superpartner particles in 5d. So the Witten index for
the threshold bounds of these particles with instantons in the Coulomb branch
\cite{Nekrasov:2002qd,Nekrasov:2003rj} will carry information on the elliptic genera of
wrapped self-dual strings. This idea has been used to study the elliptic genus of M-strings
in the 6d $(2,0)$ SCFT in \cite{Kim:2011mv,Haghighat:2013gba}. In this section, we make a
similar study for the E-strings. Since the circle reduction of the $E_8$ $(1,0)$ SCFT
is subtler than that of the $(2,0)$ theory, let us set up the problem first.

We start by considering the type IIA system consisting of $8$ D8-branes and an O8-plane
(or $16$ D8-branes in the covering space), making a type I' string background. The
D8-branes are at the tip of the half-line $\mathbb{R}^+$, formed by an O8.
The worldvolume of the $8$-branes hosts $SO(16)$ gauge symmetry. Since the
net 8-brane charges cancel, the asymptotic value of the dilaton on $\mathbb{R}^+$
is a nonzero constant. So this system admits an M-theory uplift at strong coupling, on $\mathbb{R}^{8+1}\times\mathbb{R}^+\times S^1$. The D0-branes in the type I' theory are
identified as the Kaluza-Klein modes along the M-theory circle. In the uplifted background,
an M9-plane (or the Horava-Witten wall) is located at the tip of $\mathbb{R}^+$ and wraps
$\mathbb{R}^{8+1}\times S^1$. The M9-plane hosts an $E_8$ gauge symmetry. When the M9 wraps
a circle, one can turn on nonzero $E_8$ Wilson line which reduces gauge symmetry. To get a
background which admits a weakly coupled type I' description
with unbroken $SO(16)$ gauge symmetry, one should turn on the Wilson line as follows. Let $R$ be the radius of the M-theory circle, and $A$ be the $E_8$ gauge field on the circle. $E_8$
has an $SO(16)$ subgroup, in which the adjoint representation $\mathbf{248}$ of $E_8$ decomposes
into $\mathbf{120} \oplus \mathbf{128}$. The Wilson line $RA$ that we turn on in
$SO(16)\subset E_8$ is given by \cite{Ganor:1996mu}
\begin{align}
	\label{eq:wilsonline}
    RA = (0,0,0,0,0,0,0,1)\ .
\end{align}
This is in the convention that one picks the Cartans of $SO(16)$ as rotations on the $8$ orthogonal 2-planes. The circle holonomy generated by this Wilson line is
$\exp{(2\pi i RA \cdot F)}$, with
$F=(F_1,F_2,\cdots,F_8)$ being the Cartans of $SO(16)\subset E_8$ in the same basis. The normalization is $F_l=\pm \frac{1}{2}$ for $SO(16)$ spinors. The holonomy with (\ref{eq:wilsonline}) acts on $\mathbf{128}$ as $-1$, and on $\mathbf{120}$ as $+1$. So $E_8$ symmetry breaks down
to $SO(16)$. This is the background which admits the type I' theory description for small
$R$.

Now let us consider the D4-D8-O8 system, by adding $N$ D4-branes. This uplifts in M-theory
to the M5-M9-branes wrapping the circle, in the above $E_8$ Wilson line background. On the
worldvolume of D4-branes, one obtains an $Sp(N)$ gauge theory with 1 antisymmetric and 8 fundamental hypermultiplets.\footnote{Had one been reducing the
M5-M9 system with zero Wilson line, one would have obtained the strongly interacting 5d
SCFT with $E_8$ symmetry \cite{Klemm:1996hh,Ganor:1996pc}, discovered in \cite{Seiberg:1996bd}.} This 5d gauge theory is a low-energy description of the 6d $(1,0)$ superconformal
field theory compactified on a circle with $E_8$ Wilson line.
Note that, from the worldvolume theory on D4 or M5-branes,
$SO(16)$ or $E_8$ act as global symmetries. So from the 5d/6d field theories, the Wilson line
we explained above are nondynamical background fields.

Consider the system consisting of single M5-brane and an M9-plane, compactified on a circle
with the above Wilson line. We have an $Sp(1)$ gauge theory description in 5d. Taking into
account the effect of the background Wilson line \eqref{eq:wilsonline}, we can identify
various charges of the 5d SYM theory and the 6d $(1,0)$
theory on circle as follows:
\begin{align}
	\label{eq:wilsonline_chargeshift}
    k &= 2P + n (RA \cdot RA) - 2 \left(RA\cdot \tilde{F}\right) = 2P + n - 2\tilde{F}_8\\
    F_l &= \tilde{F}_l - n(RA_l)\ \ \ \rightarrow\ \
    F_8= \tilde{F}_8-n\ .
\end{align}
Here, $k,F_l$ appearing on the left hand sides are various
charges of the 5d SYM, while $P,\tilde{F}_l$ on the right hand
sides are those of the 6d E-string theory.
$k$ is the Yang-Mills instanton charge on D4's (i.e. D0-brane number in
the type I' theory), $P$ is the momentum on E-strings along the circle,
$\tilde{F}_l$ are the $E_8$ Cartan charge in the 6d theory (which were called $F_l$
till here in this paper), and $F$ are the $SO(16)$ Cartan charges in the 5d SYM.
$n$ is the $U(1)\subset Sp(1)$ electric charge in the Coulomb phase, which is
identified with the winding number of the E-strings. This formula can be naturally
inferred by starting from the charge relations of the fundamental type I' stings
on $\mathbb{R}^{8+1}\times I$ and the heterotic strings on
$\mathbb{R}^{8+1}\times S^1$ \cite{Narain:1986am,Aharony:1997pm}, where $I$ is
a segment, and then putting an M5-brane on $I$ to decompose a heterotic string
into two E-strings \cite{Haghighat:2014pva}.

Later in this section, we shall consider an index for the E-strings, with the weight given by
\begin{align}\label{index-weight}
    q^{k} e^{2 \pi i m_8 F_8} w^{n}\prod_{l=1}^7 e^{2 \pi i m_l F_l}
    =q^{2P}(y_8^\prime)^{\tilde{F}_8}(w^\prime)^n\prod_{l=1}^7 e^{2 \pi i m_l \tilde{F}_l}
\end{align}
with $y_i \equiv e^{2 \pi i m_i}$, where
\begin{equation}\label{eq:fugacity_redef}
  y_8^\prime=y_8q^{-2}\ \ ,\ \ w^\prime=w q y_{8}^{-1}\ .
\end{equation}
The right hand side of (\ref{index-weight}), with primes and tildes for fugacities
and charges, is the natural expression for the E-strings from the 6d perspective,
while the instanton calculus will naturally use the expression on
the left hand side. After computing the instanton partition function with the
above weight, we shall express it in terms of the fugacities $y_8^\prime$, $w^\prime$
given by (\ref{eq:fugacity_redef}), which can be compared with the E-string elliptic
genus that we studied in this paper.
This redefinition of fugacities plays the role of canceling the background $E_8$
Wilson line (\ref{eq:wilsonline}), which obscures the $E_8$ symmetry in the type I'
instanton calculus.\footnote{Only in this section,
the definition of $q$ is given by $q=e^{\pi i\tau}$, instead of $q=e^{2\pi i\tau}$ used in
all other sections of this paper. This is because the single instanton carries
$q^{\frac{1}{2}}$ factor in the other convention, due to the fractional Wilson line, which
we want to change to $q^1$. This is the reason for the factor $q^{2P}$ in (\ref{index-weight}).}

Since the ADHM quantum mechanics is a UV completion of the 5d instanton quantum mechanics, it contains extra string theory degrees of freedom apart from the QFT
states. So the
partition function of the ADHM quantum mechanics may acquire contributions
from the extra string theory states in the D4-D8-O8 background. Since the 5d/6d quantum field
theories are obtained from the string theory background by taking low energy decoupling limit,
the Hilbert space of this system factorizes at low energy. In particular, in the context of the
Witten index of the ADHM quantum mechanics, one expects
\begin{align}\label{other}
    Z_\text{ADHM} = Z_\text{\rm inst} \cdot Z_\text{other}\ .
\end{align}
The quantity of our interest is the 5d instanton partition function $Z_{\rm inst}$.
The factor $Z_{\rm other}$ was
identified in \cite{Hwang:2014uwa}. For the purpose of studying the QFT spectrum, we
simply divide the ADHM quantum mechanics partition function by $Z_{\rm other}$ identified in \cite{Hwang:2014uwa}, to obtain $Z_{\rm inst}$. See section 3.4.2
of \cite{Hwang:2014uwa} for the details.

We will consider the QFT partition function $Z_\text{QFT}(q,w,,m_l,\epsilon_{1,2})$
of the 5d $Sp(1)$ gauge theory, i.e., the rank $1$ 6d $(1,0)$ SCFT compactified on
circle with $E_8$ Wilson line. The full partition function is obtained by multiplying
the 5d perturbative part $Z_{\rm pert}$ to $Z_{\rm inst}$, i.e.
\begin{equation}\label{QFT-1st}
  Z_{\rm QFT}(q,w,m_l,\epsilon_{1,2})=
  Z_{\rm pert}(w,m_l,\epsilon_{1,2})Z_{\rm inst}(q,w,m_l,\epsilon_{1,2})\ ,
\end{equation}
with
\begin{equation}
  Z_{\rm pert}\equiv
  \exp\left[\sum_{n=1}^\infty\frac{1}{n}f(w^n,nm_l,n\epsilon_{1,2})\right]\ ,\ \
  f(w,m_l,\epsilon_{1,2})\equiv
  \frac{\chi_{{\bf 16}}^{SO(16)}(m_l)w-2\cos(2\pi\epsilon_+)w^2}
  {(2i\sin\pi\epsilon_1)(2i\sin\pi\epsilon_2)}\ .
\end{equation}
The first term of $f$ comes from the quarks of the $N_f=8$ $Sp(1)$ fundamental
hypermultiplets, where $\chi_{{\bf 16}}^{SO(16)}\equiv \sum_{l=1}^8
(e^{2\pi im_l}+e^{-2\pi im_l})$ is the character of ${\bf 16}$.
The second term of $f$ comes from the $Sp(1)$ W-boson and superpartners
in the vector multiplet.
To study $Z_{\rm QFT}$ from the 6d E-string perspective, one first considers
the grand partition function of the E-string elliptic genera
$Z_n(q,m_l^\prime,\epsilon_{1,2})$ that we studied in this paper,
\begin{equation}
  Z_{\textrm{E-string}}(w^\prime,m_l^\prime,\epsilon_{1,2})=\sum_{n=0}^\infty
  (w^\prime)^nZ_n(q,m_l^\prime,\epsilon_{1,2})\ ,
\end{equation}
where $Z_0\equiv 1$.
This captures the contribution to partition function $Z_{\rm QFT}$
from the states with nonzero E-string winding number $n$.
One has to multiply the contribution from states at zero winding.
For the E-string theory in the Coulomb branch,
it comes from an $\mathcal{N}=(1,0)$ tensor multiplet, which is
\begin{equation}\label{tensor}
  Z_{\rm tensor}(q,\epsilon_{1,2})\equiv
  \exp\left[\sum_{n=1}^\infty\frac{1}{n}g(q^n,n\epsilon_{1,2})\right]\ ,\ \
  g(q,\epsilon_{1,2})\equiv -\frac{2\cos(2\pi\epsilon_-)}
  {(2i\sin\pi\epsilon_1)(2i\sin\pi\epsilon_2)}\frac{q^2}{1-q^2}\ .
\end{equation}
$g$ is the single particle index of a $(1,0)$ tensor multiplet on
a circle \cite{Kim:2011mv}.\footnote{In \cite{Hwang:2014uwa}, $Z_{\rm tensor}$
was reproduced from 5d SYM approach, in eqn.(3.78) there, with extra two terms
$\propto v+v^{-1}$ in the numerator. This part corresponds to a free 6d
hypermultiplet which in fact decouples from the 6d SCFT, but is sometimes
included into the studies for convenience to study M5-M9 system. This is similar
to sometimes including the free $(2,0)$ tensor multiplet to the $A_{N-1}$ $(2,0)$
theory, to describe N M5-branes. In this paper, the term proportional to $v+v^{-1}$
in (3.78) of \cite{Hwang:2014uwa} will be sent to $Z_{\rm other}$ of (\ref{other}).}
Then, one finds
\begin{equation}\label{QFT-2nd}
  Z_{\rm QFT}(q,w,m_l,\epsilon_{1,2})=
  Z_{\textrm{E-string}}(w^\prime,m_l^\prime,\epsilon_{1,2})
  Z_{\rm tensor}(q,\epsilon_{1,2})\ .
\end{equation}
With (\ref{eq:fugacity_redef}), this provides the second formula for $Z_{\rm QFT}$.
The expression (\ref{QFT-1st}) takes the form of series expansion in $q$, since
we know the coefficients of
$Z_{\rm inst}(q,w,m_l,\epsilon_{1,2})=\sum_{k=0}^\infty Z_k(w,m_l,\epsilon_{1,2})q^k$.
So at a given order in the modular parameter $q$, one captures the spectrum of arbitrary number of E-strings by computing $Z_k$ exactly in $w$. This is in contrast
to the formula (\ref{QFT-2nd}) obtained from the E-string elliptic genus, keeping
definite order $Z_n(q,m_l^\prime,\epsilon_{1,2})$ in $w^\prime(\sim w)$
which is exact in $q$. So to confirm that the two approaches yield the same result,
we shall make a double expansions of (\ref{QFT-1st}) and (\ref{QFT-2nd}) in
$q$, $w$ and compare, taking into account the shifts (\ref{eq:fugacity_redef}).
While making the study of instanton partition function of our $Sp(1)$ gauge theory
in \cite{Hwang:2014uwa}, $Z_k(w,m_l,\epsilon_{1,2})$ was computed up to $k=5$. So
expanding $Z_n(q,y_8^\prime,\epsilon_{1,2})=Z_n(q,y_8q^{-2},\epsilon_{1,2})$
up to $\mathcal{O}(q^5)$ at fixed $y_8=e^{2\pi im_8}$, and expanding
$Z_{\rm QFT}$ computed from
5d to $\mathcal{O}(w^n)$ for some low $n$, we shall find perfect agreement
of the two results.

\subsection{Instanton partition function}

To take into account the effect of the Wilson line which breaks $E_8$ down to $SO(16)$, we have to make a shift of the fugacities by (\ref{eq:fugacity_redef}).
We decide to express $w^\prime,y_8^\prime$ in terms of $w,y_8$.
After inserting $y_8^\prime=y_8q^{-2}$ (or
$e^{2 \pi i m_8} \rightarrow e^{2 \pi i m_8 -2 \pi i \tau}$) to the elliptic
genera $Z_n$ of section 3, one finds
\begin{align}\label{shifted-E-string}
    Z_n(q,m_l^\prime,\epsilon_{1,2})=\left(\frac{y_8}{q}\right)^n
    \tilde{Z}_\text{n}(q,m_l,\epsilon_{1,2}),
\end{align}
with
\begin{align}\label{shifted-elliptic}
    \tilde{Z}_\text{1} &= \frac{1}{2} \left( -Z_{1(1)} + Z_{1(2)} + Z_{1(3)} - Z_{1(4)} \right)\\
    \tilde{Z}_\text{2} &= \frac{1}{2} Z_\text{2(0)} + \frac{1}{4}\left(-Z_{2(1)} -
    Z_{2(2)} + Z_{2(3)} + Z_{2(4)} - Z_{2(5)} - Z_{2(6)} \right)\nonumber\\
    \tilde{Z}_\text{3} &= \frac{1}{4} \left( -Z_{3(1)} - Z_{3(2)} + Z_{3(3)} + Z_{3(4)} \right) +
    \frac{1}{8}\left(-Z_{3(1)'} - Z_{3(2)'} + Z_{3(3)'} + Z_{3(4)'} \right)\nonumber \\
    \tilde{Z}_\text{4} &= \frac{1}{8} \left(Z_{4(1)}  - Z_{4(2)} - Z_{4(2)'} - Z_{4(3)} - Z_{4(3)'} + Z_{4(4)} + Z_{4(4)'}\right) +
    \frac{1}{16} Z_{4(1)'}\nonumber\ ,
\end{align}
and so on, where $Z_{n(i)}$'s are all defined and computed in section 3 as
functions of $q,m_l,\epsilon_{1,2}$.
In all $Z_{n(i)}$ on the right hand side, the arguments are $y_8$, not $y_8^\prime$.
The overall factors $\left(y_8 q^{-1}\right)^{n}$ in (\ref{shifted-E-string})
cancel with the shift $w^\prime= w q y_8^{-1}$ in
$Z=\sum_{n=0}^\infty (w^\prime)^nZ_n$. Namely, the $E_8$ mass shift is inducing
a different value of 2d theta angle, by changing various signs in (\ref{shifted-elliptic}).
We compute $\tilde{f}(w,q,\epsilon_{1,2},m_i)$ defined by
\begin{equation}
  Z_{\rm QFT}\equiv Z_{\rm tensor}
  \sum_{n=0}^\infty w^n\tilde{Z}_n(q,\epsilon_{1,2},m_i)
  =PE\left[\tilde{f}\right]\equiv\exp\left[\sum_{n=1}^n\frac{1}{n}
  \tilde{f}(w^n,q^n,n\epsilon_1,n\epsilon_2, nm_l)\right]\ ,
\end{equation}
and expand $\tilde{f}=\sum_{n=0}^\infty w^n\tilde{f}_n(q,\epsilon_{1,2},m_i)$.
The results up to $\mathcal{O}(q^5)$ are as follows. $\tilde{f}_0$ at zero
string number has been computed from the 5d calulus in \cite{Hwang:2014uwa},
and agrees with $g$ appearing in (\ref{tensor}). So we consider
$\tilde{f}_n$ with $n\geq 1$.

Defining $t\equiv e^{i \pi \epsilon_1 + i \pi \epsilon_2}$,
$u\equiv e^{i \pi \epsilon_1 - i \pi \epsilon_2}$, $\tilde{f}_1$ is given by $\frac{t}{(1-tu)(1-t/u)}$ times
\begin{align}\label{1E-instanton}
    +q^0 &\cdot \chisixteen{16} + q^1 \cdot \chisixteen{\overline{128}}\\
    +q^2 &\Big[ (t+t^{-1})(u+u^{-1})\chisixteen{16} + \chisixteen{560} + \chisixteen{16} \Big] + q^3 \Big[ (t+t^{-1})(u+u^{-1}) \chisixteen{\overline{128}} + \chisixteen{\overline{1920}} + \chisixteen{\overline{128}} \Big] \nonumber \\
    +q^4 &\Big[ (t+t^{-1})(u+u^{-1}) (\chisixteen{560} + 2\chisixteen{16}) + \left((t^2+1+t^{-2})(u^2+1+u^{-2})-1\right) \chisixteen{16} \nonumber \\
    & +\chisixteen{4368} + \chisixteen{1344} + \chisixteen{560} + 4\chisixteen{16} \Big]
    \nonumber\\
    +q^5 &\Big[ (t+t^{-1})(u+u^{-1})( \chisixteen{\overline{1920}} + 2 \chisixteen{\overline{128}}) + \left((t^2+1+t^{-2})(u^2+1+u^{-2})-1\right)\chisixteen{\overline{128}}  \nonumber \\
    & + \chisixteen{\overline{13312}} + 2\chisixteen{\overline{1920}} + 4\chisixteen{\overline{128}}  \Big]  + \mathcal{O}(q^6) \nonumber
\end{align}
The boldfaced subscripts are the irreps of $SO(16)\subset E_8$ in the
5d $Sp(1)$ gauge theory with $8$ fundamental flavors. $\chi^{\rm SO(16)}_{\bf R}$
is the $SO(16)$ character of the representation ${\bf R}$. We computed $Z_{\rm QFT}$
of the 5d SYM, following the procedures outlined above (explained in \cite{Hwang:2014uwa}),
up to five instantons. We further expanded it in
the Coulomb VEV parameter to extract the $\mathcal{O}(w^1)$ order. This completely
agrees with (\ref{1E-instanton}).

$\tilde{f}_2$ is given by $\frac{t}{(1-tu)(1-t/u)}$ times
\begin{align}
    -q^0 & \cdot (t + t^{-1}) - q^1 \Big[ (t+t^{-1}) \chisixteen{128}\Big]\\
    -q^2 & \Big[ (t^3 + t + t^{-1} + t^{-3})(u^{2} + 1 + u^{-2}) + (u+u^{-1}) +(t^2+1+t^{-2})(u+u^{-1})(\chisixteen{120} + 1) \nonumber\\
    & + (t+t^{-1}) (\chisixteen{1820} + \chisixteen{120} + 2) \Big] \nonumber\\
    -q^3 & \Big[ (t+t^{-1})((t^{2}+t^{-2})(u^{2}+1+u^{-2})-1) \chisixteen{128}  +(u+u^{-1}) \chisixteen{128}  \nonumber \\
    &  + (t^2+1+t^{-2})(u+u^{-1})( \chisixteen{1920} + 2\chisixteen{128})+ (t+t^{-1})( \chisixteen{13312} + \chisixteen{1920} + 4\chisixteen{128})\Big] \nonumber \\
    -q^4 & \Big[ (t^{4}+t^{-4})(u+u^{-1})+ (t^{3}+t+t^{-1}+t^{-3})(u^{4}+u^{-4})\nonumber \\
    & + (t^{2}+1+t^{-2})(u^{3}+u^{-3}) + (t+t^{-1})(u^{2}+u^{-2}) + (t^5+t^{-5})(u^4+u^2+1+u^{-2}+u^{-4}) \nonumber \\
    & + (u+u^{-1})(\chisixteen{1820}+2\chisixteen{120}+3)  + \left((t^{4} + t^{2} + 1 + t^{-2} + t^{-4})(u^{3}+u^{-3})\right. \nonumber \\
    & \left. +(t^{4}+t^{-4})(u+u^{-1}) + (t^{3}+t^{-3}) + (t+t^{-1})(u^{2}+u^{-2})\right)(\chisixteen{120}+1)  \nonumber\\
    & +\left( (t^{3}+t^{-3})(u^{2}+1+u^{-2}) + (t+t^{-1})(u^{2}+u^{-2})\right)( \chisixteen{1820}+\chisixteen{135}+2\chisixteen{120}+5)  \nonumber\\
    &+ (t^{2}+1+t^{-2})(u+u^{-1}) (\chisixteen{8008}+\chisixteen{7020}+2\chisixteen{1820}+\chisixteen{135}+6\chisixteen{120}+8 ) \nonumber\\
    &+ (t+t^{-1})(\chisixteen{60060}+\chisixteen{8008}+\chisixteen{7020}+\chisixteen{\overline{6435}}+\chisixteen{5304}+4\chisixteen{1820}+3\chisixteen{135}+9\chisixteen{120}+14 ) \Big] \nonumber 
\end{align}
\begin{align}
    -q^5 & \Big[ \left((t^5+t^{-5})(u^{4}+u^{2}+1+u^{-2}+u^{-4}) + (t^{3}+t+t^{-1}+t^{-3})(u^{4}+u^{-4}) \right.\nonumber\\
    &\left.  + (t^{2}+1+t^{-2})(u^{3}+u^{-3}) + (t^{4} + t^{-4})(u+u^{-1}) + (t+t^{-1})(u^{2}+u^{-2})\right) \chisixteen{128} \nonumber\\
    & + \left( (t^{3}+t^{-3})(u^{2}+1+u^{-2}) + (t+t^{-1})(u^{2}+u^{-2}) \right)(\chisixteen{13321}+ 3\chisixteen{1920} + 7\chisixteen{128}) \nonumber\\
    &+ \left( (t^{2}+t^{-2})(u+u^{-1}) + (t+t^{-1}) + (u+u^{-1}) \right) (  \chisixteen{56320} + \chisixteen{15360} + 3\chisixteen{13312}+ 7\chisixteen{1920} + 14\chisixteen{128})\nonumber\\
    & + (u+u^{-1})( \chisixteen{13312}+2\chisixteen{1920}+5\chisixteen{128}) + \left((t^{2}+1+t^{-2})(u^{3}+u^{-3}) \right. \nonumber\\
    & \left.+ (t^{4}+t^{-4})(u^{3}+u+u^{-1}+u^{-3}) + (t+t^{-1})(u^{2}+u^{-2}) + (t^{3}+t^{-3})  \right) ( \chisixteen{1920} + 2\chisixteen{128})  \nonumber\\
    & +(t+t^{-1}) ( \chisixteen{161280} + \chisixteen{141440} + 3\chisixteen{13312}+5\chisixteen{1920}+9\chisixteen{128}  ) \Big] + \mathcal{O}(q^6) \nonumber
\end{align}
This again agrees with the result obtained from the instanton calculus of \cite{Hwang:2014uwa}.

We also computed $\tilde{f}_3$ with all $SO(16)\subset E_8$ masses turned off. It
again completely agrees with $\tilde{f}_3$ computed from 5d instanton calculus,
up to $q^5$ order that we checked. Also, for $3$ and $4$ E-strings, we have kept
all $E_8$ masses and compared our 2d elliptic genus with the instanton partition
function up to $1$ instanton order, which all show agreements.

So we saw that the instanton calculus provides the correct index for
the $E_8$ 6d SCFT. One virtue of this approach would be that, at a given order
in $q$, the index is computed exactly in $w$. In particular, the chemical potential
for the E-string number (the Coulomb VEV of 5d SYM) is an integration variable
in the curved space partition functions, which can be used to study the conformal
field theory physics. So knowing the exact form of the partition function in $w$
will be desirable to understand the curved space partition functions.

\section{Concluding remarks}

In this paper we have found a description of E-strings which
can be used to describe the IR degrees of freedom on it. This in
particular includes the information about bound states of E-strings.
The theory for $n$ E-strings involves a $(0,4)$ supersymmetric
quiver theory in 2 dimensions with $O(n)$ gauge symmetry and
some matter content.  We in particular computed the elliptic
genus of E-strings (including turning on fugacities
for the $E_8$ flavor symmetry as well as $SO(4)$ rotation
transverse to the string in 6d) for small number of E-strings.
We gave the explicit answer for $n=1,2,3,4$ and indicated how
one can use these methods to obtain arbitrary $n$ answers.
Our results successfully pass the comparison checks
with the partial results already known. Our results provide an
all genus answer for the topological string on the canonical
bundle over ${1\over 2} K3$. In addition, we explained how to compute
the same elliptic genus using the instanton partition function of
the 5d $Sp(1)$ SYM theory coupled to $8$ fundamental hypermultiplets.

We briefly discuss various physics of E-strings that we can learn 
from our gauge theories and the elliptic genus formula. Firstly, one can 
show from our contour integral expression (\ref{index-full}), 
(\ref{determinant}) and $\frac{\eta(-1/\tau)}{\theta_1(-1/\tau,z/\tau)}
=\varepsilon e^{-\frac{\pi i z^2}{\tau}}\frac{\eta(\tau)}{\theta_1(\tau,z)}$ 
(where $\varepsilon$ is a $z$ independent phase) that
\begin{equation}\label{S-modular}
  Z_{n}\left(-\frac{1}{\tau},\frac{\epsilon_{1,2}}{\tau},
  \frac{m_l}{\tau}\right)=Z_n(\tau,\epsilon_{1,2},m_l)\cdot
  \varepsilon^{-6n}\exp\left[\frac{\pi i}{\tau}
  \left(2\epsilon_1\epsilon_2 n^2-(\sum_{l=1}^8m_l^2-4\epsilon_+^2)n\right)\right]\ .
\end{equation}
This expression can be obtained by applying the S-modular transformation 
directly to the integrand (\ref{determinant}), noting that the transformation
just shuffles the discrete holonomy sectors with the same 
dimension for their Weyl groups. 
In fact, the extra exponential factor on the right hand side is 
related to the 2d 't Hooft anomaly on the strings 
\cite{Benini:2013xpa}, being 
$\exp\left[-\frac{\pi i}{\tau}\mathcal{A}^{ab}u_au_b\right]$ with chemical 
potentials $u_a$ when the 't Hooft anomaly is given by 
$\mathcal{A}^{ab}=\sum_{\rm fermions}\gamma_3 K^aK^b$. 
Thus, there are terms in the anomalies which are linear
in the string number $n$, and also a peculiar term which is proportional to
$n^2$.

The last term proportional to $n^2$ has interesting physical implications 
to the non-linear sigma models in IR that one obtains from our gauge theories.
Namely, the real $4n$ dimensional solution for $\varphi,\tilde\varphi$ 
which solves $\varphi\tilde\varphi-\tilde\varphi\varphi=0$, 
$\varphi\varphi^\dag-\tilde\varphi^\dag\tilde\varphi=0$ of section 2 is 
given by diagonal matrices for $\varphi,\tilde\varphi$. By extra modding 
out by the unbroken gauge symmetries in $O(n)$, the moduli space becomes 
the $n$'th symmetric product of $\mathbb{R}^4$, 
${\rm Sym}^n(\mathbb{R}^4)=(\mathbb{R}^4)^n/S_n$ where $S_n$ is the $n$
dimensional permutation group. Considering the non-linear sigma model on 
this target space, away from the singularity, there are no ways 
to have anomalies (or any other measures of degree of freedom) which 
scale like $n^2$, since the number of degrees of freedom visible in the 
sigma model is proportional to $n$. Therefore, the extra $n^2$ degrees of 
freedom which contribute to the first term in the anomaly should be supported 
at the orbifold singularity, where the sigma model description should break down.

This is in contrast to the dynamics of fundamental strings. Namely, if one 
wraps the fundamental string on a circle $n$ times, its dynamics on the 
transverse target space is also described by $n$'th symmetry product of 
the transverse space. So although the non-linear sigma models for our E-strings apparently looks similar to those for the fundamental strings, the way one 
treats the orbifold singularity should be very different.
For instance, for a fundamental superstring, the elliptic genus $Z_n$ for 
$n$ wrapped strings is given in terms of the elliptic genus $Z_1$ of the single 
string, by the Hecke transformation. Namely, defining the grand partition function
\begin{equation}
  Z(w,\tau,z)=\sum_{n=0}^\infty Z_n(\tau,z)w^n
\end{equation}
where $z$ collectively denotes chemical potentials, and $Z_0\equiv 1$ by definition, 
$Z(w,\tau,z)$ is given in terms of $Z_1$ by \cite{Dijkgraaf:1996xw}
\begin{equation}\label{hecke}
  Z(w,\tau,z)=\exp\left[\sum_{n=1}^\infty\frac{1}{n}w^n\sum_{ad=n;a,d\in\mathbb{Z}}
  \sum_{b(\textrm{mod }d)}Z_1\left(\frac{a\tau+b}{d},az\right)\right]
  \equiv\exp\left[\sum_{n=1}^\infty w^nT_nZ_1(\tau,z)\right]\ ,
\end{equation}
where $T_n$ are the Hecke operators. This expresses all $Z_n$'s in terms of $Z_1$.
For instance, $Z_2$ for fundamental strings is given from this relation by
\begin{equation}
  Z_2(\tau,z)=\frac{1}{2}\left[Z_1(\tau,z)^2+Z(2\tau,2z)
  +Z_1\left(\frac{\tau}{2},z\right)+Z_1\left(\frac{\tau+1}{2},z\right)\right]\ .
\end{equation}
Now with the extra anomalies on the E-strings proportional to $n^2$, it is 
easy to understand that the elliptic genera $Z_n$ at $n>1$ cannot be expressed 
in terms of $Z_1$ by Hecke transformation. This is because, from the formula 
(\ref{hecke}), the 2d anomaly has to be additive. The additive property means 
that, if $Z_1$ has the anomaly 
$\exp\left[-\frac{\pi i}{\tau}\mathcal{A}^{ab}u_au_b\right]$ under S-modular 
transformation like (\ref{S-modular}), $Z_n$ given by (\ref{hecke}) should have 
anomaly $\exp\left[-\frac{n\pi i}{\tau}\mathcal{A}^{ab}u_au_b\right]$.
However, since (\ref{S-modular}) for E-strings exhibits an anomaly proportional 
to $n^2$, (\ref{hecke}) cannot be true for E-strings. 

It is easy to see, from the viewpoint of our 2d gauge theory, how the 
non-linear sigma model description breaks down near the singularity, and 
how the $n^2$ degrees of freedom emerges at the singularity. When 
$\varphi,\tilde\varphi$ assume large nonzero values, the fermions in the 
$O(n)$ vector multiplet (which we called $\lambda_+^{\dot\alpha A}$, with $\frac{n^2-n}{2}$ components) become massive, so do not appear in the 
non-linear sigma model. However, since gauge symmetry is unbroken at 
$\varphi=\tilde\varphi=0$, these fermions become light near the orbifold singularity. 
The left-moving fermion $\lambda_+$ acquires mass only by combining with right-moving 
fermions, which are $\lambda_-^{\alpha A}$ of section 2 (superpartners of 
$\varphi,\tilde\varphi$). Both $\lambda_+,\lambda_-$ become light near the 
singularity, and the anomaly in (\ref{S-modular}) proportional to 
$\epsilon_1 \epsilon_2 n^2$ precisely comes from these fields in our UV description.
Namely, a crucial difference between fundamental strings and our E-strings 
(and more generally other self-dual strings of 6d SCFTs in the tensor branch) 
can be explained with gauge theory engineering of the latter.

As mentioned in section 3.1, another curious aspect of E-strings can be explained 
using our gauge theory descriptions. The elliptic genus of single strings have 
been computed in \cite{Klemm:1996hh} using an effective free string theory 
approach, where the GSO projection (like that of the $E_8\times E_8$ 
heterotic strings) had to be applied on R-NS sectors to get the correct results.
From our $O(1)\sim\mathbb{Z}_2$ gauge theory approach, these are simply 
the consequence of summing over all the discrete $O(1)$ flat connections on 
$T^2$. This observation generalizes to all $O(n)$ elliptic genera, as we 
have elaborated in the earlier part of section 3, by having $7$ discrete sectors 
for $n=2$ and $8$ sectors for $n\geq 3$. It is possible that understanding 
these structures more directly could be a key question to better understand 
the IR conformal field theories on these strings.

With these interesting physics in mind, let us close this paper by 
addressing a few natural extensions of the present work.
First of all it would be nice to see if we can streamline the computation
of the elliptic genus for arbitrary $n$.  Even though our methods
provide an answer, writing it explicitly is cumbersome.
Secondly, it would be interesting to see if we can find
an explicit description of the $(0,4)$ conformal theory they flow to.
Finally it would be interesting to see if we can use our
results to come up with a domain wall description of the E-string
amplitude as in \cite{Haghighat:2014pva}.  Moreover one would like
to use this to show that the partition function of a pair of $n$ E-strings
can lead to the partition function of $n$ heterotic strings
as is predicted by the Horava-Witten description of heterotic string.
Finally it would be interesting to generalize this to other $(1,0)$ superconformal
field theories in 6d, and characterize all the 2d $(0,4)$ systems that one gets
on the worldsheet of the associated strings.

\vskip 0.5cm

\hspace*{-0.8cm} {\bf\large Acknowledgements}
\vskip 0.2cm

\hspace*{-0.75cm} We are grateful to Babak Haghighat, Hee-Cheol Kim,
Jungmin Kim, Sungjay Lee, Guglielmo Lockhart for helpful discussions.
This work is supported in part by the National Research Foundation of Korea
Grants No. 2012R1A1A2042474 (JK,SK), 2012R1A2A2A02046739 (SK),
2015R1A2A2A01003124 (SK), 2006-0093850 (KL), 2009-0084601 (KL),
2012R1A1A2009117 (JP), 2012R1A2A2A06046278 (JP), 2015R1A2A2A01007058 (JP)
and by the NSF grant PHY-1067976 (CV). J.P. also
appreciates APCTP for its stimulating environment for research.

\appendix

\section{Modular forms and Jacobi forms}

A modular form $f_n (\tau)$ of weight $n$ transforms under $SL(2,\mathbb{Z})$ as
\begin{align}
    f_n \left(\frac{a \tau + b}{c \tau + d}\right) = (c \tau + d)^{n} f_n(\tau)\ \ ,\ \ \ \
    ad-bc=1\ .
\end{align}
An important class of modular forms is given by the Eisenstein series,
\begin{align}
  E_{2k}(\tau) = 1 - \frac{4k}{B_{2k}}\sum_{n=1}^\infty\sigma_{2k-1}(n) q^{n},
\end{align}
where $q = e^{2 \pi i \tau}$. The Bernoulli numbers $B_{2k}$ and the divisor functions
$\sigma_k(n)$ are defined by
\begin{align}
	\sum_{k=0}^\infty B_k\frac{x^k}{k!}=\frac{x}{e^x-1}\ ,\ \ \sigma_k(n)=\sum_{d|n}d^k.
\end{align}
$E_{2k}(\tau)$ are modular forms of weight $2k$, expect for $E_2(\tau)$ which involves an anomalous term,
\begin{align}
    E_2 \left(\frac{a \tau + b}{c \tau + d}\right) =
    (c \tau + d)^{2} E_2(\tau) + \frac{6}{i\pi}  c (c \tau + d).
\end{align}
Another example of modular form is the Dedekind eta function $\eta(\tau)$, defined by
\begin{equation}
  \eta(\tau) = q^{1/24}\prod_{n=1}^\infty(1-q^n)\ .
\end{equation}
Under the modular transformation, $\eta(\tau)$ behaves as a weight $\frac{1}{2}$ form up to a phase $\epsilon(a,b,c,d)$,
\begin{align}
    \eta \left(\frac{a \tau + b}{c \tau + d}\right) = \epsilon(a,b,c,d) \cdot (c \tau + d)^{1/2} \eta(\tau).
\end{align}

Jacobi forms have a modular parameter $\tau$ and an elliptic parameter $z$. Modular transformation for Jacobi forms $\phi_{k,m}(\tau, z)$ of weight $k$ and index $m$ is given by
\begin{align}
    \phi_{k,m} \left( \frac{a \tau + b}{c \tau + d}, \frac{z}{c \tau + d} \right) =
    (c \tau + d)^{k} e^{\frac{2\pi i m c z^2}{c \tau + d}} \phi_{k,m}(\tau, z),
\end{align}
Under the translation of the elliptic parameter $z$, they behave as
\begin{align}
    \phi_{k,m} ( \tau, z + a \tau + b) = e^{-2 \pi i m (a^2 \tau + 2 a z)} \phi_{k,m}(\tau, z).
\end{align}
where $a,b$ are integers.

The Jacobi theta function $\vartheta (\tau, z)$ is a Jacobi form of weight $\frac{1}{2}$ and index $\frac{1}{2}$, defined as
\begin{align}
	\vartheta(\tau, z) = \prod_{n=1}^\infty(1-q^n)(1+q^{n-\frac{1}{2}}y)(1+q^{n-\frac{1}{2}}y^{-1}) = \sum_{n \in \mathbb{Z}} q^{n^2/2} y^{n}
\end{align}
where $q \equiv e^{2 \pi i \tau}$ and $y \equiv e^{2 \pi i z}$. We define three other functions
which are closely related to the Jacobi theta function, and define
\begin{alignat}{3}
	\theta_1(\tau, z)&= -i q^{1/8} y^{1/2} \vartheta(\tau, z + \tfrac{1+\tau}{2}) &&= -iq^{1/8}y^{1/2}\prod_{n=1}^\infty(1-q^n)(1-q^ny)(1-q^{n-1}y^{-1})\nonumber\\
	\theta_2(\tau, z)&= q^{1/8} y^{1/2} \vartheta(\tau, z + \tfrac{\tau}{2}) &&= q^{1/8}y^{1/2}\prod_{n=1}^\infty(1-q^n)(1+q^ny)(1+q^{n-1}y^{-1})\nonumber\\
	\theta_3(\tau, z)&= \vartheta(\tau, z) &&= \prod_{n=1}^\infty(1-q^n)(1+q^{n-\frac{1}{2}}y)(1+q^{n-\frac{1}{2}}y^{-1})\nonumber\\
	\theta_4(\tau, z)&= \vartheta(\tau, z + \tfrac{1}{2}) &&= \prod_{n=1}^\infty(1-q^n)(1-q^{n-\frac{1}{2}}y)(1-q^{n-\frac{1}{2}}y^{-1}).
\end{alignat}
From here, when we omit the modular parameter in various functions, it should be understood as $\tau$.
$\theta_n (z)$'s are related to others by the half-period shifts:
\begin{align}\label{elliptic-half-period}
  &\theta_1(z+\tfrac{1}{2})=\theta_2(z) &&\theta_1(z+\tfrac{1+\tau}{2})=q^{-1/8}y^{-1/2}\theta_3(z) &&\theta_1(z+\tfrac{\tau}{2})=iq^{-1/8}y^{-1/2}\theta_4(z) \nonumber\\
  &\theta_2(z+\tfrac{1}{2})=-\theta_1(z) &&\theta_2(z+\tfrac{1+\tau}{2})
  =-iq^{-1/8}y^{-1/2}\theta_4(z) &&\theta_2(z+\tfrac{\tau}{2})=q^{-1/8}y^{-1/2}\theta_3(z) \nonumber\\
  &\theta_3(z+\tfrac{1}{2})=\theta_4(z) &&\theta_3(z+\tfrac{1+\tau}{2})
  =i q^{-1/8}y^{-1/2}\theta_1(z) &&\theta_3(z+\tfrac{\tau}{2})=q^{-1/8}y^{-1/2}\theta_2(z) \nonumber\\
  &\theta_4(z+\tfrac{1}{2})=\theta_3(z) &&\theta_4(z+\tfrac{1+\tau}{2})
  =q^{-1/8}y^{-1/2}\theta_2(z) &&\theta_4(z+\tfrac{\tau}{2})=iq^{-1/8}y^{-1/2}\theta_1(z)
\end{align}

\paragraph{Various identities:} The modular forms $E_4$, $E_6$, and $\eta$ can be expressed in terms of
Jacobi theta functions with their elliptic parameters $z$ set to zero:
\begin{eqnarray}\label{eq:Eisenstein-and-eta-using-theta}
    E_4 &=& \tfrac{1}{2} (\theta_2(0)^8 + \theta_3(0)^8 + \theta_4(0)^8)\nonumber\\
    E_6 &=& \tfrac{1}{2}(\theta_2(0)^4 + \theta_3 (0)^4) (\theta_3(0)^4 + \theta_4 (0)^4)(\theta_4(0)^4 - \theta_2 (0)^4)\nonumber\\
    2\eta^3 &=& \theta_2 (0) \theta_3 (0) \theta_4 (0).
\end{eqnarray}
$\theta_n (z)$'s also satisfy
\begin{align}\label{theta^4}
    \theta_2(z)^4 - \theta_1(z)^4 = \theta_3(z)^4 - \theta_4(z)^4\ ,\ \
    \theta_2(0)^4=\theta_3(0)^4-\theta_4(0)^4\ .
\end{align}
Further identities of $\theta_n (z)$'s with different elliptic parameters are
\begin{align}
	\label{eq:theta-addition-type1}
    &\theta_1 (a + b) \theta_1 (a - b) \theta_4(0)^2 = \theta_3(a)^2 \theta_2(b)^2 - \theta_2(a)^2 \theta_3(b)^2 = \theta_1(a)^2 \theta_4(b)^2 -\theta_4(a)^2 \theta_1(b)^2 \\
    &\theta_3 (a + b) \theta_3 (a - b) \theta_2(0)^2 = \theta_3(a)^2 \theta_2(b)^2 + \theta_4(a)^2 \theta_1(b)^2 = \theta_2(a)^2 \theta_3(b)^2 + \theta_1(a)^2 \theta_4(b)^2 \nonumber \\
    &\theta_3 (a + b) \theta_3 (a - b) \theta_3(0)^2 = \theta_1(a)^2 \theta_1(b)^2 + \theta_3(a)^2 \theta_3(b)^2 = \theta_2(a)^2 \theta_2(b)^2 + \theta_4(a)^2 \theta_4(b)^2 \nonumber \\
    &\theta_3 (a + b) \theta_3 (a - b) \theta_4(0)^2 = \theta_4(a)^2 \theta_3(b)^2 - \theta_1(a)^2 \theta_2(b)^2 = \theta_3(a)^2 \theta_4(b)^2 -\theta_2(a)^2 \theta_1(b)^2 \nonumber \\
    \label{eq:theta-duplication-identity}
 	&\theta_1 (a \pm b) \theta_2 (a \mp b) \theta_3(0) \theta_4(0) = \theta_1(a) \theta_2(a) \theta_3(b) \theta_4(b) \pm \theta_3(a) \theta_4(a) \theta_1(b) \theta_2(b)\\
 	&\theta_1 (a \pm b) \theta_3 (a \mp b) \theta_2(0) \theta_4(0) = \theta_1(a) \theta_3(a) \theta_2(b) \theta_4(b) \pm \theta_2(a) \theta_4(a) \theta_1(b) \theta_3(b) \nonumber\\
 	&\theta_1 (a \pm b) \theta_4 (a \mp b) \theta_2(0) \theta_3(0) = \theta_1(a) \theta_4(a) \theta_2(b) \theta_3(b) \pm \theta_2(a) \theta_3(a) \theta_1(b) \theta_4(b) \nonumber
\end{align}
Remaining identities of this kind can be obtained through half-period shifts on $a$.

Under the shift of modular parameter $\tau \rightarrow \tau' = \tau + 1$, the corresponding changes are
\begin{equation}\label{eq:modular-parameter-shift}
    \theta_1 (\tau+1, z) = e^{i\frac{\pi}{4}} \theta_1 (\tau, z),\
    \theta_2 (\tau+1, z) = e^{i\frac{\pi}{4}} \theta_2 (\tau, z),\
    \theta_3 (\tau+1, z) = \theta_4 (\tau, z),\
    \theta_4 (\tau+1, z) = \theta_3 (\tau, z).
\end{equation}
Watson's identities and Landen's formulas involve doubling of modular parameter $\tau$,
\begin{align}
	\label{eq:watson-identity}
    \theta_1 (\tau, z) \theta_1 (\tau, w) &= \theta_3 (2\tau, z+w) \theta_2 (2\tau, z-w) - \theta_2 (2\tau, z+w) \theta_3 (2\tau, z-w)\\
    \theta_3 (\tau, z) \theta_3 (\tau, w) &= \theta_3 (2\tau, z+w) \theta_3 (2\tau, z-w) + \theta_2 (2\tau, z+w) \theta_2 (2\tau, z-w) \nonumber \\
    \label{eq:landen-formula}
    \theta_1 (2 \tau, 2z) &= \theta_1 (\tau, z) \theta_2 (\tau, z) / \theta_4 (2\tau, 0)\\
    \theta_4 (2 \tau, 2z) &= \theta_3 (\tau, z) \theta_4 (\tau, z) / \theta_4 (2\tau, 0) \nonumber
\end{align}
Considering these identities at $z = 0$ or $z=w=0$, and also using
the second identity of (\ref{theta^4}), one obtains
\begin{align}
	\label{eq:theta-doubling-z-0}
    \theta_2 (2 \tau, 0) = \sqrt{\tfrac{\theta_3(\tau,0)^2-\theta_4(\tau,0)^2}{2}}, \,
    \theta_3 (2 \tau, 0) = \sqrt{\tfrac{\theta_3(\tau,0)^2+\theta_4(\tau,0)^2}{2}}, \,
    \theta_4 (2 \tau, 0) = \sqrt{\theta_3(\tau,0) \theta_4(\tau,0)}.
\end{align}

\paragraph{Differentiations by $\tau,z$:}
The $\tau$ derivatives of $E_2,E_4,E_6$ can be obtained from the Ramanujan identities
\begin{align}
	\label{eq:Eisenstein-Ramanujan-identity}
    q \frac{d}{dq} E_2 = \frac{1}{12} (E_2^2 - E_4),\ q \frac{d}{dq} E_4 = \frac{1}{3} (E_2 E_4 - E_6), \ q \frac{d}{dq} E_6 = \frac{1}{2} (E_2 E_6 - E_4^2).
\end{align}
The $\tau$ derivative of the eta function is given by
\begin{align}
	\label{eq:eta-derivative-eisenstein}
	q\frac{d}{dq} \eta^3 = \frac{\eta^3}{8} E_2.
\end{align}
As for the theta functions, first note that $\theta_n (z)$'s are solutions of
\begin{align}
	\label{eq:heat-equation-theta}
    \left[\frac{1}{(2\pi i)^2}\frac{\partial^2}{\partial z^2} - \frac{1}{i \pi}\frac{\partial}{\partial \tau} \right] \theta_n (\tau, z) = \left[\frac{1}{(2\pi i)^2}\frac{\partial^2}{\partial z^2} - 2 q \frac{\partial}{\partial q} \right]\theta_n (\tau, z) = 0.
\end{align}
$\theta_1$ is an odd function of $z$, while $\theta_2,\theta_3,\theta_4$ are even functions
of $z$. The lowest non-vanishing derivatives of $\theta_n$'s at $z=0$ are given by
\begin{align}
	\label{eq:theta-lowest-derivatives}
    \theta_1^{(1)} (0) &= 2 \pi \eta^3 &\theta_2^{(2)} (0) &= -\tfrac{\pi^2}{3} \theta_2 (0) \,(E_2 + \theta_3(0)^4 + \theta_4(0)^4)\\
    \theta_3^{(2)} (0) &= -\tfrac{\pi^2}{3} \theta_3 (0) \, (E_2 + \theta_2(0)^4 - \theta_4(0)^4) &\theta_4^{(2)} (0) &= -\tfrac{\pi^2}{3} \theta_4 (0) \, (E_2 - \theta_2(0)^4 - \theta_3(0)^4) \nonumber\ ,
\end{align}
where the superscript
$(n)$ denotes $n$'th derivative with respect to the elliptic parameter.
Using (\ref{eq:heat-equation-theta}), (\ref{eq:theta-lowest-derivatives}), \eqref{eq:Eisenstein-Ramanujan-identity} and (\ref{eq:eta-derivative-eisenstein}),
one can also express the higher $z$ derivatives $\theta_1^{(2n+1)}(0)$, $\theta_2^{(2n)}(0)$,
$\theta_3^{(2n)}(0)$, $\theta_4^{(2n)}(0)$ at $z=0$ in terms of $\theta_2(0)$, $\theta_3(0)$,
$\theta_4(0)$, $E_2$. See appendix C for more details, where this procedure will be illustrated
and used to prove exact properties of the E-string elliptic genera.

\section{Genus expansions of topological string amplitudes}\label{sec:genus-expansion-data}

In this appendix, we summarize some low genus results that we used in section 3.
The low genus amplitudes have been studied in \cite{Klemm:1996hh,Minahan:1997ct,Minahan:1998vr,Mohri:2001zz,Huang:2013yta}.
We list the unrefined results till $g\leq 5$
(as written in \cite{Mohri:2001zz}), and some refined results that we used to
compare with our results.

For three E-strings, the unrefined genus expansion coefficients
$F^{(0,g,3)}$ are given by
\begin{eqnarray}\label{eq:genus-3E}
  F^{(0,0,3)}&=&\frac{54E_2^2E_4^3+216E_2E_4^2E_6+109E_4^4+197E_4E_6^2}{15552\eta^{36}}\\
  F^{(0,1,3)}&=&\frac{78E_2^3E_4^3+299E_2E_4^4+360E_2^2E_4^2E_6+472E_4^3E_6
  +439E_2E_4E_6^2+80E_6^3}{62208\eta^{36}}\nonumber\\
  F^{(0,2,3)}&=&\frac{1}{2488320 \eta^{36}}\left(575 E_2^4 E_4^3 + 3040 E_2^3 E_4^2 E_6
  + 4690 E_2^2 E_4 E_6^2 + 3548 E_2^2 E_4^4 \right.\nonumber\\
  &&\hspace{2.3cm}\left.+ 1600 E_6^3 E_2 + 10176 E_6 E_4^3 E_2 + 2231 E_4^5 +
  5244 E_4^2 E_6^2\right)\nonumber\\
  F^{(0,3,3)}&=&\frac{1}{209018880\eta^{36}}\left(
  138104 E_4^4 E_6 + 224024 E_6 E_4^3 E_2^2 + 36400 E_2^4 E_4^2 E_6 +
 224456 E_4^2 E_6^2 E_2\right.\nonumber\\
 &&\left. + 49584 E_4 E_6^3 + 68460 E_2^3 E_4 E_6^2 +
 55006 E_2^3 E_4^4 + 6055 E_2^5 E_4^3 + 97431 E_4^5 E_2 +
 33600 E_6^3 E_2^2
  \right)\nonumber\\
  F^{(0,4,3)}&=&\frac{1}{75246796800 \eta^{36}}\left(3164700 E_2^4 E_4 E_6^2 +
   8993259 E_4^5 E_2^2 + 14111840 E_6^2 E_4^3 + 806400 E_6^4\right.\nonumber\\
   &&\left.+ 25171632 E_2 E_6 E_4^4 + 13855280 E_2^3 E_6 E_4^3 + 8963520 E_2 E_6^3 E_4
   + 20453520 E_2^2 E_6^2 E_4^2\right.\nonumber\\
   &&\left. + 4014627 E_4^6 + 208985 E_2^6 E_4^3 +
   2016000 E_6^3 E_2^3 + 1417920 E_2^5 E_4^2 E_6 + 2638125 E_2^4 E_4^4\right)\nonumber
\end{eqnarray}
\begin{eqnarray}
  F^{(0,5,3)}&=&\frac{1}{9932577177600 \eta^{36}}\left(935093824 E_6^2 E_4^3 E_2 +
   233170300 E_2^4 E_6 E_4^3 + 296640960 E_2^2 E_6^3 E_4\right.\nonumber\\
   &&\left. +  837550728 E_2^2 E_6 E_4^4 + 453680480 E_2^3 E_6^2 E_4^2 +
   16385600 E_2^6  E_4^2 E_6 + 42513240 E_2^5 E_4 E_6^2\right.\nonumber\\
   &&\left.  +   201151929  E_4^5 E_2^3 + 36275085 E_2^5 E_4^4 + 53222400 E_6^4 E_2
   +266767491 E_4^6 E_2\right.\nonumber\\
   &&\left. + 405268284 E_4^5 E_6 + 268326944 E_4^2 E_6^3 +
   33264000 E_6^3 E_2^4 + 2155615 E_2^7 E_4^3\right)\ .\nonumber
\end{eqnarray}
A refined coefficient $F^{(1,0,3)}$ that we studied in section 3.3 is given by
\begin{equation}\label{eq:genus-3E-refined}
  F^{(1,0,3)}=-\frac{54E_2^3E_4^3+235E_2E_4^4+216E_2^2E_4^2E_6+
  776E_4^3E_6+287E_2E_4E_6^2+160E_6^3}{124416\eta^{36}}\ .
\end{equation}

For the four E-strings, $F^{(0,g,4)}$ are given
as follows (after correcting some typos in \cite{Mohri:2001zz}):
\begin{eqnarray}\label{eq:genus-4E}
  F^{(0,0,4)}&=&\frac{1}{62208\eta^{48}} E_4 \left(272 E_4^3 E_6 + 154 E_6^3 + 109 E_2 E_4^4 +
  269 E_2 E_4 E_6^2 + 144 E_2^2 E_4^2 E_6 + 24 E_2^3 E_4^3\right)\nonumber\\
  F^{(0,1,4)}&=&\frac{1}{11943936 \eta^{48}} \left(37448 E_2^2 E_4^2 E_6^2 + 68768 E_2 E_4^4 E_6 +
  29920 E_2 E_4 E_6^3 + 13809 E_4^6\right.\nonumber\\
  &&\left. + 57750 E_4^3 E_6^2 + 17416 E_2^2  E_4^5 + 4545 E_6^4 + 16704 E_2^3  E_4^3 E_6 +
  2472 E_2^4 E_4^4\right)\nonumber\\
  F^{(0,2,4)}&=&\frac{1}{179159040\eta^{48}}\left(77280 E_2^4  E_6  E_4^3 +
   209200  E_2^2  E_6^3  E_4 + 547760 E_2^2  E_6 E_4^4 + 214811 E_4^6  E_2\right.\nonumber\\
   &&\left. +   203900 E_2^3  E_6^2  E_4^2  + 103252 E_4^5  E_2^3  +
   827230 E_6^2  E_4^3  E_2 + 10200 E_2^5  E_4^4  + 57375 E_6^4  E_2\right.\nonumber\\
   &&\left.  +   420616 E_4^5  E_6 + 314360 E_4^2  E_6^3\right)\nonumber\\
  F^{(0,3,4)}&=&\frac{1}{90296156160\eta^{48}}\left(28134630 E_4^7 + 151049093 E_4^4  E_6^2 +
   25488295 E_4 E_6^4 + 966630 E_2^6  E_4^4 \right.\nonumber\\
   &&\left.+   189296376 E_6^2  E_4^3  E_2^2 + 8172360 E_2^5  E_6 E_4^3 +
   31388000 E_2^3  E_6^3  E_4 + 88718416 E_2^3  E_6 E_4^4 \right.\nonumber\\
   &&\left. +
   24977155 E_2^4  E_6^2  E_4^2 + 13366787 E_4^5 E_2^4 +
   12119625 E_6^4  E_2^2 + 137926976 E_4^2  E_6^3  E_2 \right.\nonumber\\
   &&\left. +  51557313 E_4^6  E_2^2  + 192353224 E_4^5  E_6 E_2 \right)\nonumber\\
  F^{(0,4,4)}&=&\frac{1}{5417769369600 \eta^{48}}\left(3336940980 E_2^3  E_4^3  E_6^2 +
   7817234620 E_2 E_6^2  E_4^4  + 3248768730 E_6^3  E_4^3\right.\nonumber\\
   &&\left. +   5085796952 E_2^2 E_4^5  E_6 + 101280375 E_6^5 +
   3550525000 E_2^2  E_4^2  E_6^3 + 1290318725 E_2 E_4 E_6^4 \right.\nonumber\\
   &&\left.+   936363912 E_4^6  E_2^3 + 1481276055 E_4^7 E_2 + 2912603799 E_4^6  E_6 +
   1216807640 E_2^4  E_4^4  E_6 \right.\nonumber\\
   &&\left.+ 152620090 E_2^5  E_4^5 +
   78676080 E_2^6  E_6 E_4^3 + 410158000 E_2^4 E_6^3  E_4 +
   274844990 E_2^5  E_6^2  E_4^2\right.\nonumber\\
   &&\left. + 8381520 E_2^7  E_4^4 + 202702500 E_6^4  E_2^3 \right)\nonumber\\
  F^{(0,5,4)}&=&\frac{1}{2860582227148800\eta^{48}}\left(12207942670 E_2^6  E_4^5
  + 523849095 E_2^8  E_4^4 +
   156150752805 E_4^8 \right.\nonumber\\
   &&\left.+ 113811930320 E_2^5 E_4^4 E_6 +
   1311485716360 E_4^6 E_6 E_2 + 1760563778482 E_2^2  E_6^2  E_4^4\right.\nonumber\\
   &&\left. +   286289201000 E_2^2  E_4 E_6^4 + 381058740370 E_2^4  E_4^3  E_6^2 +
   1449394307792 E_6^3  E_4^3  E_2\right.\nonumber\\
   &&\left. + 1106487740990 E_6^2 E_4^5 +
   44575839000 E_6^5  E_2 + 109025587484 E_4^6  E_2^4\right.\nonumber\\
   &&\left. +
   774483173328 E_2^3  E_4^5  E_6 + 531170439360 E_2^3  E_4^2  E_6^3 +
   5431290480 E_2^7 E_6 E_4^3\right.\nonumber\\
   &&\left. + 37160939200 E_2^5  E_6^3  E_4 +
   337421738130 E_4^7  E_2^2 + 21439577390 E_2^6  E_6^2  E_4^2 \right.\nonumber\\
   &&\left.+
   22344052500 E_6^4  E_2^4 + 344998537324 E_6^4  E_4^2\right)\ .
\end{eqnarray}

\section{Exact properties of the E-string elliptic genus}

We explain the details on how we checked various exact properties of our E-string elliptic genera,
using the identities of appendix A. We made lots of symbolic computations using computer.
Below, we explain how one can simplify various expressions which can be put on a computer for
further simplifications.

\paragraph{2 E-strings} We compare the two expressions for the elliptic genus of 2 E-strings,
\eqref{2E-gauge} and \eqref{2E-E8}. Let us denote them by $Z_2$ and $Z_2^\text{E8}$ respectively,
in the sense that the latter expression shows manifest $E_8$ symmetry. After setting
$\epsilon_1 = - \epsilon_2 = \epsilon$ for simplicity, $Z_2$ is given by
\begin{align}
    &Z_2 = \sum_{n=1}^4\frac{\prod_{l=1}^{8} \theta_n (m_l \pm \frac{\epsilon}{2})}{2 \eta^{12} \theta_1(\epsilon)^2 \theta_1(2\epsilon)^2} + \frac{1}{4 \eta^{12} \theta_1(\epsilon)^4}\Bigg[ \frac{\theta_2(0)^2}{\theta_2(\epsilon)^2}\Big( \prod_{l=1}^{8} \theta_1 (m_l)\theta_2 (m_l) + \prod_{l=1}^{8} \theta_3 (m_l)\theta_4 (m_l) \Big) \\
    &+ \frac{\theta_4(0)^2}{\theta_4(\epsilon)^2}\Big( \prod_{l=1}^{8} \theta_1 (m_l)\theta_4 (m_l) + \prod_{l=1}^{8} \theta_2 (m_l)\theta_3 (m_l) \Big) + \frac{\theta_3(0)^2}{\theta_3(\epsilon)^2}\Big( \prod_{l=1}^{8} \theta_1 (m_l)\theta_3 (m_l) + \prod_{l=1}^{8} \theta_2 (m_l)\theta_4 (m_l) \Big)   \Bigg] \nonumber.
\end{align}
Using the identity \eqref{eq:theta-duplication-identity} with $a = b$, one can write
$Z_2 = \frac{N(\tau,z,m_l)}{\eta^{12} \theta_1(\epsilon)^2 \theta_1(2\epsilon)^2}$ with
\begin{align}
    &N= \sum_{n=1}^4\frac{1}{2}\prod_{l=1}^{8} \theta_n (m_l \pm \frac{\epsilon}{2}) + \frac{\theta_3(\epsilon)^2 \theta_4(\epsilon)^2}{\theta_3(0)^2 \theta_4(0)^2} \Big( \prod_{l=1}^{8} \theta_1 (m_l)\theta_2 (m_l) + \prod_{l=1}^{8} \theta_3 (m_l)\theta_4 (m_l) \Big) +\\
    & \frac{\theta_2(\epsilon)^2 \theta_3(\epsilon)^2}{\theta_2(0)^2 \theta_3(0)^2}
    \Big(\!\prod_{l=1}^{8} \theta_1 (m_l)\theta_4 (m_l)\! +\! \prod_{l=1}^{8} \theta_2 (m_l)\theta_3 (m_l) \!\Big)
    +  \frac{\theta_2(\epsilon)^2 \theta_4(\epsilon)^2}{\theta_2(0)^2 \theta_4(0)^2}
    \Big(\! \prod_{l=1}^{8} \theta_1 (m_l)\theta_3 (m_l)\! +\! \prod_{l=1}^{8} \theta_2 (m_l)\theta_4 (m_l) \!\Big). \nonumber
\end{align}
We apply \eqref{eq:theta-addition-type1} to the first term of $N$,
where we take $a=m_l$, $b=\epsilon/2$. Then $N$ can be expressed as a polynomial
of $\theta_n (m_l)$, $\theta_n(\epsilon)$ and $\theta_n (\epsilon/2)$, with coefficients given by
$\theta_n(0)$.

On the other hand, expressing \eqref{2E-E8} as
$Z_2^\text{E8} = N^\text{E8} / (\eta^{12} \theta_1(\epsilon)^2 \theta_1(2\epsilon)^2)$, we consider
\begin{align}\label{E8-numerator}
    N^\text{E8} &= \frac{1}{72} A_1^2 (\phi_{0,1}(\epsilon)^2 - E_4 \phi_{-2,1}(\epsilon)^2) + \frac{1}{96} A_2 (E_4^2 \phi_{-2,1}(\epsilon)^2 - E_6 \phi_{-2,1}(\epsilon)\phi_{0,1}(\epsilon)) \nonumber \\
    &+ \frac{5}{288} B_2 (E_6 \phi_{-2,1}(\epsilon)^2 - E_4 \phi_{-2,1}(\epsilon)\phi_{0,1}(\epsilon)).
\end{align}
We first insert \eqref{eq:Eisenstein-and-eta-using-theta} to replace $E_4$, $E_6$, $\eta$
by expressions containing $\theta_2(0)$, $\theta_3(0)$, $\theta_4(0)$ only.
Looking at the definition of $A_2$ and $B_2$ in \eqref{eq:e8weyl-inv}, there appear $\theta_n (\tfrac{\tau}{2}, m_l)$ and $\theta_n (\tfrac{\tau+1}{2}, m_l)$. To simplify them,
we first consider the identities,
\begin{align}\label{tau/2}
    \hspace*{-.2cm}\theta_1 (\tfrac{\tau}{2},m_1)\theta_1 (\tfrac{\tau}{2},m_2) &= \theta_3 (\tau,m_1\!+\!m_2) \theta_2 (\tau,m_1\!-\!m_2) - \theta_2 (\tau,m_1+m_2) \theta_3 (\tau,m_1-m_2)\\
    \hspace*{-.2cm}\theta_1 (\tfrac{\tau+1}{2},m_1)\theta_1 (\tfrac{\tau+1}{2},m_2) &= e^{i \pi/4}\theta_4 (\tau,m_1+m_2) \theta_2 (\tau,m_1-m_2) - e^{i \pi/4}\theta_2 (\tau,m_1+m_2) \theta_4 (\tau,m_1-m_2)\nonumber\ .
\end{align}
The first identity can be obtained by replacing $\tau,z,w$ in (\ref{eq:watson-identity}) by
$\frac{\tau}{2},m_1,m_2$, respectively, and the second one is obtained from the first identity
by using \eqref{eq:modular-parameter-shift}. One can also obtain three more copies of similar
identities, replacing $\theta_1$ on the left hand sides by $\theta_2,\theta_3,\theta_4$,
by using (\ref{elliptic-half-period}). The expressions appearing on the right hand sides of
(\ref{tau/2}) can be written as polynomials of $\theta_n(\tau,m_l)$ by using
(\ref{eq:theta-duplication-identity}). We apply these identities, and also those with $(m_1,m_2)$
replaced by $(m_3,m_4)$, $(m_5,m_6)$, $(m_7,m_8)$, to (\ref{E8-numerator}). Then one can express all
theta functions with modular parameters $\frac{\tau}{2}$ or $\frac{\tau+1}{2}$ in terms of
$\theta_n(\tau,m_l)$. Other terms including $\theta_n (2\tau, 2m_l)$ can be reorganized using \eqref{eq:landen-formula} and \eqref{eq:theta-doubling-z-0}, in terms of $\theta_n(\tau,m_l)$
and $\theta_n(\tau,0)$. So finally, $N^{E_8}$ is written as a polynomial of
$\theta_n(\tau,m_l)$, $\theta_n(\tau,\epsilon)$, with coefficients given by $\theta_n(\tau,0)$.

Finally, to straightforwardly compare $N$ and $N^{E_8}$, we want to express
$\theta_n(\epsilon)$'s in terms of $\theta_n(\epsilon/2)$'s. Plugging $b=\frac{\epsilon}{2}$
and $a=\frac{\epsilon}{2}+\frac{p}{2}$ (with $p=0,1,\tau,\tau+1$) into \eqref{eq:theta-addition-type1} and \eqref{eq:theta-duplication-identity}, one obtains
the desired formulae. Then inserting them into
$N,N^{E_8}$, we obtain polynomials of $\theta_n(\tau,m_l)$, $\theta_n(\tau,\frac{\epsilon}{2})$
with coefficients given by $\theta_n(\tau,0)$. Now we can evaluate $N^{E_8}-N$ on computer, by eliminating $\theta_1(m_l)$, $\theta_1(\epsilon/2)$, $\theta_2(0)$ by using (\ref{theta^4}). Then one finds $N^{E_8}-N=0$,
proving the equivalence of \eqref{2E-gauge} and \eqref{2E-E8}.

\paragraph{3 and 4 E-strings} We compare our elliptic genera \eqref{eq:elliptic-genus-3E} and \eqref{eq:elliptic-genus-4E} against the known results summarized in Appendix \ref{sec:genus-expansion-data}. The free energy is expanded as
\begin{align}
    F = \log{Z} = \sum_{n_b=1}^{\infty} w^{n_b} F_{n_b} = \sum_{n,g,n_b} (\epsilon_1 + \epsilon_2)^{2n} (\epsilon_1 \epsilon_2)^{g-1} w^{n_b} F^{(n,g,n_b)}\ ,
\end{align}
where $F_1=Z_1$, $F_2=Z_2-\frac{1}{2}Z_1^2$,
$F_3 = Z_3 - Z_1 Z_2 + \tfrac{1}{3} Z_1^{3}$ and
$F_4 = Z_4 - Z_1 Z_3 - \tfrac{1}{2} Z_2^{2} + Z_1^2 Z_2 - \tfrac{1}{4} Z_1^{4}$.
The coefficients $F^{(n,g,n_b)}$ computed from topological strings, summarized
in appendix B, depend on $\eta$, $E_2$, $E_4$, $E_6$. Using
(\ref{eq:Eisenstein-and-eta-using-theta}), these can be arranged into expressions
involving $E_2$ and $\theta_n(0)$ only.

On the other hand, if we set $m_l=0$ and compute $F^{(n,g,n_b)}$
from our gauge theory indices,
they will be rational functions of $\theta_n(0)$, $\eta$, $\theta^{(k)}_n(0)$. The derivatives $\theta_n^{(k)}(0)$ appear because we are expanding the index with $\epsilon_1,\epsilon_2$.
We want to express our gauge theory expressions for $F^{(n,g,n_b)}$ in terms of $\theta_n(0)$'s and $E_2$ only, to compare with the results summarized in appendix B. Firstly, \eqref{eq:Eisenstein-and-eta-using-theta} can be used to eliminate $\eta$. The remaining
task is to write $\theta^{(k)}_{1,2,3,4}(0)$ in terms of $\theta_n(0)$'s and $E_2$,
which can be done in the following way.

Starting from the lowest non-vanishing derivatives \eqref{eq:theta-lowest-derivatives}
at $z=0$, we can iteratively obtain
$\theta^{(k)}_{n}(0)$ for higher $k$'s. For example,
\begin{align}
    (\partial_z)^3 \theta_1 (\tau,z) |_{z=0} &= -8 \pi^2 (\partial_z) (q \partial_q) \theta_1 (\tau,z) |_{z=0} = -8 \pi^2 (q \partial_q) (\partial_z  \theta_1 (\tau,z)) |_{z=0} \nonumber\\
     &= - 16 \pi^3 (q \partial_q) \eta^3 = - 2 \pi^3 \eta^3 E_2
\end{align}
where \eqref{eq:heat-equation-theta} and \eqref{eq:eta-derivative-eisenstein} are applied at the last step. If we look at another example,
\begin{align}
    (\partial_z)^4 \theta_2 (\tau,z) |_{z=0} &= -8 \pi^2  (\partial_z)^2 (q \partial_q) \theta_2 (\tau,z) |_{z=0} = -8 \pi^2 (q \partial_q) (\partial_z^2  \theta_2 (\tau,z)) |_{z=0} \nonumber\\
     &= \tfrac{8 }{3}\pi^4\, q \partial_q [\theta_2(0) \cdot (E_2 + \theta_3(0)^4 + \theta_4(0)^4)] \nonumber\\
     &= \tfrac{1}{9} \pi^4 \theta_2 (0) [\alpha_2^2 + 4 \theta_3(0)^4 \alpha_3 + 4 \theta_4(0)^4 \alpha_4 + \tfrac{1}{12}(E_2^2 - E_4)].
\end{align}
for $\alpha_2 \equiv E_2 + \theta_3(0)^4 + \theta_4(0)^4$, $\alpha_3 \equiv E_2 + \theta_2(0)^4 - \theta_4(0)^4$, and $\alpha_4 \equiv E_2 - \theta_2(0)^4 - \theta_3(0)^4$. At the last step, we applied \eqref{eq:heat-equation-theta} and \eqref{eq:Eisenstein-Ramanujan-identity}. Going to higher derivatives involves no more difficulty, and this way we can always express $F^{(n,g,n_b)}$ in terms of $\theta_n(0)$'s
and $E_2$ only.

So we find two expressions for $F^{(n,g,n_b)}$, depending on $\theta_n(0)$'s and $E_2$
only, one from the topological string calculus and another from our gauge theories.
In particular, we focus on the 3 and 4 E-strings, obtained by expanding (\ref{eq:elliptic-genus-3E}), \ (\ref{eq:elliptic-genus-4E}). We computed the
differences of the two expressions for $F^{(0,0,3)}$, $F^{(0,1,3)}$, $F^{(1,0,3)}$,
$F^{(0,0,4)}$, $F^{(0,1,4)}$, $F^{(0,2,4)}$ on computer, substituting $\theta_2(0)^4=\theta_3(0)^4 - \theta_4(0)^4$, and found zero in all cases.
Of course, further analytic tests can also be easily made on computer for higher genus results.

\end{document}